\documentclass[preprint,12pt]{elsarticle}

\usepackage{amssymb}
\usepackage{amsthm}

\usepackage{float}
\usepackage{color}
\usepackage{amsmath}

\usepackage{graphicx, subfigure}
\usepackage{pslatex}
\usepackage{relsize}
\usepackage{bm}
\usepackage{verbatim}

\usepackage{booktabs}
\usepackage{xcolor}
\usepackage{soul}

\usepackage[utf8]{inputenc}
\usepackage[T1]{fontenc}

\biboptions{numbers,sort&compress}

\newcounter{bla}

\journal{Computer Physics Communications}

\begin{document}

\begin{frontmatter}



\title{MAELAS 2.0: A new version of a computer program  for the calculation of magneto-elastic properties}


\author[a]{P.~Nieves\corref{author}}
\author[a]{S.~Arapan}
\author[b,c]{S.~H.~Zhang}
\author[a]{A.~P.~K\k{a}dzielawa}
\author[b,c]{R.~F.~Zhang }
\author[a]{D.~Legut}

\cortext[author] {Corresponding author.\\\textit{E-mail address:} pablo.nieves.cordones@vsb.cz}
\address[a]{IT4Innovations, V\v{S}B - Technical University of Ostrava, 17. listopadu 2172/15, 70800 Ostrava-Poruba, Czech Republic}
\address[b]{School of Materials Science and Engineering, Beihang University, Beijing 100191, PR China}
\address[c]{Center for Integrated Computational Materials Engineering, International Research Institute for Multidisciplinary Science, Beihang University, Beijing
100191, PR China}
%\address[d]{Instyut Fizyki Teoretycznej, Uniwersytet Jagielloński, \L{}ojasiewicza 11, 30-348 Krak\'ow, Poland}

\begin{abstract}

MAELAS is a computer program for the calculation of magnetocrystalline anisotropy energy, anisotropic magnetostrictive coefficients and magnetoelastic constants in an automated way. The method originally implemented in version 1.0 of MAELAS was based on the length optimization of the unit cell, proposed by Wu and Freeman, to calculate the anisotropic magnetostrictive coefficients. We present here a revised and updated version (v2.0) of MAELAS, where we added a new methodology to compute anisotropic magnetoelastic constants from a linear fitting of the energy versus applied strain. We analyze and compare the accuracy of both methods showing that the new approach is more reliable and robust than the one implemented in version 1.0, especially for non-cubic crystal symmetries. This analysis also help us to find that the accuracy of the method implemented in version 1.0 could be improved by using deformation gradients derived from the equilibrium magnetoelastic strain tensor, as well as potential future alternative methods like the strain optimization method. Additionally, we clarify the role of the demagnetized state in the fractional change in length, and derive the expression for saturation magnetostriction for polycrystals with trigonal, tetragonal and orthorhombic crystal symmetry. In this new version, we also fix some issues related to trigonal crystal symmetry found in version 1.0.

\end{abstract}


\end{frontmatter}

% \linenumbers

% All CPiP articles must contain the following
% PROGRAM SUMMARY.

{\bf PROGRAM SUMMARY}
  %Delete as appropriate.

\begin{small}
\noindent
{\em Program Title:} MAELAS 
\\
{\em Developer's respository link:} https://github.com/pnieves2019/MAELAS 
\\
{\em Licensing provisions:} BSD 3-clause 
\\
{\em Programming language:} Python3 
\\
{\em Journal reference of previous version:} P. Nieves, S. Arapan, S.H. Zhang, A.P. Kądzielawa, R.F. Zhang and D. Legut, Comput. Phys. Commun. 264, 107964 (2021)
\\
  %Only required for a New Version summary, otherwise leave out.
{\em Does the new version supersede the previous version?:} Yes   
\\
  %Only required for a New Version summary, otherwise leave out.
{\em Reasons for the new version:}
To implement a more accurate methodology to compute magnetoelastic constants and magnetostrictive coefficients, and fix some issues related to trigonal crystal symmetry.
\\
  %Only required for a New Version summary, otherwise leave out.
{\em Summary of revisions: } 
\begin{itemize}
    \item New method to calculate magnetoelastic constants and magnetostrictive coefficients derived from the magnetoelastic energy. 
    \item Correction of the trigonal crystal symmetry. 
    \item Implementation of the saturation magnetostriction for polycrystals with trigonal, tetragonal and orthorhombic crystal symmetry.
    \item In the visualization tool MAELASviewer, we included the possibility to choose the type of reference demagnetized state in the calculation of the fractional change in length
\end{itemize}
%\\
  %Only required for a New Version summary, otherwise leave out.
{\em Nature of problem:} To calculate anisotropic magnetostrictive coefficients and magnetoelastic constants in an automated way based on Density Functional Theory methods.
\\
{\em Solution method:} In the first stage, the unit cell is relaxed through a spin-polarized calculation without spin-orbit coupling (SOC).  Next, after a crystal symmetry analysis, a set of deformed lattice and spin configurations are  generated using the pymatgen library \cite{pymatgenlib}. The energy of these states is calculated by the Vienna Ab-initio Simulation Package (VASP) \cite{VASPcode}, including SOC. The anisotropic magnetoelastic constants are derived from the fitting of these energies to a linear polynomial. Finally, if the elastic tensor is provided \cite{AELAScode}, then the magnetostrictive coefficients are also calculated from the theoretical relations between elastic and magnetoelastic constants. 
\\
{\em Additional comments including restrictions and unusual features:} This version supports the following crystal systems: Cubic (point groups $432$, $\bar{4}3m$, $m\bar{3}m$), Hexagonal ($6mm$, $622$, $\bar{6}2m$, $6/mmm$), Trigonal ($32$, $3m$, $\bar{3}m$), Tetragonal ($4mm$, $422$, $\bar{4}2m$, $4/mmm$) and Orthorhombic ($222$, $2mm$, $mmm$).
\\
  %Provide any additional comments here.
   \\

\end{small}

\section{Introduction}
\label{section:Intro}

MAELAS aims to provide an efficient and reliable computational program for the study of magnetostriction by automated first-principles calculations. The version 1.0 of MAELAS was originally published in Ref. \cite{maelas_publication2021}, where we implemented and generalized the length optimization method proposed by Wu and Freeman to calculate the magnetostrictive coefficients ($\lambda$) \cite{Wu1996}. In version 1.0, we also proposed to compute the magnetoelastic constants ($b$) indirectly from the theoretical relations between the elastic constants ($C_{ij}$) and magnetostrictive coefficients using the calculated values of these quantities. The present release of the MAELAS program (version 2.0) aims to add a more robust, rigorous and accurate alternative methodology to calculate the magnetoelastic constants with respect to the previous version. This is accomplished by implementing cell deformations and spin directions theoretically derived from the magnetoelastic energy for each crystal symmetry. In this new approach, the magnetoelastic constants are directly obtained from a linear fitting of the energy versus strain data.

A quantitative analysis of the accuracy of these methods can be very useful to know their reliability and to optimize them. In every calculation with MAELAS we can identify two main sources of errors. The first source of errors comes from the methodology used to compute the magnetostrictive coefficients and magnetoelastic constants, while the second source is the evaluation of the total energies performed with Density Functional Theory (DFT) itself. For instance, in some cases DFT gives energy values that can not fitted well to the polynomial used in these methods, due to numerical or physical reasons. Frequently, the lack of available experimental data makes it difficult to estimate the reliability and precision of these calculations. Aiming to overcome these limitations, we propose here to analyze the systematic error coming from the methodologies implemented in MAELAS by evaluating the exact energy from the theory of magnetostriction. This strategy allows us to compare the accuracy between the length optimization method (originally implemented in version 1.0) and the new approach (added in version 2.0). It also helps us to identify some issues with the trigonal crystal symmetry in the previous version of MAELAS, which is fixed in the present version and discussed here.  

Typically, in studies of magnetostriction the change in length is referred to an initial
demagnetized state. However, this fact can lead to difficulties in the interpretation of the results since the demagnetized state may not correspond to a unique reference state \cite{Birss}. Aiming to clarify this point, we also study the role of the demagnetized state in the fractional change in length. The paper is organized as follows. In Section \ref{section:updates}, we describe the main updates in the new  version, while the accuracy of MAELAS is analyzed in Section \ref{section:accuracy}. The new implemented method is benchmarked in Section \ref{section:test}. The paper ends with a summary of the main conclusions and future perspectives (Section \ref{section:con}).

\section{Main updates in new version}
\label{section:updates}

In this section, we explain in detail three major updates implemented in the new version of MAELAS. An introduction to the theory of magnetostriction and overview of the theoretical background were already provided in the publication of version 1.0 of MAELAS \cite{maelas_publication2021}. Here, we use the same notation, definitions and conventions as in Ref. \cite{maelas_publication2021}.

\subsection{New method for direct calculation of anisotropic magnetoelastic constants}
\label{subsection:new_method}

We have added an alternative method to compute anisotropic magnetoelastic constants in version 2.0. This approach is derived from the total energy ($E$) including elastic ($E_{el}$), magnetoelastic ($E_{me}$) and unstrained magnetocrystalline anisotropy ($E_{K}^{0}$) terms 
%%%%%%%%%%%%%%%%%%%%%%%%%%%%%%%%%%
\begin{equation}
    E(\boldsymbol{\epsilon},\boldsymbol{\alpha})=E_{el}(\boldsymbol{\epsilon})+E_{me}(\boldsymbol{\epsilon},\boldsymbol{\alpha})+E_{K}^{0}(\boldsymbol{\alpha}),
\label{eq:E_tot}
\end{equation}
%%%%%%%%%%%%%%%%%%%%%%%%%%%%%%%%%%
where $\boldsymbol{\epsilon}$ is the strain tensor and $\boldsymbol{\alpha}$ is the normalized magnetization ($\vert\boldsymbol{\alpha}\vert =1$). The basic idea of this method is to subtract the total energy of two magnetization directions $\boldsymbol{\alpha}_1$ and $\boldsymbol{\alpha}_2$ for a deformed unit cell in such a way that we can get the i-th anisotropic magnetoelastic constant $b_i$  from a linear fitting of the energy versus strain data
%%%%%%%%%%%%%%%%%%%%%%%%%%%%%%%%%%
\begin{equation}
   \frac{1}{V_0}\left[ E(\boldsymbol{\epsilon}^i(s),\boldsymbol{\alpha}^i_1)-E(\boldsymbol{\epsilon}^i(s),\boldsymbol{\alpha}^i_2)\right]=\Gamma_i b_i s + \Phi_i(K_1,K_2),
\label{eq:dE_tot}
\end{equation}
%%%%%%%%%%%%%%%%%%%%%%%%%%%%%%%%%%
where $V_0$ is the equilibrium volume, $\Gamma_i$ is a real number, $\Phi_i$ can depend on the magnetocrystalline anisotropy constants $K_1$ and $K_2$, and $s$ is the parameter used to parameterize the strain tensor ${\epsilon}^i(s)$. In practice, Eq. \ref{eq:dE_tot} is fitted to a linear polynomial
%%%%%%%%%%%%%%%%%%%%%%%%%%%%%%%%%%
\begin{equation}
    f(s)=As+B,
\label{eq:dE_tot_fit}
\end{equation}
%%%%%%%%%%%%%%%%%%%%%%%%%%%%%%%%%%
where $A$ and $B$ are fitting parameters, so that the i-th anisotropic magnetoelastic constant $b_i$ is given by
%%%%%%%%%%%%%%%%%%%%%%%%%%%%%%%%%%
\begin{equation}
    b_i=\frac{A}{\Gamma_i}.
\label{eq:b_i_sol}
\end{equation}
%%%%%%%%%%%%%%%%%%%%%%%%%%%%%%%%%%
Additionally, we note that in some cases from the fitting parameter $B$ is possible to estimate the magnetocrystalline anisotropy constants since $B=\Phi_i(K_1,K_2)$. In  \ref{app_matrix} we describe the procedure to generate the deformations for each magnetoelastic constant, while in Table \ref{tab:beta_alpha_data} we show the selected set of  $\boldsymbol{\alpha}_1^i$ and $\boldsymbol{\alpha}_2^i$, and the corresponding value of $\Gamma_i$ that fulfils Eq. \ref{eq:dE_tot}. We implemented this method for all supported crystal symmetries in version 1.0 \cite{maelas_publication2021}.

%%%%%%%%%%%%%%%%%%%%%%%%%%%%%%%%%%%%%%%%%%%%%%%%%%
\begin{table}[]
\centering
\caption{Selected  magnetization directions ($\boldsymbol{\alpha}_1$, $\boldsymbol{\alpha}_2$) in the new method implemented in MAELAS version 2.0 to calculate the anisotropic magnetoelastic constants. The first column shows the crystal system and the corresponding lattice convention set in MAELAS based on the IEEE format \cite{AELAS}.  In the fifth and sixth columns we show the values of the parameters $\Gamma$ and $\Phi$ that are defined in Eq.\ref{eq:dE_tot}. Last column presents the equation of the deformation gradient $F_{ij}$ that we used in Eq.\ref{eq:dE_tot} for the calculation of each magnetoelastic constant. The symbols $a,b,c$ correspond to the lattice parameters of the relaxed (not distorted) unit cell.}
\label{tab:beta_alpha_data}
\resizebox{\textwidth}{!}{%
\begin{tabular}{@{}ccccccc@{}}
\toprule
Crystal system  &
  \multicolumn{1}{c}{\begin{tabular}[c]{@{}c@{}}Magnetoelastic \\ constant\end{tabular}}  &
  \multicolumn{1}{c}{$\boldsymbol{\alpha}_1$} &
  \multicolumn{1}{c}{$\boldsymbol{\alpha}_2$} &
  \multicolumn{1}{c}{$\Gamma$} & \multicolumn{1}{c}{$\Phi$} & \multicolumn{1}{c}{$\boldsymbol{F}$} \\ \hline\hline
Cubic (I)      & \multicolumn{1}{c}{$b_{1}$} &  \multicolumn{1}{c}{$(1,0,0)$}& \multicolumn{1}{c}{$\left(\frac{1}{\sqrt{2}},\frac{1}{\sqrt{2}},0\right)$} & \multicolumn{1}{c}{$\frac{1}{2}$} & \multicolumn{1}{c}{$-\frac{K_1}{4}$} & \multicolumn{1}{c}{Eq.\ref{eq:strain_cub_I_b1}}\\
  $\boldsymbol{a}\|\hat{\boldsymbol{x}}$, $\boldsymbol{b}\|\hat{\boldsymbol{y}}$, $\boldsymbol{c}\|\hat{\boldsymbol{z}}$          &   \multicolumn{1}{c}{$b_{2}$} & \multicolumn{1}{c}{$\left(\frac{1}{\sqrt{2}},\frac{1}{\sqrt{2}},0\right)$} & \multicolumn{1}{c}{$\left(\frac{-1}{\sqrt{2}},\frac{1}{\sqrt{2}},0\right)$} & \multicolumn{1}{c}{$2$} & \multicolumn{1}{c}{$0$} &  \multicolumn{1}{c}{Eq.\ref{eq:strain_cub_I_b2}} \\ \hline
Hexagonal (I)   & \multicolumn{1}{c}{$b_{21}$}    & \multicolumn{1}{c}{$\left(0,0,1\right)$}                              &  \multicolumn{1}{c}{$\left(\frac{1}{\sqrt{2}},\frac{1}{\sqrt{2}},0\right)$} & \multicolumn{1}{c}{$1$} & \multicolumn{1}{c}{$-K_1-K_2$} & \multicolumn{1}{c}{Eq.\ref{eq:strain_hex_I_b21}} \\
   $\boldsymbol{a}\|\hat{\boldsymbol{x}}$,  $\boldsymbol{c}\|\hat{\boldsymbol{z}}$        &    \multicolumn{1}{c}{$b_{22}$}    & \multicolumn{1}{c}{$\left(0,0,1\right)$}                              &  \multicolumn{1}{c}{$\left(\frac{1}{\sqrt{2}},\frac{1}{\sqrt{2}},0\right)$} & \multicolumn{1}{c}{$1$} & \multicolumn{1}{c}{$-K_1-K_2$} & \multicolumn{1}{c}{Eq.\ref{eq:strain_hex_I_b22}} \\
    $\boldsymbol{b}=\left(-\frac{a}{2},\frac{\sqrt{3}a}{2},0\right)$        & \multicolumn{1}{c}{$b_{3}$}    & \multicolumn{1}{c}{$\left(1,0,0\right)$}                              &  \multicolumn{1}{c}{$\left(0,1,0\right)$} & \multicolumn{1}{c}{$1$} & \multicolumn{1}{c}{$0$} & \multicolumn{1}{c}{Eq.\ref{eq:strain_hex_I_b3}}  \\
    $a=b\neq c$      & \multicolumn{1}{c}{$b_{4}$}    & \multicolumn{1}{c}{$\left(\frac{1}{\sqrt{2}},0,\frac{1}{\sqrt{2}}\right)$}                              &  \multicolumn{1}{c}{$\left(\frac{-1}{\sqrt{2}},0,\frac{1}{\sqrt{2}}\right)$} & \multicolumn{1}{c}{$2$} & \multicolumn{1}{c}{$0$} & \multicolumn{1}{c}{Eq.\ref{eq:strain_hex_I_b4}}\\ \hline
Trigonal (I)   & \multicolumn{1}{c}{$b_{21}$}    & \multicolumn{1}{c}{$\left(0,0,1\right)$}                              &  \multicolumn{1}{c}{$\left(\frac{1}{\sqrt{2}},\frac{1}{\sqrt{2}},0\right)$} & \multicolumn{1}{c}{$1$} & \multicolumn{1}{c}{$-K_1-K_2$} & \multicolumn{1}{c}{Eq.\ref{eq:strain_hex_I_b21}} \\
   $\boldsymbol{a}\|\hat{\boldsymbol{x}}$,  $\boldsymbol{c}\|\hat{\boldsymbol{z}}$        &    \multicolumn{1}{c}{$b_{22}$}    & \multicolumn{1}{c}{$\left(0,0,1\right)$}                              &  \multicolumn{1}{c}{$\left(\frac{1}{\sqrt{2}},\frac{1}{\sqrt{2}},0\right)$} & \multicolumn{1}{c}{$1$} & \multicolumn{1}{c}{$-K_1-K_2$} & \multicolumn{1}{c}{Eq.\ref{eq:strain_hex_I_b22}} \\
    $\boldsymbol{b}=\left(-\frac{a}{2},\frac{\sqrt{3}a}{2},0\right)$        & \multicolumn{1}{c}{$b_{3}$}    & \multicolumn{1}{c}{$\left(1,0,0\right)$}                              &  \multicolumn{1}{c}{$\left(0,1,0\right)$} & \multicolumn{1}{c}{$1$} & \multicolumn{1}{c}{$0$} & \multicolumn{1}{c}{Eq.\ref{eq:strain_hex_I_b3}}  \\
    $a=b\neq c$      & \multicolumn{1}{c}{$b_{4}$}    & \multicolumn{1}{c}{$\left(\frac{1}{\sqrt{2}},0,\frac{1}{\sqrt{2}}\right)$}                              &  \multicolumn{1}{c}{$\left(\frac{-1}{\sqrt{2}},0,\frac{1}{\sqrt{2}}\right)$} & \multicolumn{1}{c}{$2$} & \multicolumn{1}{c}{$0$} & \multicolumn{1}{c}{Eq.\ref{eq:strain_hex_I_b4}}\\
           & \multicolumn{1}{c}{$b_{14}$}    & \multicolumn{1}{c}{$\left(\frac{1}{\sqrt{2}},\frac{1}{\sqrt{2}},0\right)$}                              &  \multicolumn{1}{c}{$\left(\frac{-1}{\sqrt{2}},\frac{1}{\sqrt{2}},0\right)$} & \multicolumn{1}{c}{$2$} & \multicolumn{1}{c}{$0$} & \multicolumn{1}{c}{Eq.\ref{eq:strain_trig_I_b14}} \\
            & \multicolumn{1}{c}{$b_{34}$}    & \multicolumn{1}{c}{$\left(\frac{1}{\sqrt{2}},0,\frac{1}{\sqrt{2}}\right)$}                              &  \multicolumn{1}{c}{$\left(\frac{-1}{\sqrt{2}},0,\frac{1}{\sqrt{2}}\right)$} & \multicolumn{1}{c}{$2$} & \multicolumn{1}{c}{$0$} & \multicolumn{1}{c}{Eq.\ref{eq:strain_trig_I_b34}}  \\ \hline
Tetragonal (I) &  \multicolumn{1}{c}{$b_{21}$}    & \multicolumn{1}{c}{$\left(0,0,1\right)$}                              &  \multicolumn{1}{c}{$\left(\frac{1}{\sqrt{2}},\frac{1}{\sqrt{2}},0\right)$} & \multicolumn{1}{c}{$1$} & \multicolumn{1}{c}{$-K_1-K_2$} & \multicolumn{1}{c}{Eq.\ref{eq:strain_hex_I_b21}}   \\
    $\boldsymbol{a}\|\hat{\boldsymbol{x}}$, $\boldsymbol{b}\|\hat{\boldsymbol{y}}$, $\boldsymbol{c}\|\hat{\boldsymbol{z}}$        & \multicolumn{1}{c}{$b_{22}$}    & \multicolumn{1}{c}{$\left(0,0,1\right)$}                              &  \multicolumn{1}{c}{$\left(\frac{1}{\sqrt{2}},\frac{1}{\sqrt{2}},0\right)$} & \multicolumn{1}{c}{$1$} & \multicolumn{1}{c}{$-K_1-K_2$} & \multicolumn{1}{c}{Eq.\ref{eq:strain_hex_I_b22}} \\
     $a=b\neq c$      & \multicolumn{1}{c}{$b_{3}$}    & \multicolumn{1}{c}{$\left(1,0,0\right)$}                              &  \multicolumn{1}{c}{$\left(0,1,0\right)$} & \multicolumn{1}{c}{$1$} & \multicolumn{1}{c}{$0$} & \multicolumn{1}{c}{Eq.\ref{eq:strain_hex_I_b3}}  \\
           & \multicolumn{1}{c}{$b_{4}$}    & \multicolumn{1}{c}{$\left(\frac{1}{\sqrt{2}},0,\frac{1}{\sqrt{2}}\right)$}                              &  \multicolumn{1}{c}{$\left(\frac{-1}{\sqrt{2}},0,\frac{1}{\sqrt{2}}\right)$} & \multicolumn{1}{c}{$2$} & \multicolumn{1}{c}{$0$} & \multicolumn{1}{c}{Eq.\ref{eq:strain_hex_I_b4}}  \\
         & \multicolumn{1}{c}{$b'_{3}$}    & \multicolumn{1}{c}{$\left(\frac{1}{\sqrt{2}},\frac{1}{\sqrt{2}},0\right)$}                              &  \multicolumn{1}{c}{$\left(\frac{-1}{\sqrt{2}},\frac{1}{\sqrt{2}},0\right)$} & \multicolumn{1}{c}{$2$} & \multicolumn{1}{c}{$0$} & \multicolumn{1}{c}{Eq.\ref{eq:strain_tet_I_bp3}} \\ \hline
Orthorhombic & \multicolumn{1}{c}{$b_{1}$}    & \multicolumn{1}{c}{$\left(1,0,0\right)$}                              &  \multicolumn{1}{c}{$\left(0,0,1\right)$} & \multicolumn{1}{c}{$1$} & \multicolumn{1}{c}{$K_1$} & \multicolumn{1}{c}{Eq.\ref{eq:strain_orto_b1}} \\
    $c<a<b$        & \multicolumn{1}{c}{$b_{2}$}    & \multicolumn{1}{c}{$\left(0,1,0\right)$}                              &  \multicolumn{1}{c}{$\left(0,0,1\right)$} & \multicolumn{1}{c}{$1$} & \multicolumn{1}{c}{$K_2$} & \multicolumn{1}{c}{Eq.\ref{eq:strain_orto_b1}} \\
            & \multicolumn{1}{c}{$b_{3}$}    & \multicolumn{1}{c}{$\left(1,0,0\right)$}                              &  \multicolumn{1}{c}{$\left(0,0,1\right)$} & \multicolumn{1}{c}{$1$} & \multicolumn{1}{c}{$K_1$} & \multicolumn{1}{c}{Eq.\ref{eq:strain_orto_b3}} \\
             & \multicolumn{1}{c}{$b_{4}$}    & \multicolumn{1}{c}{$\left(0,1,0\right)$}                              &  \multicolumn{1}{c}{$\left(0,0,1\right)$} & \multicolumn{1}{c}{$1$} & \multicolumn{1}{c}{$K_2$} & \multicolumn{1}{c}{Eq.\ref{eq:strain_orto_b3}} \\
            & \multicolumn{1}{c}{$b_{5}$}    & \multicolumn{1}{c}{$\left(1,0,0\right)$}                              &  \multicolumn{1}{c}{$\left(0,0,1\right)$} & \multicolumn{1}{c}{$1$} & \multicolumn{1}{c}{$K_1$} & \multicolumn{1}{c}{Eq.\ref{eq:strain_orto_b5}}\\
             & \multicolumn{1}{c}{$b_{6}$}    & \multicolumn{1}{c}{$\left(0,1,0\right)$}                              &  \multicolumn{1}{c}{$\left(0,0,1\right)$} & \multicolumn{1}{c}{$1$} & \multicolumn{1}{c}{$K_2$} & \multicolumn{1}{c}{Eq.\ref{eq:strain_orto_b5}} \\
              & \multicolumn{1}{c}{$b_{7}$}    & \multicolumn{1}{c}{$\left(\frac{1}{\sqrt{2}},\frac{1}{\sqrt{2}},0\right)$}                              &  \multicolumn{1}{c}{$\left(0,0,1\right)$} & \multicolumn{1}{c}{$1$} & \multicolumn{1}{c}{$\frac{K_1}{2}+\frac{K_2}{2}$} & \multicolumn{1}{c}{Eq.\ref{eq:strain_orto_b7}}  \\
              & \multicolumn{1}{c}{$b_{8}$}    & \multicolumn{1}{c}{$\left(\frac{1}{\sqrt{2}},0,\frac{1}{\sqrt{2}}\right)$}                              &  \multicolumn{1}{c}{$\left(0,0,1\right)$} & \multicolumn{1}{c}{$1$} & \multicolumn{1}{c}{$\frac{K_1}{2}$} & \multicolumn{1}{c}{Eq.\ref{eq:strain_orto_b8}} \\
               & \multicolumn{1}{c}{$b_{9}$}    & \multicolumn{1}{c}{$\left(0,\frac{1}{\sqrt{2}},\frac{1}{\sqrt{2}}\right)$}                              &  \multicolumn{1}{c}{$\left(0,0,1\right)$} & \multicolumn{1}{c}{$1$} & \multicolumn{1}{c}{$\frac{K_2}{2}$} & \multicolumn{1}{c}{Eq.\ref{eq:strain_orto_b9}} \\\bottomrule
\end{tabular}
}
\end{table}
%%%%%%%%%%%%%%%%%%%%%%%%%%%%%%%%%%%%%%%%%%%%%%%%%%%%%%%

To illustrate this method, let's apply it for the calculation of $b_1$ of cubic symmetry since it is easy to handle. For cubic (I) systems (point groups $432$, $\bar{4}3m$, $m\bar{3}m$) the energy terms in Eq.\ref{eq:E_tot} read \cite{maelas_publication2021}
%%%%%%%%%%%%%%%%%%%%%%%%%55
\begin{equation}
\begin{aligned}
\frac{1}{V_0} E_{el}^{cub}(\boldsymbol{\epsilon})& = \frac{c_{xxxx}}{2}(\epsilon_{xx}^2+\epsilon_{yy}^2+\epsilon_{zz}^2)+c_{xxyy}(\epsilon_{xx}\epsilon_{yy}+\epsilon_{xx}\epsilon_{zz}+\epsilon_{yy}\epsilon_{zz})\\
& + 2c_{yzyz}(\epsilon_{xy}^2+\epsilon_{yz}^2+\epsilon_{xz}^2),\\
\frac{1}{V_0} E_{me}^{cub(I)}(\boldsymbol{\epsilon},\boldsymbol{\alpha}) & =  b_0(\epsilon_{xx}+\epsilon_{yy}+\epsilon_{zz})+b_1(\alpha_x^2\epsilon_{xx}+\alpha_y^2\epsilon_{yy}+\alpha_z^2\epsilon_{zz})\\
    & +  2b_2(\alpha_x\alpha_y\epsilon_{xy}+\alpha_x\alpha_z\epsilon_{xz}+\alpha_y\alpha_z\epsilon_{yz}), \\
    \frac{1}{V_0} E_{K}^{0,cub}(\boldsymbol{\alpha})& = K_0+K_1(\alpha_x^2\alpha_y^2+\alpha_x^2\alpha_z^2+\alpha_y^2\alpha_z^2)+K_2\alpha_x^2\alpha_y^2\alpha_z^2, 
\end{aligned}
\label{eq:E_cub}
\end{equation}
%%%%%%%%%%%%%%%%%%%%%%%%%%%%
where  $c_{xxxx}=C_{11}$, $c_{xxyy}=C_{12}$ and $c_{yzyz}=C_{44}$ are the elastic constants, $b_0$, $b_1$ and $b_2$ are the magnetoelastic constants, and $K_0$, $K_1$ and $K_2$ are the magnetocrystalline anisotropy constants. In Table \ref{tab:beta_alpha_data}, we see that the deformation gradient $\boldsymbol{F}$ to calculate $b_1$ is given by Eq. \ref{eq:strain_cub_I_b1}, which leads to the following strain tensor through Eq.\ref{eq:strain_deform0}
%%%%%%%%%%%%%%%%%%%%%%%%%%%%%
\begin{eqnarray}
\boldsymbol{\epsilon}^{b_{1}}(s)  = 
\begin{pmatrix}
\epsilon_{xx} & \epsilon_{xy} & \epsilon_{xz}\\
\epsilon_{yx} & \epsilon_{yy} & \epsilon_{yz}\\
\epsilon_{zx} & \epsilon_{zy} & \epsilon_{zz} \\
\end{pmatrix}
= \begin{pmatrix}
s & 0 & 0\\
0 & 0 & 0\\
0 & 0 & 0 \\
\end{pmatrix}
.
\label{eq:strain_cub_I_b1_example}
\end{eqnarray}
%%%%%%%%%%%%%%%%%%%%%%%%%%%
We also see in Table \ref{tab:beta_alpha_data} that the two magnetization directions to calculate $b_1$ are $\boldsymbol{\alpha}_1=(1,0,0)$ and $\boldsymbol{\alpha}_2=(1/\sqrt{2},1/\sqrt{2},0)$. Hence, Eq. \ref{eq:dE_tot} becomes
%%%%%%%%%%%%%%%%%%%%%%%%%%%%%%%%%%
\begin{equation}
\frac{1}{V_0}\left[    E(\boldsymbol{\epsilon}^{b_1}(s),1,0,0)-E\left(\boldsymbol{\epsilon}^{b_1}(s),\frac{1}{\sqrt{2}},\frac{1}{\sqrt{2}},0\right)\right]=\frac{1}{2}b_1 s - \frac{1}{4}K_1 ,
\label{eq:dE_tot_cub}
\end{equation}
%%%%%%%%%%%%%%%%%%%%%%%%%%%%%%%%%%
where we recover $\Gamma=1/2$, as shown in Table \ref{tab:beta_alpha_data}. Therefore, $b_1$ can be estimated from the fitting parameter $A$ of the linear polynomial (Eq. \ref{eq:b_i_sol}) as $b_1=A/\Gamma=2A$. We also observe that $\Phi=-K_1/4$, so the first magnetocrystalline anisotropy constant $K_1$ for the unstrained cubic state can be estimated as well from the fitting parameter $B$ of the linear polynomial ($K_1=-4B$), see Eq.\ref{eq:dE_tot_fit}.

\begin{figure}[h!]
\centering
\includegraphics[width=\columnwidth ,angle=0]{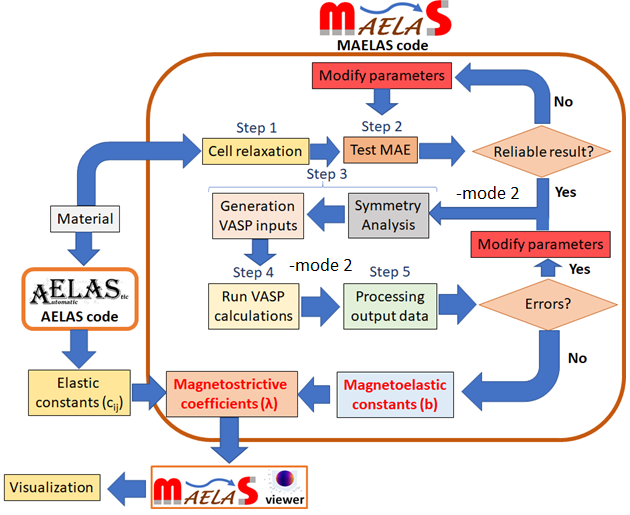}
\caption{Workflow of the new methodology for direct calculation of anisotropic magnetoelastic constants implemented in MAELAS v2.0. This new method is executed by tag -mode 2. The method originally implemented in version 1.0 for direct calculation of anisotropic magnetostrictive coefficients \cite{maelas_publication2021} based on the length optimization \cite{Wu1996} can also be executed in version 2.0 by using tag -mode 1.}
\label{fig:workflow_mode2}
\end{figure}

\begin{figure}[h!]
\centering
\includegraphics[width=\columnwidth ,angle=0]{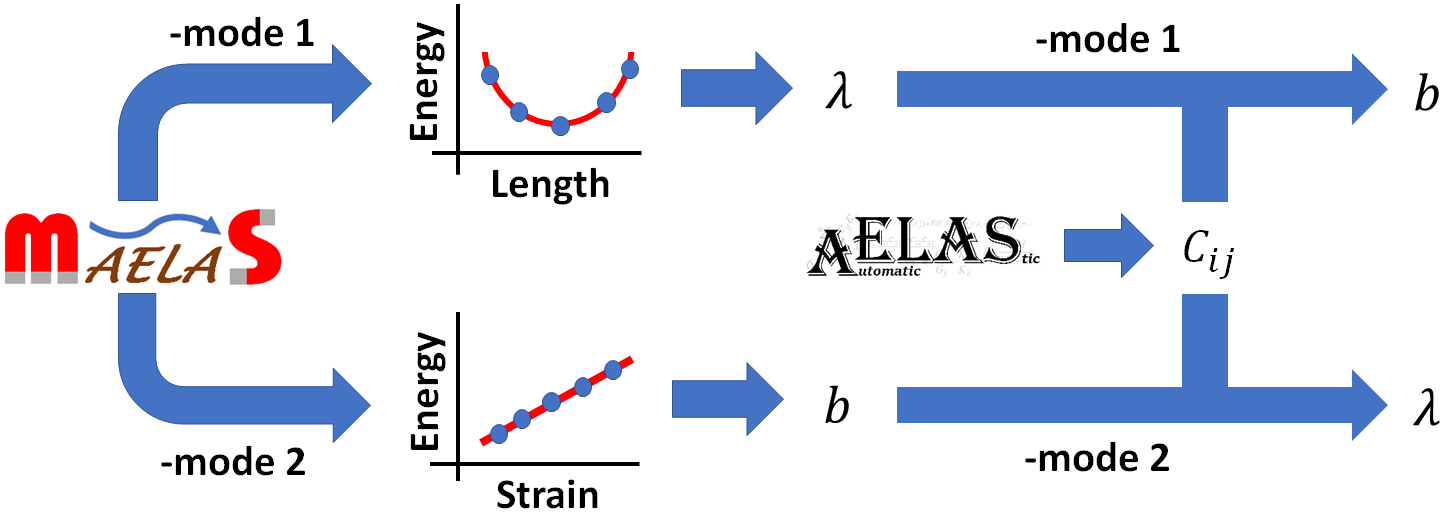}
\caption{Diagram showing the two methods available in MAELAS v2.0 to calculate anisotropic magnetostrictive coefficients ($\lambda$) and magnetoelastic constants ($b$). The method -mode 1 corresponds to the approach originally implemented in version 1.0 \cite{maelas_publication2021} based on the length optimization \cite{Wu1996}, while -mode 2 is the new method added in version 2.0 (Section \ref{subsection:new_method}).}
\label{fig:diagram_2_modes}
\end{figure}

The corresponding workflow for this new methodology is depicted in Fig.\ref{fig:workflow_mode2}. It contains the same steps as the workflow for the method implemented in version 1.0 for direct calculation of anisotropic magnetostrictive coefficients \cite{maelas_publication2021}. However, here the generated deformations and spin configurations for the Vienna Ab initio Simulation Package (VASP)  \cite{vasp_1,vasp_2,vasp_3} in step 3 corresponds to the new method. In step 5, after processing VASP output data, MAELAS v2.0 will calculate directly the magnetoelastic constants. Additionally, if the elastic constants are also provided in AELAS format \cite{AELAS}, then it will also calculate the anisotropic magnetostrictive coefficients ($\lambda$) from the theoretical equations $\lambda(b_{k},C_{ij})$ given in Ref. \cite{maelas_publication2021}. The calculated magnetostrictive coefficients can be analyzed and visualized with the online tool MAELASviewer \cite{maelasviewer}. This new method is executed by using tag -mode 2 in the command line. The length optimization method \cite{Wu1996} originally implemented in version 1.0 for direct calculation of anisotropic magnetostrictive coefficients \cite{maelas_publication2021} can also be executed in version 2.0 by using tag -mode 1, see Fig.\ref{fig:diagram_2_modes}.

%------------------------------
%------------------------------

%------------------------------
%------------------------------

\subsection{Revision of the trigonal (I) symmetry}
\label{subsection:correct_trigonal}

The analysis of the accuracy of MAELAS that is presented in Section \ref{section:accuracy} helped us to identify some issues in version 1.0 \cite{maelas_publication2021} related to the implementation of the length optimization method (originally proposed by Wu and Freeman \cite{Wu1996}) for the trigonal (I) symmetry. First of all, we point out a misprint in the publication of version 1.0 \cite{maelas_publication2021} in the elastic energy that should read
%%%%%%%%%%%%%%%%%%%%%%%%%55
\begin{equation}
\begin{aligned}
  \frac{E_{el}^{trig(I)}-E_0}{V_0} & =  \frac{1}{2}c_{xxxx}(\epsilon_{xx}^2+\epsilon_{yy}^2)+c_{xxyy}\epsilon_{xx}\epsilon_{yy}+c_{xxzz}(\epsilon_{xx}+\epsilon_{yy})\epsilon_{zz}+\frac{1}{2}c_{zzzz}\epsilon_{zz}^2 \\
  & +2c_{yzyz}(\epsilon_{xz}^2+\epsilon_{yz}^2)+(c_{xxxx}-c_{xxyy})\epsilon_{xy}^2\\
  & +c_{xxyz}(4\epsilon_{xy}\epsilon_{xz}+2\epsilon_{xx}\epsilon_{yz}-2\epsilon_{yy}\epsilon_{yz}).
\end{aligned}
\label{eq:E_el_trig_correct}
\end{equation}
%%%%%%%%%%%%%%%%%%%%%%%%%%%%
where $c_{xxxx}=C_{11}$, $c_{xxyy}=C_{12}$, $c_{xxzz}=C_{13}$, $c_{xxyz}=C_{14}$, $c_{zzzz}=C_{33}$, and $c_{yzyz}=C_{44}$. Secondly, the correct theoretical relations between the magnetostrictive coefficients $\lambda^{\gamma,1}$, $\lambda^{\gamma,2}$ and $\lambda_{21}$, and the elastic and magnetoelastic constants are
%%%%%%%%%%%%%%%%%%%%%%%%%%%%%
\begin{equation}
    \begin{aligned}
        \lambda^{\gamma,1} & = \frac{\frac{1}{2}C_{14}b_{14}-\frac{1}{2}C_{44}b_{3}}{\frac{1}{2}C_{44}(C_{11}-C_{12})-C_{14}^2},\\
        \lambda^{\gamma,2} & = \frac{-\frac{1}{2}b_4(C_{11}-C_{12})+b_{34}C_{14}}{\frac{1}{2}C_{44}(C_{11}-C_{12})-C_{14}^2},\\
        \lambda_{21} & = \frac{-\frac{1}{2}b_{14}(C_{11}-C_{12})+b_{3}C_{14}}{\frac{1}{2}C_{44}(C_{11}-C_{12})-C_{14}^2}.
    \label{eq:lamb_trig_correct}
    \end{aligned}
\end{equation}
%%%%%%%%%%%%%%%%%%%%%%%%%%%
Lastly, based on the equation of the relative length change $\Delta l/l_0$, for the calculation $\lambda_{12}$ in version 1.0 we proposed to compute the cell length in the direction $\boldsymbol{\beta}=(a/\sqrt{a^2+c^2},0,c/\sqrt{a^2+c^2})$ using the deformation gradient \cite{maelas_publication2021}
%%%%%%%%%%%%%%%%%%%%%%%%%%%%%
\begin{equation}
\begin{aligned}
\boldsymbol{F}\Big\vert_{\boldsymbol{\beta}=\frac{\left(a,0,c\right)}{\sqrt{a^2+c^2}}}^{\lambda_{12}(version1.0)}(s)=\Omega
\begin{pmatrix}
1 & 0 & \frac{s c}{2a}\\
0 & 1 & 0\\
\frac{s a}{2c} & 0& 1 \\
\end{pmatrix}.
\label{eq:strain_trig_I_old}
\end{aligned}
\end{equation}
%%%%%%%%%%%%%%%%%%%%%%%%%%%
where $\Omega=\sqrt[3]{4/(4-s^2)}$, $a$ and $c$ are the lattice parameters of the relaxed (not deformed) unit cell. However, this choice leads to the same theoretical total energy for the selected magnetization directions $\boldsymbol{\alpha}_1=(0,1/\sqrt{2},1/\sqrt{2})$ and $\boldsymbol{\alpha}_2=(0,1/\sqrt{2},-1/\sqrt{2})$, that is, $E(\boldsymbol{\epsilon}^{\lambda_{12}},\boldsymbol{\alpha}_1)=E(\boldsymbol{\epsilon}^{\lambda_{12}},\boldsymbol{\alpha}_2)$. Consequently, in practice this would always give very similar equilibrium lengths $l_1$ and $l_2$ for the cases with $\boldsymbol{\alpha}_1$ and $\boldsymbol{\alpha}_2$, respectively. Thus, a negligible magnetostrictive coefficient $\lambda_{12}$ ($\lambda_{12}\cong 0$) will be obtained always. To fix this problem in version 2.0, we use the same magnetization directions $\boldsymbol{\alpha}_1=(0,1/\sqrt{2},1/\sqrt{2})$ and $\boldsymbol{\alpha}_2=(0,1/\sqrt{2},-1/\sqrt{2})$, and replace the measuring length direction by $\boldsymbol{\beta}=(1,0,0)$, and deformation gradient by 
%%%%%%%%%%%%%%%%%%%%%%%%%%%%%
\begin{equation}
\begin{aligned}
\boldsymbol{F}\Big\vert_{\boldsymbol{\beta}=(1,0,0)}^{\lambda_{12}(version2.0)}(s)=
\begin{pmatrix}
1+s & 0 & 0\\
0 & \frac{1}{\sqrt{1+s}} & 0\\
0 & 0 & \frac{1}{\sqrt{1+s}} \\
\end{pmatrix}.
\label{eq:strain_trig_I_new}
\end{aligned}
\end{equation}
%%%%%%%%%%%%%%%%%%%%%%%%%%%
In this case, we have
%%%%%%%%%%%%%%%%%%%%%%%%%%%%%%%%%%%%%%%%%%%%
\begin{equation}
     \lambda_{12}=\frac{2 (l_1 -l_2)}{\rho (l_1+l_2)}=\frac{4 (l_1 -l_2)}{(l_1+l_2)},
    \label{eq:lambda_sol_trig}
\end{equation}
%%%%%%%%%%%%%%%%%%%%%%%%%%%%%%%%%%%%%%%%%%%%
where $\rho=1/2$, and $l_1$ and $l_2$ are the equilibrium cell length along $\boldsymbol{\beta}=(1,0,0)$ when the magnetization points to $\boldsymbol{\alpha}_1=(0,1/\sqrt{2},1/\sqrt{2})$ and $\boldsymbol{\alpha}_2=(0,1/\sqrt{2},-1/\sqrt{2})$, respectively.

\subsection{The demagnetized state and its role in the fractional change in length}
\label{subsection:poly}

The magnetostrictive coefficients calculated with MAELAS correspond to the values at the magnetic saturated state. In experiment, these coefficients can be obtained by measuring the change in length of the material under saturation with respect to an initial length at a reference demagnetized state. Thus, the measured change in length, as well as the form of the equation describing the fractional change in length, depend on the initial demagnetized state that is used as a reference state. Unfortunately, this state may not correspond to a unique reference state since it could have different distribution of magnetic domains \cite{Birss}. This fact is important to take into account in order to compare  experiment and theory properly, see the analysis of L1$_0$ FePd in Section \ref{section:FePd}. In this section, based on Birss's work \cite{Birss}, we analyze the influence of the demagnetized state on the fractional change in length for the crystals supported by MAELAS. To facilitate the comparison between theory and experiment we included the possibility to choose the type of reference demagnetized state in the calculation of the fractional change in length in the visualization tool MAELASviewer \cite{maelasviewer}. On the other hand, in version 1.0 we already implemented the expressions of fractional change in length for polycrystals with cubic (I) and hexagonal (I) symmetries \cite{Birss,maelas_publication2021}.  Here, we derive the corresponding equation to compute the saturated magnetostriction for polycrystals with trigonal (I), tetragonal (I) and orthorhombic symmetries,  which are implemented in version 2.0. 

\subsubsection{Cubic (I)}
\label{section:demagentized_cub}

\begin{figure}[h!]
\centering
\includegraphics[width=\columnwidth ,angle=0]{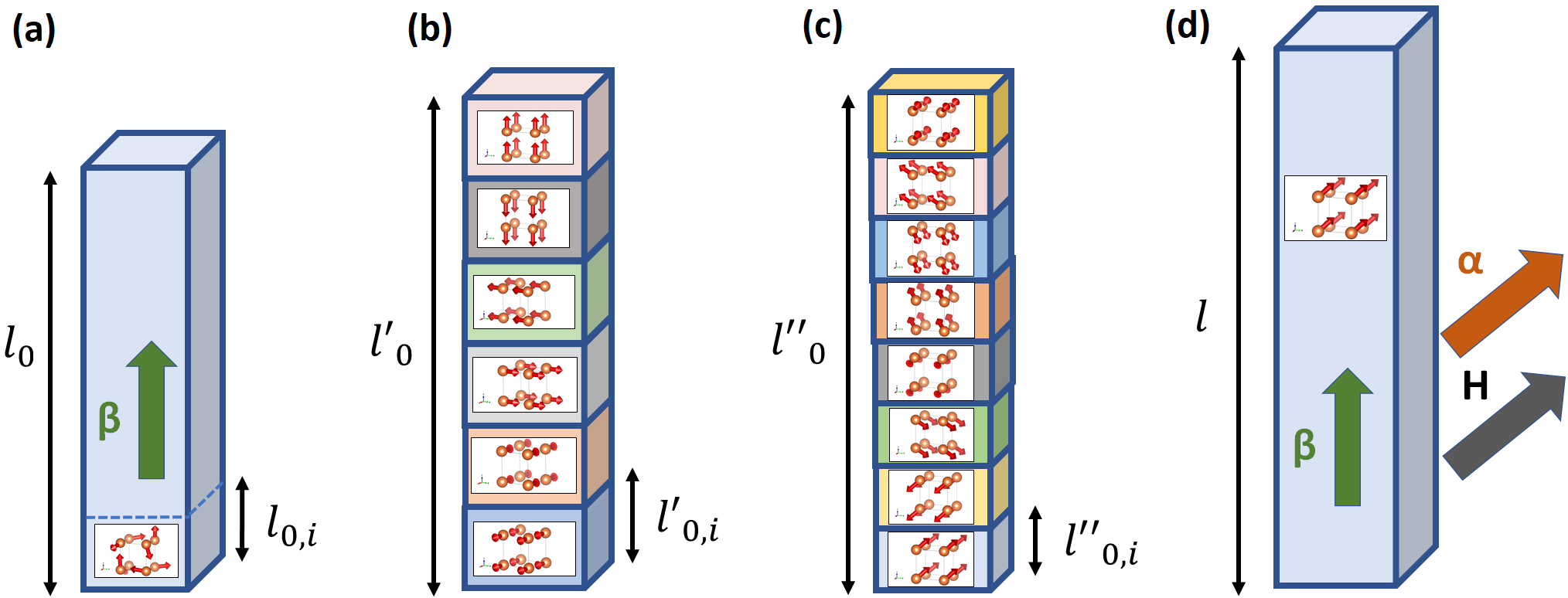}
\caption{(a) Ideal demagnetized state of a cubic single crystal with randomly oriented atomic magnetic moments. Demagnetized state with magnetic domains along all possible easy directions (b) $<100>$ and (c) $<111>$. (d) Saturated magnetic state.}
\label{fig:demag_cub}
\end{figure}

Let's consider a single crystal with Cubic (I) symmetry with length $l_0$ at an arbitrary direction $\boldsymbol{\beta}$ in an initial demagnetized state with randomly oriented atomic magnetic moments. After the magnetization of the crystal is saturated in the direction $\boldsymbol{\alpha}$,  the length in the direction $\boldsymbol{\beta}$ is $l$, see Fig.\ref{fig:demag_cub}. In this case we have the following fractional change in length \cite{maelas_publication2021}
%%%%%%%%%%%%%%%%%%%%%%%%%%%%%%%%%%%%%%%%%%%%
\begin{equation}
\begin{aligned}
     \frac{l-l_0}{l_0}\Bigg\vert_{\boldsymbol{\beta}}^{\boldsymbol{\alpha}} & =\lambda^\alpha+\frac{3}{2}\lambda_{001}\left(\alpha_x^2\beta_{x}^2+\alpha_y^2\beta_{y}^2+\alpha_z^2\beta_{z}^2-\frac{1}{3}\right)\\
     & + 3\lambda_{111}(\alpha_x\alpha_y\beta_{x}\beta_{y}+\alpha_y\alpha_z\beta_{y}\beta_{z}+\alpha_x\alpha_z\beta_{x}\beta_{z}).
    \label{eq:delta_l_cub_I}
\end{aligned}
\end{equation}
%%%%%%%%%%%%%%%%%%%%%%%%%%%%%%%%%%%%%%%%%%%%
The ideal demagnetized state with randomly oriented atomic magnetic moments can not be achieved in experiment below the Curie temperature ($T_C$) due to the ferromagnetic order. The lattice parameters for single crystal at this hypothetical demagnetized state below $T_C$ can be experimentally estimated by extrapolating the lattice parameters above $T_C$ via the Debye theory and the Gr\"{u}neisen relation  \cite{ANDREEV199559}.  In practice, the real experimental demagnetized states below $T_C$ exhibit different distribution of magnetic domains that depend on the demagnetization method \cite{Birss}. Thus, it is convenient to define a standard reference state as a demagnetized state with many ferromagnetic domains, where there are the same number of  domains oriented in each of the crystallographically equivalent directions of easy magnetization \cite{Birss}. For instance, if the easy
directions are the quaternary axes $<100>$, as in BCC Fe, then we have six easy directions $\boldsymbol{\alpha}=(\pm1,0,0)$, $\boldsymbol{\alpha}=(0,\pm1,0)$ and $\boldsymbol{\alpha}=(0,0,\pm1)$. On the other hand, if the easy
directions are the ternary axes $<111>$, as in FCC Ni, then we have eight easy directions $\boldsymbol{\alpha}=(\pm1/\sqrt{3},\pm1/\sqrt{3},\pm1/\sqrt{3})$, $\boldsymbol{\alpha}=(\mp1/\sqrt{3},\pm1/\sqrt{3},\pm1/\sqrt{3})$, $\boldsymbol{\alpha}=(\pm1/\sqrt{3},\mp1/\sqrt{3},\pm1/\sqrt{3})$ and $\boldsymbol{\alpha}=(\pm1/\sqrt{3},\pm1/\sqrt{3},\mp1/\sqrt{3})$. These demagnetized states are schematically represented in Fig. \ref{fig:demag_cub}, where the length in the direction $\boldsymbol{\beta}$ is labeled with  $l'_0$ and $l''_0$ for easy ternary and quaternary axes, respectively. 
%------------------------------
%------------------------------

 If the initial state is the ideal demagnetized state with randomly oriented atomic magnetic moments and the final state is the demagnetized state with  the same number of  domains oriented in each of the easy ternary directions, then the fractional change in length can be written as
%%%%%%%%%%%%%%%%%%%%%%%%%%%%%%%%%%%%%%%%%%%%
\begin{equation}
\begin{aligned}
     \frac{l'_0-l_0}{l_0}\Bigg\vert_{\boldsymbol{\beta}} & =  \frac{\sum\limits_{i=1}^{N} l'_{0,i}-\sum\limits_{i=1}^{N}l_{0,i}}{\sum\limits_{i=1}^{N}l_{0,i}}\Bigg\vert_{\boldsymbol{\beta}}= \sum_{i=1}^{N} \frac{l_{0,i}}{l_0}\left(\frac{l'_{0,i}-l_{0,i}}{l_{0,i}}\Bigg\vert_{\boldsymbol{\beta}}^{\boldsymbol{\alpha}_i}\right)\\
     & =\sum_{j=1}^{6}\sum_{i=1}^{N(\boldsymbol{\alpha}_j)} \frac{l_{0,i}}{l_0}\left(\frac{l'_{0,i}-l_{0,i}}{l_{0,i}}\Bigg\vert_{\boldsymbol{\beta}}^{\boldsymbol{\alpha}_j}\right),
    \label{eq:sum_delta_l_cub_I}
\end{aligned}
\end{equation}
%%%%%%%%%%%%%%%%%%%%%%%%%%%%%%%%%%%%%%%%%%%%
where $N$ is the total number of magnetic domains within a cross section along $\boldsymbol{\beta}$, as shown in Fig.\ref{fig:demag_cub}, and $N(\boldsymbol{\alpha}_j)$ is the number of magnetic domains with the same easy magnetization direction $\boldsymbol{\alpha_j}$. In the last step, we grouped the domains with equal easy magnetization direction together. Each term in the last summation can be computed by applying Eq.\ref{eq:delta_l_cub_I} to each magnetic domain individually 
%%%%%%%%%%%%%%%%%%%%%%%%%%%%%%%%%%%%%%%%%%%%
\begin{equation}
\begin{aligned}
     \sum_{j=1}^{6}\sum_{i=1}^{N(\boldsymbol{\alpha}_j)} \frac{l_{0,i}}{l_0}\left(\frac{l'_{0,i}-l_{0,i}}{l_{0,i}}\Bigg\vert_{\boldsymbol{\beta}}^{\boldsymbol{\alpha}_j}\right) & =\lambda^\alpha+\frac{3}{2}\lambda_{001}(\beta_x^2\left[\sum_{i=1}^{N(\boldsymbol{\alpha}_1)} \frac{l_{0,i}}{l_0}\right]+\beta_x^2\left[\sum_{i=1}^{N(\boldsymbol{\alpha}_2)} \frac{l_{0,i}}{l_0}\right]\\
    & + \beta_y^2\left[\sum_{i=1}^{N(\boldsymbol{\alpha}_3)} \frac{l_{0,i}}{l_0}\right]+\beta_y^2\left[\sum_{i=1}^{N(\boldsymbol{\alpha}_4)} \frac{l_{0,i}}{l_0}\right]\\
   & + \beta_z^2\left[\sum_{i=1}^{N(\boldsymbol{\alpha}_5)} \frac{l_{0,i}}{l_0}\right]+\beta_z^2\left[\sum_{i=1}^{N(\boldsymbol{\alpha}_6)} \frac{l_{0,i}}{l_0}\right]-\frac{1}{3}),
    \label{eq:sum_delta_l_cub_I_2b}
\end{aligned}
\end{equation}
%%%%%%%%%%%%%%%%%%%%%%%%%%%%%%%%%%%%%%%%%%%%
where $\boldsymbol{\alpha}_1=(1,0,0)$, $\boldsymbol{\alpha}_2=(-1,0,0)$, $\boldsymbol{\alpha}_3=(0,1,0)$, $\boldsymbol{\alpha}_4=(0,-1,0)$, $\boldsymbol{\alpha}_5=(0,0,1)$ and $\boldsymbol{\alpha}_6=(0,0,-1)$. Next, we assume the same contribution of the sum  of all magnetic domains with equal easy magnetization direction to the total length  (equally weighted domain distribution \cite{Birss}), that is,
%%%%%%%%%%%%%%%%%%%%%%%%%%%%%%%%%%%%%%%%%%%%
\begin{equation}
\begin{aligned}
     \sum_{i=1}^{N(\boldsymbol{\alpha}_j)} \frac{l_{0,i}}{l_0}=\frac{1}{N_e},\quad j=1,..,N_e
    \label{eq:sum_delta_l_cub_I_3}
\end{aligned}
\end{equation}
%%%%%%%%%%%%%%%%%%%%%%%%%%%%%%%%%%%%%%%%%%%%
where $N_e$ is the number of easy directions for magnetization. For example, for quaternary axes we have $N_e=6$. Using Eq.\ref{eq:sum_delta_l_cub_I_3} in Eq.\ref{eq:sum_delta_l_cub_I_2b} gives
%%%%%%%%%%%%%%%%%%%%%%%%%%%%%%%%%%%%%%%%%%%%
\begin{equation}
\begin{aligned}
     \sum_{j=1}^{6}\sum_{i=1}^{N(\boldsymbol{\alpha}_j)} \frac{l_{0,i}}{l_0}\left(\frac{l'_{0,i}-l_{0,i}}{l_{0,i}}\Bigg\vert_{\boldsymbol{\beta}}^{\boldsymbol{\alpha}_j}\right) & =\lambda^\alpha.
    \label{eq:sum_delta_l_cub_I_2c}
\end{aligned}
\end{equation}
%%%%%%%%%%%%%%%%%%%%%%%%%%%%%%%%%%%%%%%%%%%%
And, replacing Eq. \ref{eq:sum_delta_l_cub_I_2c} in Eq.\ref{eq:sum_delta_l_cub_I} yields
%%%%%%%%%%%%%%%%%%%%%%%%%%%%%%%%%%%%%%%%%%%%
\begin{equation}
\begin{aligned}
     \frac{l'_0-l_0}{l_0}\Bigg\vert_{\boldsymbol{\beta}} & = \lambda^{\alpha}.
    \label{eq:sum_delta_l_cub_I_4}
\end{aligned}
\end{equation}
%%%%%%%%%%%%%%%%%%%%%%%%%%%%%%%%%%%%%%%%%%%%
Hence, only the isotropic magnetostriction related to the exchange interaction (volume magnetostriction $\omega_s\simeq 3\lambda^\alpha$ \cite{ANDREEV199559,nieves2021spinlattice_prb}) takes place effectively. We may write this calculation in the following general and compact form
%%%%%%%%%%%%%%%%%%%%%%%%%%%%%%%%%%%%%%%%%%%%
\begin{equation}
\begin{aligned}
     \frac{l'_0-l_0}{l_0}\Bigg\vert_{\boldsymbol{\beta}} &  =\frac{1}{N_e}\sum_{j=1}^{N_e}\frac{l-l_{0}}{l_{0}}\Bigg\vert_{\boldsymbol{\beta}}^{\boldsymbol{\alpha}_j},
    \label{eq:sum_delta_l_cub_I_4b}
\end{aligned}
\end{equation}
%%%%%%%%%%%%%%%%%%%%%%%%%%%%%%%%%%%%%%%%%%%%
where each term inside the summation corresponds to the Eq.\ref{eq:delta_l_cub_I} evaluated at the easy direction $\boldsymbol{\alpha}_j$. On the other hand, if the initial state is the demagnetized state with  the same number of  domains oriented in each of the easy ternary directions and the final state is the saturated state with magnetization along $\boldsymbol{\alpha}$, then the fractional change in length is
%%%%%%%%%%%%%%%%%%%%%%%%%%%%%%%%%%%%%%%%%%%%
\begin{equation}
\begin{aligned}
     \frac{l-l'_0}{l'_0}\Bigg\vert_{\boldsymbol{\beta}}^{\boldsymbol{\alpha}} & = \frac{l-l'_0}{l_0}\cdot\frac{1}{1-\left(\frac{l_0-l'_0}{l_0}\right)} \Bigg\vert_{\boldsymbol{\beta}}^{\boldsymbol{\alpha}}= \frac{l-l'_0}{l_0}\left[1+\left(\frac{l_0-l'_0}{l_0}\right)+...\right] \Bigg\vert_{\boldsymbol{\beta}}^{\boldsymbol{\alpha}}\\&\cong \frac{l-l'_0}{l'_0}\Bigg\vert_{\boldsymbol{\beta}}^{\boldsymbol{\alpha}} = \frac{l-l_0}{l_0}\Bigg\vert_{\boldsymbol{\beta}}^{\boldsymbol{\alpha}}-\frac{l'_0-l_0}{l_0}\Bigg\vert_{\boldsymbol{\beta}}\\&=\frac{3}{2}\lambda_{001}\left(\alpha_x^2\beta_{x}^2+\alpha_y^2\beta_{y}^2+\alpha_z^2\beta_{z}^2-\frac{1}{3}\right)\\
     & + 3\lambda_{111}(\alpha_x\alpha_y\beta_{x}\beta_{y}+\alpha_y\alpha_z\beta_{y}\beta_{z}+\alpha_x\alpha_z\beta_{x}\beta_{z}).
    \label{eq:sum_delta_l_cub_I_5}
\end{aligned}
\end{equation}
%%%%%%%%%%%%%%%%%%%%%%%%%%%%%%%%%%%%%%%%%%%%
Thus, in this case only the anisotropic magnetostriction related to the SOC and crystal field takes place effectively. The same result is found for the case with easy quaternary directions
%%%%%%%%%%%%%%%%%%%%%%%%%%%%%%%%%%%%%%%%%%%%
\begin{equation}
\begin{aligned}
     \frac{l''_0-l_0}{l_0}\Bigg\vert_{\boldsymbol{\beta}} & = \lambda^{\alpha},\\
     \frac{l-l''_0}{l''_0}\Bigg\vert_{\boldsymbol{\beta}}^{\boldsymbol{\alpha}} & =\frac{3}{2}\lambda_{001}\left(\alpha_x^2\beta_{x}^2+\alpha_y^2\beta_{y}^2+\alpha_z^2\beta_{z}^2-\frac{1}{3}\right)\\
     & + 3\lambda_{111}(\alpha_x\alpha_y\beta_{x}\beta_{y}+\alpha_y\alpha_z\beta_{y}\beta_{z}+\alpha_x\alpha_z\beta_{x}\beta_{z}).
    \label{eq:sum_delta_l_cub_I_4b}
\end{aligned}
\end{equation}
%%%%%%%%%%%%%%%%%%%%%%%%%%%%%%%%%%%%%%%%%%%%
However, we point out that including  higher order terms in Eq.\ref{eq:delta_l_cub_I} leads to different results between easy ternary and quaternary directions \cite{Birss}. We see that using the initial demagnetized state with domains pointing to all easy directions as a reference state allows to characterize the fractional change in length without the need to measure or calculate the isotropic magnetostrictive coefficient $\lambda^\alpha$.

To compute the fractional change in length for a polycrystal, one needs to perform the following average of the corresponding fractional change in length for a single crystal (uniform stress approximation)\cite{Birss}
%%%%%%%%%%%%%%%%%%%%%%%%%%%%%%%%%%%%%%%%%%%%
\begin{equation}
\begin{aligned}
     <f(\boldsymbol{\alpha},\boldsymbol{\beta}) > & =\frac{1}{8\pi}\int_{0}^{\pi}\int_{0}^{2\pi}\int_{0}^{2\pi}f(\boldsymbol{\alpha}',\boldsymbol{\beta}')\sin\theta d\theta d\phi d\psi,
    \label{eq:sum_delta_l_cub_I_6}
\end{aligned}
\end{equation}
%%%%%%%%%%%%%%%%%%%%%%%%%%%%%%%%%%%%%%%%%%%%
where the magnetization and measuring length directions inside the integral are parameterized as
%%%%%%%%%%%%%%%%%%%%%%%%%%%%%%%%%%%%%%%%%%%%
\begin{equation}
\begin{aligned}
     \alpha'_{x} & =\cos \Theta\sin\theta\cos\phi+\sin\Theta(\cos\theta\cos\phi\cos\psi+\sin\phi\sin\psi),\\
     \alpha'_{y} & =\cos \Theta\sin\theta\sin\phi+\sin\Theta(\cos\theta\sin\phi\cos\psi-\cos\phi\sin\psi),\\
     \alpha'_{z} & =\cos \Theta\cos\theta-\sin\Theta\sin\theta\cos\psi,\\
     \beta'_{x} & =\sin\theta\cos\phi,\\
     \beta'_{y} & =\sin\theta\sin\phi,\\
     \beta'_{z} & =\cos\theta,
    \label{eq:sum_delta_l_cub_I_7}
\end{aligned}
\end{equation}
%%%%%%%%%%%%%%%%%%%%%%%%%%%%%%%%%%%%%%%%%%%%
where $\Theta$ is the angle between $\boldsymbol{\alpha}$ and $\boldsymbol{\beta}$. For instance, inserting the Eq.\ref{eq:delta_l_cub_I} in Eq.\ref{eq:sum_delta_l_cub_I_6} yields
%%%%%%%%%%%%%%%%%%%%%%%%%%%%%%%%%%%%%%%%%%%%
\begin{equation}
\begin{aligned}
     <\frac{l-l_0}{l_0}\Bigg\vert_{\boldsymbol{\beta}}^{\boldsymbol{\alpha}} > & =\xi+\eta(\boldsymbol{\alpha}\cdot\boldsymbol{\beta})^2 =\lambda^\alpha+\frac{3}{2}\lambda_S\left[(\boldsymbol{\alpha}\cdot\boldsymbol{\beta})^2-\frac{1}{3}\right],
    \label{eq:sum_delta_l_cub_I_8}
\end{aligned}
\end{equation}
%%%%%%%%%%%%%%%%%%%%%%%%%%%%%%%%%%%%%%%%%%%%
where
%%%%%%%%%%%%%%%%%%%%%%%%%%%%%%%%%%%%%%%%%%%%%%
\begin{equation}
\begin{aligned}
\xi & = \lambda^{\alpha}-\frac{1}{2}\lambda_S,\\
\eta & = \frac{3}{2}\lambda_S,\\
\lambda_S & =\frac{2}{5}\lambda_{001}+\frac{3}{5}\lambda_{111}.
\end{aligned}
\end{equation}
%%%%%%%%%%%%%%%%%%%%%%%%%%%%%%%%%%%%%%%%%%%%%
 If we consider that the initial state of the polycrystal is the demagnetized state with  the same number of  domains oriented in each of the easy ternary or quaternary directions, then using Eq.\ref{eq:sum_delta_l_cub_I_6} we find
%%%%%%%%%%%%%%%%%%%%%%%%%%%%%%%%%%%%%%%%%%%%
\begin{equation}
\begin{aligned}
     <\frac{l-l'_0}{l'_0}\Bigg\vert_{\boldsymbol{\beta}}^{\boldsymbol{\alpha}} >= <\frac{l-l''_0}{l''_0}\Bigg\vert_{\boldsymbol{\beta}}^{\boldsymbol{\alpha}} > & =\xi+\eta(\boldsymbol{\alpha}\cdot\boldsymbol{\beta})^2 =\frac{3}{2}\lambda_S\left[(\boldsymbol{\alpha}\cdot\boldsymbol{\beta})^2-\frac{1}{3}\right],
    \label{eq:sum_delta_l_cub_I_9}
\end{aligned}
\end{equation}
%%%%%%%%%%%%%%%%%%%%%%%%%%%%%%%%%%%%%%%%%%%%
where
%%%%%%%%%%%%%%%%%%%%%%%%%%%%%%%%%%%%%%%%%%%%%%
\begin{equation}
\begin{aligned}
\xi & = -\frac{1}{2}\lambda_S,\\
\eta & = \frac{3}{2}\lambda_S,\\
\lambda_S & =\frac{2}{5}\lambda_{001}+\frac{3}{5}\lambda_{111}.
\end{aligned}
\end{equation}
%%%%%%%%%%%%%%%%%%%%%%%%%%%%%%%%%%%%%%%%%%%%%
Recently, we verified the Eq.\ref{eq:sum_delta_l_cub_I_8} by means of finite element methods \cite{nieves2021influence}.

\subsubsection{Hexagonal (I)}
\label{section:demagentized_hex}

\begin{figure}[h!]
\centering
\includegraphics[width=0.55\columnwidth ,angle=0]{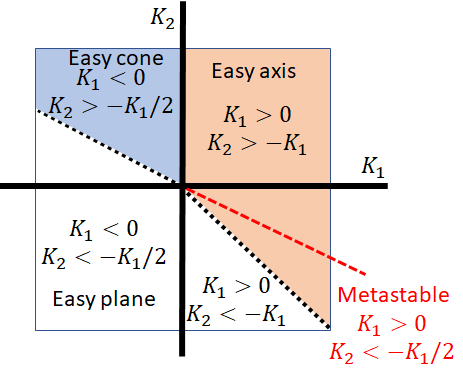}
\caption{Phase diagram of magnetocrystalline anisotropy energy for uniaxial crystals. A metastable anisotropy occurs when $K_1>0$ and $K_2<-K_1/2$. In this case, easy axis and easy plane are in coexistence, one is an absolute minimum and the other one is a local minimum.}
\label{fig:typeani_hex}
\end{figure}

\begin{figure}[h!]
\centering
\includegraphics[width=\columnwidth ,angle=0]{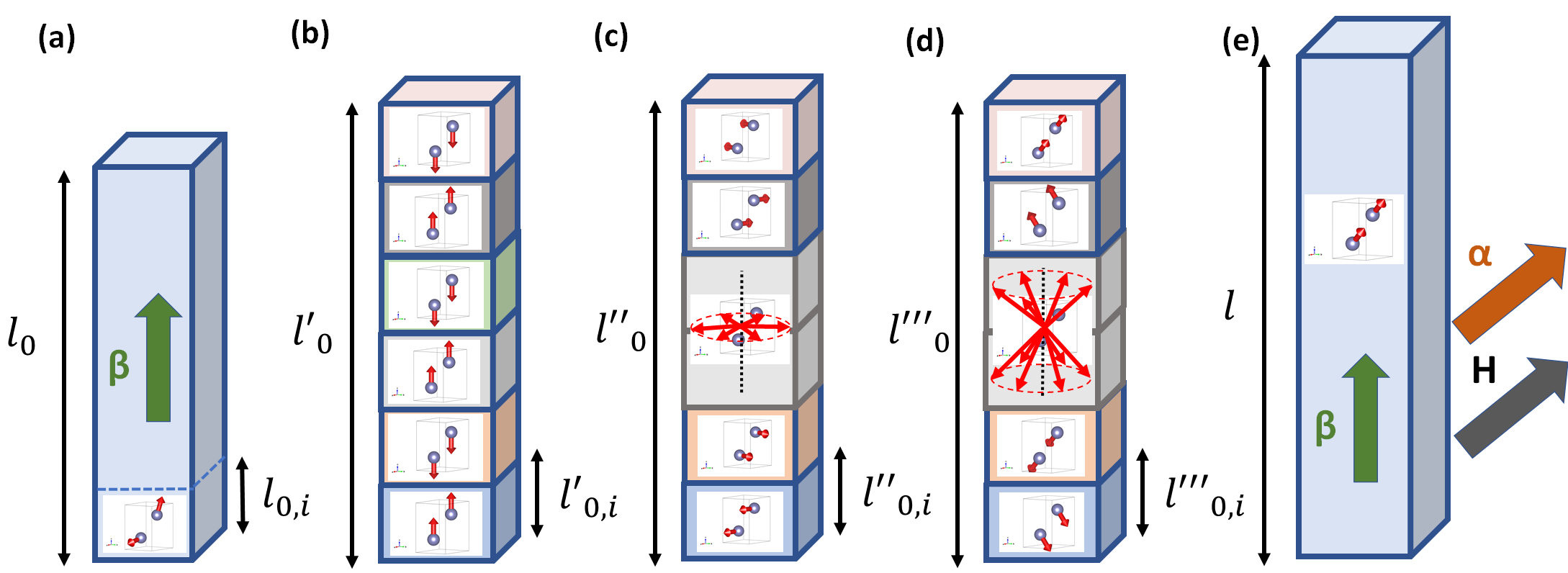}
\caption{(a) Ideal demagnetized state of a hexagonal single crystal with randomly oriented atomic magnetic moments. Demagnetized state with magnetic domains along all possible easy directions: (b) easy axis, (c) easy plane and (d) easy cone. (e) Saturated magnetic state.}
\label{fig:demag_hex}
\end{figure}

Let's now consider a single crystal with Hexagonal (I) symmetry with length $l_0$ in an arbitrary direction $\boldsymbol{\beta}$ at an initial demagnetized state with randomly oriented atomic magnetic moments. After the magnetization of the crystal is saturated in the direction $\boldsymbol{\alpha}$,  the length in the direction $\boldsymbol{\beta}$ is $l$, see Fig.\ref{fig:demag_hex}. In this case we have the following fractional change in length using Clark's notation \cite{maelas_publication2021,Clark}
%%%%%%%%%%%%%%%%%%%%%%%%%%%%%%%%%%%%%%%%%%%%
\begin{equation}
\begin{aligned}
     \frac{l- l_0}{l_0}\Bigg\vert_{\boldsymbol{\beta}}^{\boldsymbol{\alpha}} & =\lambda^{\alpha1,0}(\beta_x^2+\beta_y^2)+\lambda^{\alpha2,0}\beta_z^2+\lambda^{\alpha1,2}\left(\alpha_z^2-\frac{1}{3}\right)(\beta_x^2+\beta_y^2)\\
     & + \lambda^{\alpha2,2}\left(\alpha_z^2-\frac{1}{3}\right)\beta_z^2+\lambda^{\gamma,2}\left[\frac{1}{2}(\alpha_x^2-\alpha_y^2)(\beta_x^2-\beta_y^2)+2\alpha_x\alpha_y\beta_x\beta_y\right]\\
     & + 2\lambda^{\epsilon,2}(\alpha_x\alpha_z\beta_x\beta_z+\alpha_y\alpha_z\beta_y\beta_z),
    \label{eq:delta_l_hex_0}
\end{aligned}
\end{equation}
%%%%%%%%%%%%%%%%%%%%%%%%%%%%%%%%%%%%%%%%%%%%
As in the cubic case, we define a standard reference state as a demagnetized state with many ferromagnetic domains, where there are the same number of  domains oriented in each of the crystallographically equivalent directions of easy magnetization \cite{Birss}. For uniaxial crystals (hexagonal, trigonal and tetragonal symmetries) there are three types of easy directions: easy axis, easy plane and easy cone \cite{Coeybook}, see Fig.\ref{fig:typeani_hex}. 
%------------------------------
%------------------------------

The fractional change in length for hexagonal (I) crystals assuming an initial demagnetized state with domains pointing to all possible easy directions for easy axis and easy plane was previously derived by Birss \cite{Birss}. Here, we reproduce Birss's results using Clark's notation, and include the case with easy cone.  The considered demagnetized states are schematically represented in Fig. \ref{fig:demag_hex}, where the length in the direction $\boldsymbol{\beta}$ is labeled with  $l'_0$, $l''_0$ and $l'''_0$ for the cases with easy axis, easy plane and easy cone, respectively.
%------------------------------
%------------------------------

For the case with easy axis, there are two easy directions ($N_e=2$) along the lattice vector $\boldsymbol{c}$, parallel to the z-axis, $\boldsymbol{\alpha}_1=(0,0,1)$ and $\boldsymbol{\alpha}_2=(0,0,-1)$. Hence, applying the general formula given by Eq.\ref{eq:sum_delta_l_cub_I_4b} we obtain 
%%%%%%%%%%%%%%%%%%%%%%%%%%%%%%%%%%%%%%%%%%%%
\begin{equation}
\begin{aligned}
     \frac{l'_0-l_0}{l_0}\Bigg\vert_{\boldsymbol{\beta}}  & =\frac{1}{2}\sum_{j=1}^{2}\frac{l-l_{0}}{l_{0}}\Bigg\vert_{\boldsymbol{\beta}}^{\boldsymbol{\alpha}_j} \\ & =\lambda^{\alpha1,0}(\beta_x^2+\beta_y^2)+\lambda^{\alpha2,0}\beta_z^2+\frac{2}{3}\lambda^{\alpha1,2}(\beta_x^2+\beta_y^2) + \frac{2}{3}\lambda^{\alpha2,2}\beta_z^2.
    \label{eq:sum_delta_l_hex_1}
\end{aligned}
\end{equation}
%%%%%%%%%%%%%%%%%%%%%%%%%%%%%%%%%%%%%%%%%%%%
Hence, if the initial state is the demagnetized state with  the same number of  domains oriented in each of the easy  directions $\boldsymbol{\alpha}_1=(0,0,1)$ and $\boldsymbol{\alpha}_2=(0,0,-1)$, and the final state is the saturated state with magnetization along $\boldsymbol{\alpha}$, then the fractional change in length is
%%%%%%%%%%%%%%%%%%%%%%%%%%%%%%%%%%%%%%%%%%%%
\begin{equation}
\begin{aligned}
     \frac{l-l'_0}{l'_0}\Bigg\vert_{\boldsymbol{\beta}}^{\boldsymbol{\alpha}} & \cong \frac{l-l_0}{l_0}\Bigg\vert_{\boldsymbol{\beta}}^{\boldsymbol{\alpha}}-\frac{l'_0-l_0}{l_0}\Bigg\vert_{\boldsymbol{\beta}} =\lambda^{\alpha1,2}\left(\alpha_z^2-1\right)(\beta_x^2+\beta_y^2)\\
     & + \lambda^{\alpha2,2}\left(\alpha_z^2-1\right)\beta_z^2+\lambda^{\gamma,2}\left[\frac{1}{2}(\alpha_x^2-\alpha_y^2)(\beta_x^2-\beta_y^2)+2\alpha_x\alpha_y\beta_x\beta_y\right]\\
     & + 2\lambda^{\epsilon,2}(\alpha_x\alpha_z\beta_x\beta_z+\alpha_y\alpha_z\beta_y\beta_z).
    \label{eq:delta_l_hex_2}
\end{aligned}
\end{equation}
%%%%%%%%%%%%%%%%%%%%%%%%%%%%%%%%%%%%%%%%%%%%

In the case of easy plane, the number of easy directions is $N_e=2\pi$. Thus, we should formally replace the summation in Eq.\ref{eq:sum_delta_l_cub_I_4b} by an integral as
%%%%%%%%%%%%%%%%%%%%%%%%%%%%%%%%%%%%%%%%%%%%
\begin{equation}
\begin{aligned}
     \frac{l''_0-l_0}{l_0}\Bigg\vert_{\boldsymbol{\beta}}  & =\frac{1}{2\pi}\int_{0}^{2\pi}\frac{l-l_{0}}{l_{0}}\Bigg\vert_{\boldsymbol{\beta}}^{\boldsymbol{\alpha}=(\cos\theta,\sin\theta,0)}d\theta \\ & =\lambda^{\alpha1,0}(\beta_x^2+\beta_y^2)+\lambda^{\alpha2,0}\beta_z^2-\frac{1}{3}\lambda^{\alpha1,2}(\beta_x^2+\beta_y^2) - \frac{1}{3}\lambda^{\alpha2,2}\beta_z^2.
    \label{eq:sum_delta_l_hex_3}
\end{aligned}
\end{equation}
%%%%%%%%%%%%%%%%%%%%%%%%%%%%%%%%%%%%%%%%%%%%
Therefore, if the initial state is the demagnetized state with  the same number of  domains oriented in each of the easy  plane directions, and the final state is the saturated state with magnetization along $\boldsymbol{\alpha}$, then the fractional change in length is
%%%%%%%%%%%%%%%%%%%%%%%%%%%%%%%%%%%%%%%%%%%%
\begin{equation}
\begin{aligned}
     \frac{l-l''_0}{l''_0}\Bigg\vert_{\boldsymbol{\beta}}^{\boldsymbol{\alpha}} & \cong \frac{l-l_0}{l_0}\Bigg\vert_{\boldsymbol{\beta}}^{\boldsymbol{\alpha}}-\frac{l''_0-l_0}{l_0}\Bigg\vert_{\boldsymbol{\beta}} =\lambda^{\alpha1,2}\alpha_z^2(\beta_x^2+\beta_y^2) + \lambda^{\alpha2,2}\alpha_z^2\beta_z^2 \\
     & +\lambda^{\gamma,2}\left[\frac{1}{2}(\alpha_x^2-\alpha_y^2)(\beta_x^2-\beta_y^2)+2\alpha_x\alpha_y\beta_x\beta_y\right]\\
     & + 2\lambda^{\epsilon,2}(\alpha_x\alpha_z\beta_x\beta_z+\alpha_y\alpha_z\beta_y\beta_z).
    \label{eq:delta_l_hex_4}
\end{aligned}
\end{equation}
%%%%%%%%%%%%%%%%%%%%%%%%%%%%%%%%%%%%%%%%%%%%
The results obtained by Birss\cite{Birss} up to second order are recovered by using in the above equations the conversion formulas between Clark and Birss definition of the magnetostrictive coefficients \cite{Clark,maelas_publication2021}
%%%%%%%%%%%%%%%%%%%%%%%%%%%%%%%%%%%%%%%%%%%%
\begin{equation}
\begin{aligned}
     Q_0 & = \lambda^{\alpha1,0}+\frac{2}{3}\lambda^{\alpha1,2}\\
     Q_1 & =  \lambda^{\alpha2,0}+\frac{2}{3}\lambda^{\alpha2,2}-\lambda^{\alpha1,0}-\frac{2}{3}\lambda^{\alpha1,2}\\
     Q_2 & =  -\lambda^{\alpha1,2}-\frac{1}{2}\lambda^{\gamma,2}\\
     Q_4 & =  \lambda^{\alpha1,2}+\frac{1}{2}\lambda^{\gamma,2}-\lambda^{\alpha2,2}\\
     Q_6 & =  2\lambda^{\epsilon,2}\\
     Q_8 & = \lambda^{\gamma,2}.
    \label{eq:lamb_Birss_hex}
\end{aligned}
\end{equation}
%%%%%%%%%%%%%%%%%%%%%%%%%%%%%%%%%%%%%%%%%%%%

The magnetocrystalline anisotropy with easy cone is obtained when $K_1<0$ and $K_2>-K_1/2$ \cite{Coeybook}. The cone angle $\Omega$ with respect to the z-axis is given by \cite{Coeybook}
%%%%%%%%%%%%%%%%%%%%%%%%%%%%%%%%%%%%%%%%%%%%
\begin{equation}
\begin{aligned}
     \Omega=\arcsin{\sqrt{\frac{\vert K_1\vert}{2K_2}}}
    \label{eq:cone_angle}
\end{aligned}
\end{equation}
%%%%%%%%%%%%%%%%%%%%%%%%%%%%%%%%%%%%%%%%%%%%
In this case, the total number of easy directions is $N_e=4\pi\sin\Omega$. Formally, we should replace the summation in Eq.\ref{eq:sum_delta_l_cub_I_4b} by an integral as
%%%%%%%%%%%%%%%%%%%%%%%%%%%%%%%%%%%%%%%%%%%%
\begin{equation}
\begin{aligned}
     \frac{l'''_0-l_0}{l_0}\Bigg\vert_{\boldsymbol{\beta}}  & =\frac{1}{4\pi\sin\Omega}\int_{0}^{2\pi}\frac{l-l_{0}}{l_{0}}\Bigg\vert_{\boldsymbol{\beta}}^{\boldsymbol{\alpha}=(\sin\Omega\cos\theta,\sin\Omega\sin\theta,\cos\Omega)}\sin\Omega d\theta \\ 
     &+\frac{1}{4\pi\sin\Omega}\int_{0}^{2\pi}\frac{l-l_{0}}{l_{0}}\Bigg\vert_{\boldsymbol{\beta}}^{\boldsymbol{\alpha}=(\sin\Omega\cos\theta,\sin\Omega\sin\theta,-\cos\Omega)}\sin\Omega d\theta\\
     &=\lambda^{\alpha1,0}(\beta_x^2+\beta_y^2)+\lambda^{\alpha2,0}\beta_z^2\\
     & +\left(\cos^2\Omega-\frac{1}{3}\right)\lambda^{\alpha1,2}(\beta_x^2+\beta_y^2) +\left(\cos^2\Omega-\frac{1}{3}\right)\lambda^{\alpha2,2}\beta_z^2.
    \label{eq:sum_delta_l_hex_5}
\end{aligned}
\end{equation}
%%%%%%%%%%%%%%%%%%%%%%%%%%%%%%%%%%%%%%%%%%%%
 We see that setting the cone angle to zero ($\Omega=0$) in Eq.\ref{eq:sum_delta_l_hex_5} leads to the same result as in Eq.\ref{eq:sum_delta_l_hex_1} for case with the easy axis. Similarly, if we set the cone angle $\Omega=\pi/2$, then we recover the result for the case with easy plane given by  Eq.\ref{eq:sum_delta_l_hex_3}. Next, if the initial state is the demagnetized state with  the same number of  domains oriented in each of the easy  cone directions, and the final state is the saturated state with magnetization along $\boldsymbol{\alpha}$, then the fractional change in length is
%%%%%%%%%%%%%%%%%%%%%%%%%%%%%%%%%%%%%%%%%%%%
\begin{equation}
\begin{aligned}
     \frac{l-l'''_0}{l'''_0}\Bigg\vert_{\boldsymbol{\beta}}^{\boldsymbol{\alpha}} & \cong \frac{l-l_0}{l_0}\Bigg\vert_{\boldsymbol{\beta}}^{\boldsymbol{\alpha}}-\frac{l'''_0-l_0}{l_0}\Bigg\vert_{\boldsymbol{\beta}} \\
  & =  \left(\alpha_z^2-\cos^2\Omega\right) \lambda^{\alpha1,2}(\beta_x^2+\beta_y^2) + \left(\alpha_z^2-\cos^2\Omega\right)\lambda^{\alpha2,2}\beta_z^2\\
     & +\lambda^{\gamma,2}\left[\frac{1}{2}(\alpha_x^2-\alpha_y^2)(\beta_x^2-\beta_y^2)+2\alpha_x\alpha_y\beta_x\beta_y\right]\\
     & + 2\lambda^{\epsilon,2}(\alpha_x\alpha_z\beta_x\beta_z+\alpha_y\alpha_z\beta_y\beta_z).
    \label{eq:delta_l_hex_6}
\end{aligned}
\end{equation}
%%%%%%%%%%%%%%%%%%%%%%%%%%%%%%%%%%%%%%%%%%%%
We see that assuming easy cone directions in the initial demagnetized state allows to derive general and compact formulas of the fractional change in length for uniaxial crystals that  include the easy axis and easy plane as particular cases when the cone angle is $\Omega=0$ and $\Omega=\pi/2$, respectively. For instance, Eqs.\ref{eq:delta_l_hex_2} and \ref{eq:delta_l_hex_4} are recovered by setting $\Omega=0$ and $\Omega=\pi/2$ in Eq.\ref{eq:delta_l_hex_6}, respectively. Similarly, using Mason's definitions of the magnetostrictive coefficients\cite{Mason,maelas_publication2021}, we find
%%%%%%%%%%%%%%%%%%%%%%%%%%%%%%%%%%%%%%%%%%%%
\begin{equation}
\begin{aligned}
     \frac{l- l'''_0}{l'''_0}\Bigg\vert_{\boldsymbol{\beta}}^{\boldsymbol{\alpha}} & =\lambda_A[(\alpha_x\beta_x+\alpha_y\beta_y)^2-(\alpha_x\beta_x+\alpha_y\beta_y)\alpha_z\beta_z-\frac{1}{2} (\beta_x^2 + \beta_y^2) \sin^2\Omega]\\
     & + \lambda_B[(\cos^2\Omega-\alpha_z^2)(1-\beta_z^2)-(\alpha_x\beta_x+\alpha_y\beta_y)^2+\frac{1}{2} (\beta_x^2 + \beta_y^2) \sin^2\Omega]\\
     & +\lambda_C[(\cos^2\Omega-\alpha_z^2)\beta_z^2-(\alpha_x\beta_x+\alpha_y\beta_y)\alpha_z\beta_z] + 4\lambda_D(\alpha_x\beta_x+\alpha_y\beta_y)\alpha_z\beta_z,
    \label{eq:delta_l_Mason_hex_I}
\end{aligned}
\end{equation}
%%%%%%%%%%%%%%%%%%%%%%%%%%%%%%%%%%%%%%%%%%%%
where the original result derived by Mason\cite{Mason},  assuming a demagnetized condition with equal
numbers of domains directed along the z-axis (easy axis), is recovered by setting the cone angle $\Omega=0$.

Lastly, let's compute the fractional change in length for polycrystals with hexagonal (I) crystal symmetry. If we consider a reference initial demagnetized state with randomly oriented atomic magnetic moments and final saturated state,  then we should replace Eq.\ref{eq:delta_l_hex_0} in Eq.\ref{eq:sum_delta_l_cub_I_6}
%%%%%%%%%%%%%%%%%%%%%%%%%%%%%%%%%%%%%%%%%%%
\begin{equation}
     <\frac{l-l_0}{l_0}\Bigg\vert_{\boldsymbol{\beta}}^{\boldsymbol{\alpha}}> = \xi+\eta(\boldsymbol{\alpha}\cdot\boldsymbol{\beta})^2,
    \label{eq:delta_l_hex_poly}
\end{equation}
%%%%%%%%%%%%%%%%%%%%%%%%%%%%%%%%%%%%%%%%%%%%
where  
%%%%%%%%%%%%%%%%%%%%%%%%%%%%%%%%%%%%%%%%%%%%
\begin{eqnarray}
 \xi & = & \frac{10}{15}\lambda^{\alpha 1,0}+\frac{1}{3}\lambda^{\alpha 2,0}+\frac{2}{45}\lambda^{\alpha 1,2}-\frac{2}{45}\lambda^{\alpha 2,2}-\frac{2}{15}\lambda^{\epsilon ,2}-\frac{2}{15}\lambda^{\gamma,2},\\
 \eta & = & -\frac{2}{15}\lambda^{\alpha1,2}+\frac{2}{15}\lambda^{\alpha2,2}+\frac{2}{5}\lambda^{\epsilon ,2}+\frac{2}{5}\lambda^{\gamma,2}.
\label{eq:eta_hex_0}
\end{eqnarray}
%%%%%%%%%%%%%%%%%%%%%%%%%%%%%%%%%%%
For the case with a reference initial demagnetized state made by domains aligned to all possible easy directions and final saturated state, we only need to compute the crystal with easy cone because it includes the easy axis and easy plane as particular cases of the cone angle. Thus, inserting Eq.\ref{eq:delta_l_hex_6} in Eq.\ref{eq:sum_delta_l_cub_I_6} gives
%%%%%%%%%%%%%%%%%%%%%%%%%%%%%%%%%%%%%%%%%%%
\begin{equation}
     <\frac{l-l'''_0}{l'''_0}\Bigg\vert_{\boldsymbol{\beta}}^{\boldsymbol{\alpha}}> = \xi(\Omega)+\eta(\boldsymbol{\alpha}\cdot\boldsymbol{\beta})^2,
    \label{eq:delta_l_hex_poly_2}
\end{equation}
%%%%%%%%%%%%%%%%%%%%%%%%%%%%%%%%%%%%%%%%%%%%
where  
%%%%%%%%%%%%%%%%%%%%%%%%%%%%%%%%%%%%%%%%%%%%
\begin{eqnarray}
 \xi(\Omega) & = & \frac{4}{15}\lambda^{\alpha 1,2}+\frac{1}{15}\lambda^{\alpha 2,2}-\frac{2}{15}\lambda^{\epsilon ,2}-\frac{2}{15}\lambda^{\gamma,2}-\frac{1}{3}(2\lambda^{\alpha 1,2}+\lambda^{\alpha 2,2})\cos^2\Omega,\nonumber\\
 \eta & = & -\frac{2}{15}\lambda^{\alpha1,2}+\frac{2}{15}\lambda^{\alpha2,2}+\frac{2}{5}\lambda^{\epsilon ,2}+\frac{2}{5}\lambda^{\gamma,2}.
\label{eq:eta_hex_1b}
\end{eqnarray}
%%%%%%%%%%%%%%%%%%%%%%%%%%%%%%%%%%%
The cases with easy axis and easy plane are obtained by setting the cone angle $\Omega=0$ and $\Omega=\pi/2$ in Eq.\ref{eq:eta_hex_1b}, respectively. By doing so, one recovers the results derived by Birss \cite{Birss,maelas_publication2021} with the help of the conversion formulas between Clark and Birss definitions given by Eq.\ref{eq:lamb_Birss_hex}.

\subsubsection{Trigonal (I)}
\label{subsubsection:trig_I_poly}

The case with trigonal (I) symmetry is very similar to the hexagonal one since both are uniaxial crystals. Thus, we use the same notation as shown in Fig.\ref{fig:demag_hex}. Let's first consider a single crystal with Trigonal (I) symmetry with length $l_0$ in an arbitrary direction $\boldsymbol{\beta}$ in an initial demagnetized state with randomly oriented atomic magnetic moments. After the magnetization of the crystal is saturated in the direction $\boldsymbol{\alpha}$,  the length in the direction $\boldsymbol{\beta}$ is $l$. In this case we have the following fractional change in length  \cite{maelas_publication2021,Cullen}
%%%%%%%%%%%%%%%%%%%%%%%%%%%%%%%%%%%%%%%%%%%%
\begin{equation}
\begin{aligned}
     \frac{l- l_0}{l_0}\Bigg\vert_{\boldsymbol{\beta}}^{\boldsymbol{\alpha}} & =\lambda^{\alpha1,0}(\beta_x^2+\beta_y^2)+\lambda^{\alpha2,0}\beta_z^2+\lambda^{\alpha1,2}\left(\alpha_z^2-\frac{1}{3}\right)(\beta_x^2+\beta_y^2)\\
     & + \lambda^{\alpha2,2}\left(\alpha_z^2-\frac{1}{3}\right)\beta_z^2+\lambda^{\gamma,1}\left[\frac{1}{2}(\alpha_x^2-\alpha_y^2)(\beta_x^2-\beta_y^2)+2\alpha_x\alpha_y\beta_x\beta_y\right]\\
     & + \lambda^{\gamma,2}(\alpha_x\alpha_z\beta_x\beta_z+\alpha_y\alpha_z\beta_y\beta_z)+\lambda_{12}\left[\frac{1}{2}\alpha_y\alpha_z(\beta_x^2-\beta_y^2)+\alpha_x\alpha_z\beta_x\beta_y\right]\\
     & +\lambda_{21}\left[\frac{1}{2}(\alpha_x^2-\alpha_y^2)\beta_y\beta_z+\alpha_x\alpha_y\beta_x\beta_z\right].
    \label{eq:delta_l_trig_I}
\end{aligned}
\end{equation}
%%%%%%%%%%%%%%%%%%%%%%%%%%%%%%%%%%%%%%%%%%%%
Following the same analysis as in the case of hexagonal (I) symmetry, we find that assuming a reference initial demagnetized state with domains pointing to all possible easy cone directions and a final saturated state leads to the fractional change in length 
%%%%%%%%%%%%%%%%%%%%%%%%%%%%%%%%%%%%%%%%%%%%
\begin{equation}
\begin{aligned}
     \frac{l- l'''_0}{l'''_0}\Bigg\vert_{\boldsymbol{\beta}}^{\boldsymbol{\alpha}} & =\lambda^{\alpha1,2}\left(\alpha_z^2-\cos^2\Omega\right)(\beta_x^2+\beta_y^2) + \lambda^{\alpha2,2}\left(\alpha_z^2-\cos^2\Omega\right)\beta_z^2\\
     &+\lambda^{\gamma,1}\left[\frac{1}{2}(\alpha_x^2-\alpha_y^2)(\beta_x^2-\beta_y^2)+2\alpha_x\alpha_y\beta_x\beta_y\right]\\
     & + \lambda^{\gamma,2}(\alpha_x\alpha_z\beta_x\beta_z+\alpha_y\alpha_z\beta_y\beta_z)+\lambda_{12}\left[\frac{1}{2}\alpha_y\alpha_z(\beta_x^2-\beta_y^2)+\alpha_x\alpha_z\beta_x\beta_y\right]\\
     & +\lambda_{21}\left[\frac{1}{2}(\alpha_x^2-\alpha_y^2)\beta_y\beta_z+\alpha_x\alpha_y\beta_x\beta_z\right],
    \label{eq:delta_l_trig_Ib}
\end{aligned}
\end{equation}
%%%%%%%%%%%%%%%%%%%%%%%%%%%%%%%%%%%%%%%%%%%%
where the cases with easy axis and easy plane are obtained by setting the cone angle $\Omega=0$ and $\Omega=\pi/2$, respectively.

In the case of polycrystals, if we consider an initial demagnetized state with randomly oriented atomic magnetic moments and final saturated state, then we find 
%%%%%%%%%%%%%%%%%%%%%%%%%%%%%%%%%%%%%%%%%%%
\begin{equation}
     <\frac{l- l_0}{l_0}\Bigg\vert_{\boldsymbol{\beta}}^{\boldsymbol{\alpha}}> = \xi+\eta(\boldsymbol{\alpha}\cdot\boldsymbol{\beta})^2,
    \label{eq:delta_l_trig_poly}
\end{equation}
%%%%%%%%%%%%%%%%%%%%%%%%%%%%%%%%%%%%%%%%%%%%
where  
%%%%%%%%%%%%%%%%%%%%%%%%%%%%%%%%%%%%%%%%%%%%
\begin{eqnarray}
 \xi & = & \frac{10}{15}\lambda^{\alpha 1,0}+\frac{1}{3}\lambda^{\alpha 2,0}+\frac{2}{45}\lambda^{\alpha 1,2}-\frac{2}{45}\lambda^{\alpha 2,2}-\frac{2}{15}\lambda^{\gamma ,1}-\frac{1}{15}\lambda^{\gamma,2},\\
 \eta & = & -\frac{2}{15}\lambda^{\alpha1,2}+\frac{2}{15}\lambda^{\alpha2,2}+\frac{2}{5}\lambda^{\gamma ,1}+\frac{1}{5}\lambda^{\gamma,2}.
\label{eq:eta_trig_poly_0}
\end{eqnarray}
%%%%%%%%%%%%%%%%%%%%%%%%%%%%%%%%%%%
 On the other hand, if the reference initial state of the polycrystal is a demagnetized state with domains pointing to all easy cone directions and the final state is the saturated state, then we obtain
%%%%%%%%%%%%%%%%%%%%%%%%%%%%%%%%%%%%%%%%%%%
\begin{equation}
     <\frac{l-l'''_0}{l'''_0}\Bigg\vert_{\boldsymbol{\beta}}^{\boldsymbol{\alpha}}> = \xi(\Omega)+\eta(\boldsymbol{\alpha}\cdot\boldsymbol{\beta})^2,
    \label{eq:delta_l_trig_poly_2}
\end{equation}
%%%%%%%%%%%%%%%%%%%%%%%%%%%%%%%%%%%%%%%%%%%%
where  
%%%%%%%%%%%%%%%%%%%%%%%%%%%%%%%%%%%%%%%%%%%%
\begin{eqnarray}
 \xi(\Omega) & = & \frac{4}{15}\lambda^{\alpha 1,2}+\frac{1}{15}\lambda^{\alpha 2,2}-\frac{2}{15}\lambda^{\gamma ,1}-\frac{1}{15}\lambda^{\gamma,2}-\frac{1}{3}(2\lambda^{\alpha 1,2}+\lambda^{\alpha 2,2})\cos^2\Omega,\nonumber\\
 \eta & = & -\frac{2}{15}\lambda^{\alpha1,2}+\frac{2}{15}\lambda^{\alpha2,2}+\frac{2}{5}\lambda^{\gamma ,1}+\frac{1}{5}\lambda^{\gamma,2}.
\label{eq:eta_trig_1b}
\end{eqnarray}
%%%%%%%%%%%%%%%%%%%%%%%%%%%%%%%%%%%
Again, the corresponding cases with easy axis and easy plane are obtained by setting the cone angle $\Omega=0$ and $\Omega=\pi/2$, respectively. Interestingly, we see that $\xi$ and $\eta$ do  not depend on $\lambda_{12}$ and $\lambda_{21}$ since their angular dependence gives zero in the integration of Eq.\ref{eq:sum_delta_l_cub_I_6}. Further studies are needed to clarify the validity of this result.

\subsubsection{Tetragonal (I)}
\label{subsubsection:tetra_I_poly}

The case with tetragonal (I) symmetry is also very similar to the hexagonal and trigonal ones. Again, we use the same notation as shown in Fig.\ref{fig:demag_hex}. Let's start by considering a single crystal with Tetragonal (I) symmetry with length $l_0$ in an arbitrary direction $\boldsymbol{\beta}$ in an initial demagnetized state with randomly oriented atomic magnetic moments. After the magnetization of the crystal is saturated in the direction $\boldsymbol{\alpha}$,  the length in the direction $\boldsymbol{\beta}$ is $l$. In this case we have the following fractional change in length  \cite{maelas_publication2021,Cullen}
%%%%%%%%%%%%%%%%%%%%%%%%%%%%%%%%%%%%%%%%%%%%
\begin{equation}
\begin{aligned}
     \frac{l- l_0}{l_0}\Bigg\vert_{\boldsymbol{\beta}}^{\boldsymbol{\alpha}} & =\lambda^{\alpha1,0}(\beta_x^2+\beta_y^2)+\lambda^{\alpha2,0}\beta_z^2+\lambda^{\alpha1,2}\left(\alpha_z^2-\frac{1}{3}\right)(\beta_x^2+\beta_y^2)\\
     & + \lambda^{\alpha2,2}\left(\alpha_z^2-\frac{1}{3}\right)\beta_z^2+\frac{1}{2}\lambda^{\gamma,2}(\alpha_x^2-\alpha_y^2)(\beta_x^2-\beta_y^2)+2\lambda^{\delta,2}\alpha_x\alpha_y\beta_x\beta_y\\
     & + 2\lambda^{\epsilon,2}(\alpha_x\alpha_z\beta_x\beta_z+\alpha_y\alpha_z\beta_y\beta_z),
    \label{eq:delta_l_tet_I}
\end{aligned}
\end{equation}
%%%%%%%%%%%%%%%%%%%%%%%%%%%%%%%%%%%%%%%%%%%%
Following the same analysis as in the case of hexagonal (I) crystal, we find that assuming a reference initial demagnetized state with domains pointing to all possible easy cone directions and a final saturated state leads to the fractional change in length 
%%%%%%%%%%%%%%%%%%%%%%%%%%%%%%%%%%%%%%%%%%%%
\begin{equation}
\begin{aligned}
     \frac{l- l'''_0}{l'''_0}\Bigg\vert_{\boldsymbol{\beta}}^{\boldsymbol{\alpha}} & =\lambda^{\alpha1,2}\left(\alpha_z^2-\cos^2\Omega\right)(\beta_x^2+\beta_y^2) + \lambda^{\alpha2,2}\left(\alpha_z^2-\cos^2\Omega\right)\beta_z^2\\
     & +\frac{1}{2}\lambda^{\gamma,2}(\alpha_x^2-\alpha_y^2)(\beta_x^2-\beta_y^2)+2\lambda^{\delta,2}\alpha_x\alpha_y\beta_x\beta_y\\
     & + 2\lambda^{\epsilon,2}(\alpha_x\alpha_z\beta_x\beta_z+\alpha_y\alpha_z\beta_y\beta_z),
    \label{eq:delta_l_tet_I_2}
\end{aligned}
\end{equation}
%%%%%%%%%%%%%%%%%%%%%%%%%%%%%%%%%%%%%%%%%%%%
where the cases with easy axis and easy plane are obtained by setting the cone angle $\Omega=0$ and $\Omega=\pi/2$, respectively.

In the case of polycrystals, if we consider an initial demagnetized state with randomly oriented atomic magnetic moments and final saturated state, then we find 
%%%%%%%%%%%%%%%%%%%%%%%%%%%%%%%%%%%%%%%%%%%
\begin{equation}
     <\frac{l- l_0}{l_0}\Bigg\vert_{\boldsymbol{\beta}}^{\boldsymbol{\alpha}}> = \xi+\eta(\boldsymbol{\alpha}\cdot\boldsymbol{\beta})^2,
    \label{eq:delta_l_tetra_poly}
\end{equation}
%%%%%%%%%%%%%%%%%%%%%%%%%%%%%%%%%%%%%%%%%%%%
where  
%%%%%%%%%%%%%%%%%%%%%%%%%%%%%%%%%%%%%%%%%%%%
\begin{eqnarray}
 \xi & = & \frac{10}{15}\lambda^{\alpha 1,0}+\frac{1}{3}\lambda^{\alpha 2,0}+\frac{2}{45}\lambda^{\alpha 1,2}-\frac{2}{45}\lambda^{\alpha 2,2}-\frac{2}{15}\lambda^{\epsilon ,2}-\frac{1}{15}\lambda^{\gamma,2}-\frac{1}{15}\lambda^{\delta,2},\\
 \eta & = & -\frac{2}{15}\lambda^{\alpha1,2}+\frac{2}{15}\lambda^{\alpha2,2}+\frac{2}{5}\lambda^{\epsilon ,2}+\frac{1}{5}\lambda^{\gamma,2}+\frac{1}{5}\lambda^{\delta,2}.
\label{eq:eta_tetra_poly_0}
\end{eqnarray}
%%%%%%%%%%%%%%%%%%%%%%%%%%%%%%%%%%%
 On the other hand, if the reference initial state of the polycrystal is a demagnetized state with domains pointing to all easy cone directions and the final state is the saturated state, then we obtain
%%%%%%%%%%%%%%%%%%%%%%%%%%%%%%%%%%%%%%%%%%%
\begin{equation}
     <\frac{l-l'''_0}{l'''_0}\Bigg\vert_{\boldsymbol{\beta}}^{\boldsymbol{\alpha}}> = \xi(\Omega)+\eta(\boldsymbol{\alpha}\cdot\boldsymbol{\beta})^2,
    \label{eq:delta_l_tetra_poly_2}
\end{equation}
%%%%%%%%%%%%%%%%%%%%%%%%%%%%%%%%%%%%%%%%%%%%
where  
%%%%%%%%%%%%%%%%%%%%%%%%%%%%%%%%%%%%%%%%%%%%
\begin{eqnarray}
\xi(\Omega) & = & \frac{4}{15}\lambda^{\alpha 1,2}+\frac{1}{15}\lambda^{\alpha 2,2}-\frac{2}{15}\lambda^{\epsilon ,2}-\frac{1}{15}\lambda^{\gamma,2}-\frac{1}{15}\lambda^{\delta,2}\nonumber\\
& - & \frac{1}{3}(2\lambda^{\alpha 1,2}+\lambda^{\alpha 2,2})\cos^2\Omega,\nonumber\\
  \eta & = & -\frac{2}{15}\lambda^{\alpha1,2}+\frac{2}{15}\lambda^{\alpha2,2}+\frac{2}{5}\lambda^{\epsilon ,2}+\frac{1}{5}\lambda^{\gamma,2}+\frac{1}{5}\lambda^{\delta,2}.
\label{eq:eta_tetra_1b}
\end{eqnarray}
%%%%%%%%%%%%%%%%%%%%%%%%%%%%%%%%%%%
Again, the corresponding cases with easy axis and easy plane are obtained by setting the cone angle $\Omega=0$ and $\Omega=\pi/2$, respectively.

\subsubsection{Orthorhombic}
\label{subsubsection:orto_poly}

\begin{figure}[h!]
\centering
\includegraphics[width=\columnwidth ,angle=0]{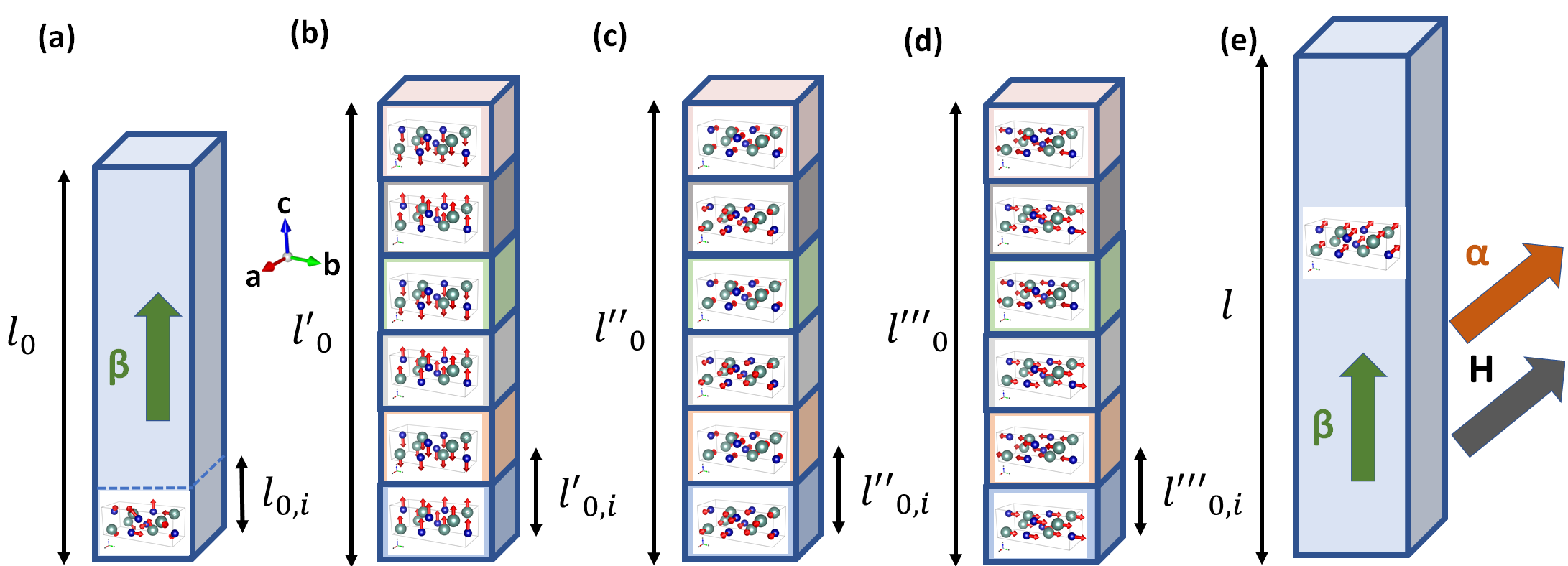}
\caption{(a) Ideal demagnetized state of a orthorhombic single crystal with randomly oriented atomic magnetic moments. Demagnetized state with magnetic domains along all possible easy directions: (b) easy axis along the lattice vector $\boldsymbol{c}$ parallel to z-axis, (c) easy axis along the lattice vector $\boldsymbol{a}$ parallel to x-axis and (d) easy axis along the lattice vector $\boldsymbol{b}$ parallel to y-axis.  (e) Saturated magnetic state. MAELAS uses the IEEE lattice convention $c<a<b$.}
\label{fig:demag_orto}
\end{figure}

Now, we analyze the case of Orthorhombic crystals. Let's  consider a single crystal with Orthorhombic symmetry and  length $l_0$ in an arbitrary direction $\boldsymbol{\beta}$ in an initial demagnetized state with randomly oriented atomic magnetic moments. After the magnetization of the crystal is saturated in the direction $\boldsymbol{\alpha}$,  the length in the direction $\boldsymbol{\beta}$ is $l$, see Fig.\ref{fig:demag_orto}. In this case, we have the following fractional change in length  \cite{maelas_publication2021,Mason}
%%%%%%%%%%%%%%%%%%%%%%%%%%%%%%%%%%%%%%%%%%%%
\begin{equation}
\begin{aligned}
     \frac{l- l_0}{l_0}\Bigg\vert_{\boldsymbol{\beta}}^{\boldsymbol{\alpha}} & =\lambda^{\alpha1,0}\beta_x^2+\lambda^{\alpha2,0}\beta_y^2+\lambda^{\alpha3,0}\beta_z^2+\lambda_1(\alpha_x^2\beta_x^2-\alpha_x\alpha_y\beta_x\beta_y-\alpha_x\alpha_z\beta_x\beta_z)\\
     & +\lambda_2(\alpha_y^2\beta_x^2-\alpha_x\alpha_y\beta_x\beta_y)+\lambda_3(\alpha_x^2\beta_y^2-\alpha_x\alpha_y\beta_x\beta_y)\\
     & +\lambda_4(\alpha_y^2\beta_y^2-\alpha_x\alpha_y\beta_x\beta_y-\alpha_y\alpha_z\beta_y\beta_z)+\lambda_5(\alpha_x^2\beta_z^2-\alpha_x\alpha_z\beta_x\beta_z)\\
     & + \lambda_6(\alpha_y^2\beta_z^2-\alpha_y\alpha_z\beta_y\beta_z)+4\lambda_7\alpha_x\alpha_y\beta_x\beta_y+4\lambda_8\alpha_x\alpha_z\beta_x\beta_z+4\lambda_9\alpha_y\alpha_z\beta_y\beta_z.
    \label{eq:delta_l_ortho}
\end{aligned}
\end{equation}
%%%%%%%%%%%%%%%%%%%%%%%%%%%%%%%%%%%%%%%%%%%%
As in the previous crystal symmetries, we define a standard reference state as a demagnetized state with many ferromagnetic domains, where there are the same number of  domains oriented in each of the crystallographically equivalent directions of easy magnetization \cite{Birss}. 
%------------------------------
\begin{figure}[h!]
\centering
\includegraphics[width=0.7\columnwidth ,angle=0]{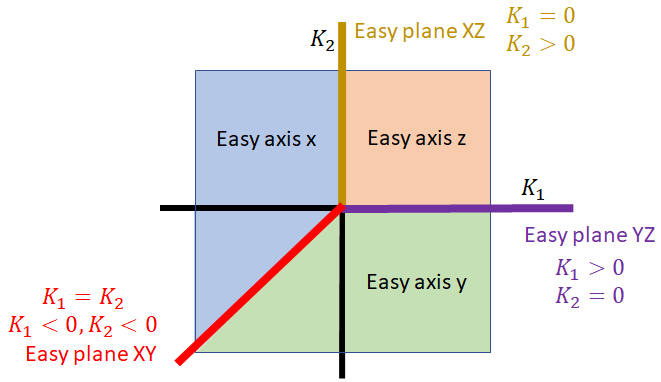}
\caption{Phase diagram of magnetocrystalline anisotropy energy for orthorhombic crystals.}
\label{fig:typeani_ort}
\end{figure}
%------------------------------

The magnetocrystalline anisotropy energy in an unstrained orthorhombic crystal up to second-order in $\alpha$ is\cite{Mason,maelas_publication2021}
%%%%%%%%%%%%%%%%%%%%%%%%%%%%%%%%%
\begin{equation}
\begin{aligned}
    \frac{E_{K}^0}{V_0}= K_0+K_1\alpha_x^2+K_2\alpha_y^2. 
\label{eq:E_mca_ortho}     
\end{aligned}
\end{equation}
%%%%%%%%%%%%%%%%%%%%%%%%%%%%%%%%%
The analysis of this equation shows that there are two types of easy directions: easy axis and easy plane. However, the case with easy axis is more relevant since the easy plane is a singular transition state between two different easy axis directions, see Fig.\ref{fig:typeani_ort}. Thus, here we only study the three possible easy axis directions: easy axis along the lattice vector $\boldsymbol{c}$ parallel to z-axis, easy axis along the lattice vector $\boldsymbol{a}$ parallel to x-axis, and easy axis along the lattice vector $\boldsymbol{b}$ parallel to y-axis. Hence, in each case there are two easy directions $N_e=2$. The considered demagnetized states are schematically represented in Fig. \ref{fig:demag_orto}, where the length in the direction $\boldsymbol{\beta}$ is labeled with  $l'_0$, $l''_0$ and $l'''_0$ for the cases with easy axis along the lattice vector $\boldsymbol{c}$, $\boldsymbol{a}$ and $\boldsymbol{b}$, respectively. We assume the same IEEE lattice convention ($c<a<b$) that is used in MAELAS and AELAS \cite{AELAS}.
%------------------------------
%------------------------------

Applying the same procedure used for cubic and hexagonal crystals, we obtain the following fractional change in length assuming an initial demagnetized state with domains pointing to all easy directions parallel to the lattice vector $\boldsymbol{c}$ and final saturated state
%%%%%%%%%%%%%%%%%%%%%%%%%%%%%%%%%%%%%%%%%%%%
\begin{equation}
\begin{aligned}
     \frac{l- l'_0}{l'_0}\Bigg\vert_{\boldsymbol{\beta}}^{\boldsymbol{\alpha}} & \cong \frac{l- l_0}{l_0}\Bigg\vert_{\boldsymbol{\beta}}^{\boldsymbol{\alpha}}-\frac{l'_0- l_0}{l_0}\Bigg\vert_{\boldsymbol{\beta}}\\& =\lambda_1(\alpha_x^2\beta_x^2-\alpha_x\alpha_y\beta_x\beta_y-\alpha_x\alpha_z\beta_x\beta_z)\\
     & +\lambda_2(\alpha_y^2\beta_x^2-\alpha_x\alpha_y\beta_x\beta_y)+\lambda_3(\alpha_x^2\beta_y^2-\alpha_x\alpha_y\beta_x\beta_y)\\
     & +\lambda_4(\alpha_y^2\beta_y^2-\alpha_x\alpha_y\beta_x\beta_y-\alpha_y\alpha_z\beta_y\beta_z)+\lambda_5(\alpha_x^2\beta_z^2-\alpha_x\alpha_z\beta_x\beta_z)\\
     & + \lambda_6(\alpha_y^2\beta_z^2-\alpha_y\alpha_z\beta_y\beta_z)+4\lambda_7\alpha_x\alpha_y\beta_x\beta_y+4\lambda_8\alpha_x\alpha_z\beta_x\beta_z+4\lambda_9\alpha_y\alpha_z\beta_y\beta_z.
    \label{eq:delta_l_ortho_1}
\end{aligned}
\end{equation}
%%%%%%%%%%%%%%%%%%%%%%%%%%%%%%%%%%%%%%%%%%%%
Thus, we recover the result originally derived by Mason \cite{Mason}. In the case with easy axis parallel to the lattice vector $\boldsymbol{a}$ we have
%%%%%%%%%%%%%%%%%%%%%%%%%%%%%%%%%%%%%%%%%%%%
\begin{equation}
\begin{aligned}
     \frac{l- l''_0}{l''_0}\Bigg\vert_{\boldsymbol{\beta}}^{\boldsymbol{\alpha}} & \cong \frac{l- l_0}{l_0}\Bigg\vert_{\boldsymbol{\beta}}^{\boldsymbol{\alpha}}-\frac{l''_0- l_0}{l_0}\Bigg\vert_{\boldsymbol{\beta}}\\& =\lambda_1([\alpha_x^2-1]\beta_x^2-\alpha_x\alpha_y\beta_x\beta_y-\alpha_x\alpha_z\beta_x\beta_z)\\
     & +\lambda_2(\alpha_y^2\beta_x^2-\alpha_x\alpha_y\beta_x\beta_y)+\lambda_3([\alpha_x^2-1]\beta_y^2-\alpha_x\alpha_y\beta_x\beta_y)\\
     & +\lambda_4(\alpha_y^2\beta_y^2-\alpha_x\alpha_y\beta_x\beta_y-\alpha_y\alpha_z\beta_y\beta_z)+\lambda_5([\alpha_x^2-1]\beta_z^2-\alpha_x\alpha_z\beta_x\beta_z)\\
     & + \lambda_6(\alpha_y^2\beta_z^2-\alpha_y\alpha_z\beta_y\beta_z)+4\lambda_7\alpha_x\alpha_y\beta_x\beta_y+4\lambda_8\alpha_x\alpha_z\beta_x\beta_z+4\lambda_9\alpha_y\alpha_z\beta_y\beta_z,
    \label{eq:delta_l_ortho_2}
\end{aligned}
\end{equation}
%%%%%%%%%%%%%%%%%%%%%%%%%%%%%%%%%%%%%%%%%%%%
while for the case of easy axis parallel to the lattice vector $\boldsymbol{b}$ we find
%%%%%%%%%%%%%%%%%%%%%%%%%%%%%%%%%%%%%%%%%%%%
\begin{equation}
\begin{aligned}
     \frac{l- l'''_0}{l'''_0}\Bigg\vert_{\boldsymbol{\beta}}^{\boldsymbol{\alpha}} & \cong \frac{l- l_0}{l_0}\Bigg\vert_{\boldsymbol{\beta}}^{\boldsymbol{\alpha}}-\frac{l'''_0- l_0}{l_0}\Bigg\vert_{\boldsymbol{\beta}}\\& =\lambda_1(\alpha_x^2\beta_x^2-\alpha_x\alpha_y\beta_x\beta_y-\alpha_x\alpha_z\beta_x\beta_z)\\
     & +\lambda_2([\alpha_y^2-1]\beta_x^2-\alpha_x\alpha_y\beta_x\beta_y)+\lambda_3(\alpha_x^2\beta_y^2-\alpha_x\alpha_y\beta_x\beta_y)\\
     & +\lambda_4([\alpha_y^2-1]\beta_y^2-\alpha_x\alpha_y\beta_x\beta_y-\alpha_y\alpha_z\beta_y\beta_z)+\lambda_5(\alpha_x^2\beta_z^2-\alpha_x\alpha_z\beta_x\beta_z)\\
     & + \lambda_6([\alpha_y^2-1]\beta_z^2-\alpha_y\alpha_z\beta_y\beta_z)+4\lambda_7\alpha_x\alpha_y\beta_x\beta_y+4\lambda_8\alpha_x\alpha_z\beta_x\beta_z\\
     & +4\lambda_9\alpha_y\alpha_z\beta_y\beta_z.
    \label{eq:delta_l_ortho_3}
\end{aligned}
\end{equation}
%%%%%%%%%%%%%%%%%%%%%%%%%%%%%%%%%%%%%%%%%%%%

In the case of polycrystals, if we consider an initial demagnetized state with randomly oriented atomic magnetic moments and final saturated state, then we find 
%%%%%%%%%%%%%%%%%%%%%%%%%%%%%%%%%%%%%%%%%%%
\begin{equation}
     <\frac{l- l_0}{l_0}\Bigg\vert_{\boldsymbol{\beta}}^{\boldsymbol{\alpha}}> = \xi+\eta(\boldsymbol{\alpha}\cdot\boldsymbol{\beta})^2,
    \label{eq:delta_l_orto_poly_0}
\end{equation}
%%%%%%%%%%%%%%%%%%%%%%%%%%%%%%%%%%%%%%%%%%%%
where  
%%%%%%%%%%%%%%%%%%%%%%%%%%%%%%%%%%%%%%%%%%%%
\begin{eqnarray}
 \xi & = & \frac{1}{3}(\lambda^{\alpha1,0}+\lambda^{\alpha2,0}+\lambda^{\alpha3,0})+\frac{2}{15}(\lambda_1+\lambda_4-\lambda_7-\lambda_8-\lambda_9)\nonumber\\ & + &\frac{1}{6}(\lambda_2+\lambda_3+\lambda_5+\lambda_6),\\
 \eta & = & -\frac{1}{15}\lambda_1-\frac{1}{15}\lambda_4-\frac{1}{6}(\lambda_2 + \lambda_3 + \lambda_5 + \lambda_6)+\frac{2}{5}(\lambda_7+\lambda_8+\lambda_9).
\label{eq:eta_orto_poly_0}
\end{eqnarray}
%%%%%%%%%%%%%%%%%%%%%%%%%%%%%%%%%%%
 On the other hand, if the reference initial state of the polycrystal is a demagnetized state with domains pointing to all easy directions along the lattice vector $\boldsymbol{c}$ and the final state is the saturated state, then we obtain
%%%%%%%%%%%%%%%%%%%%%%%%%%%%%%%%%%%%%%%%%%%
\begin{equation}
     <\frac{l- l'_0}{l'_0}\Bigg\vert_{\boldsymbol{\beta}}^{\boldsymbol{\alpha}}> = \xi+\eta(\boldsymbol{\alpha}\cdot\boldsymbol{\beta})^2,
    \label{eq:delta_l_orto_poly_1}
\end{equation}
%%%%%%%%%%%%%%%%%%%%%%%%%%%%%%%%%%%%%%%%%%%%
where  
%%%%%%%%%%%%%%%%%%%%%%%%%%%%%%%%%%%%%%%%%%%%
\begin{eqnarray}
 \xi & = & \frac{2}{15}(\lambda_1+\lambda_4-\lambda_7-\lambda_8-\lambda_9)+\frac{1}{6}(\lambda_2+\lambda_3+\lambda_5+\lambda_6),\\
 \eta & = & -\frac{1}{15}\lambda_1-\frac{1}{15}\lambda_4-\frac{1}{6}(\lambda_2 + \lambda_3 + \lambda_5 + \lambda_6)+\frac{2}{5}(\lambda_7+\lambda_8+\lambda_9).
\label{eq:eta_orto_poly_1}
\end{eqnarray}
%%%%%%%%%%%%%%%%%%%%%%%%%%%%%%%%%%%
In the case with easy axis parallel to the lattice vector $\boldsymbol{a}$ we have
%%%%%%%%%%%%%%%%%%%%%%%%%%%%%%%%%%%%%%%%%%%
\begin{equation}
     <\frac{l- l''_0}{l''_0}\Bigg\vert_{\boldsymbol{\beta}}^{\boldsymbol{\alpha}}> = \xi+\eta(\boldsymbol{\alpha}\cdot\boldsymbol{\beta})^2,
    \label{eq:delta_l_orto_poly_2}
\end{equation}
%%%%%%%%%%%%%%%%%%%%%%%%%%%%%%%%%%%%%%%%%%%%
where  
%%%%%%%%%%%%%%%%%%%%%%%%%%%%%%%%%%%%%%%%%%%%
\begin{eqnarray}
 \xi & = & \frac{2}{15}\left(-\frac{3}{2}\lambda_1+\lambda_4-\lambda_7-\lambda_8-\lambda_9\right)+\frac{1}{6}(\lambda_2-\lambda_3-\lambda_5+\lambda_6),\\
 \eta & = & -\frac{1}{15}\lambda_1-\frac{1}{15}\lambda_4-\frac{1}{6}(\lambda_2 + \lambda_3 + \lambda_5 + \lambda_6)+\frac{2}{5}(\lambda_7+\lambda_8+\lambda_9),
\label{eq:eta_orto_poly_2}
\end{eqnarray}
%%%%%%%%%%%%%%%%%%%%%%%%%%%%%%%%%%%
while for the case of easy axis parallel to the lattice vector $\boldsymbol{b}$ we obtain
%%%%%%%%%%%%%%%%%%%%%%%%%%%%%%%%%%%%%%%%%%%
\begin{equation}
     <\frac{l- l'''_0}{l'''_0}\Bigg\vert_{\boldsymbol{\beta}}^{\boldsymbol{\alpha}}> = \xi+\eta(\boldsymbol{\alpha}\cdot\boldsymbol{\beta})^2,
    \label{eq:delta_l_orto_poly_3}
\end{equation}
%%%%%%%%%%%%%%%%%%%%%%%%%%%%%%%%%%%%%%%%%%%%
where  
%%%%%%%%%%%%%%%%%%%%%%%%%%%%%%%%%%%%%%%%%%%%
\begin{eqnarray}
 \xi & = & \frac{2}{15}\left(\lambda_1-\frac{3}{2}\lambda_4-\lambda_7-\lambda_8-\lambda_9\right)-\frac{1}{6}(\lambda_2-\lambda_3-\lambda_5+\lambda_6),\\
 \eta & = & -\frac{1}{15}\lambda_1-\frac{1}{15}\lambda_4-\frac{1}{6}(\lambda_2 + \lambda_3 + \lambda_5 + \lambda_6)+\frac{2}{5}(\lambda_7+\lambda_8+\lambda_9).
\label{eq:eta_orto_poly_3}
\end{eqnarray}
%%%%%%%%%%%%%%%%%%%%%%%%%%%%%%%%%%%

\section{Analysis of the accuracy of the methods implemented in MAELAS}
\label{section:accuracy}

 There are two main sources of errors in any calculation with MAELAS, (i) one coming from the used methodology to compute the magnetostrictive coefficients and magnetoelastic constants and other from (ii) the DFT calculation of the  total energy for each deformed state. Unfortunately, the lack of available experimental data for some crystal symmetries makes it difficult to estimate the reliability and precision of these calculations. In this Section, we try to estimate the systematic error coming from the methodologies implemented in MAELAS by evaluating the exact energy of each deformed state from the theory of magnetostriction. In this way, we can get rid of possible errors from DFT,  and reveal the accuracy of the implemented methods, as well as possible issues. To do so, we first consider a theoretical unit cell for each supported crystal symmetry in MAELAS that is characterized by an imposed set of lattice parameters, elastic ($C^{exact}_{ij}$) and magnetoelastic ($b^{exact}_k$) constants. The imposed values for these quantities are shown in the first, third and fifth columns of Table \ref{tab:accuracy}. Next, using these elastic  and magnetoelastic  constants, we calculate the exact magnetostrictive coefficients (shown in the seventh column of Table \ref{tab:accuracy}) from the theoretical relations between these quantities $\lambda^{exact}(b^{exact}_k,C^{exact}_{ij}$) \cite{maelas_publication2021}. Finally, we also use these elastic  and magnetoelastic  constants to evaluate the exact elastic and magnetoelastic energies \cite{maelas_publication2021} 
  %%%%%%%%%%%%%%%%%%%%%%%%%%%%%%%%%%%%%%%%%%%%
\begin{eqnarray}
E^{exact}(\boldsymbol{\epsilon},\boldsymbol{\alpha})=E^{exact}_{el}(\boldsymbol{\epsilon})+E^{exact}_{me}(\boldsymbol{\epsilon},\boldsymbol{\alpha}),
\label{eq:E_exact}
\end{eqnarray}
%%%%%%%%%%%%%%%%%%%%%%%%%%%%%%%%%%%%%%%%%%%%
for each deformed state and magnetization  direction generated by the two implemented methods in MAELAS. Then these exact energies are used as inputs in the step 5 of MAELAS workflow (see Fig.\ref{fig:workflow_mode2}) to compute the magnetoelastic constants and magnetostrictive coefficients. The exact expression of the elastic and magnetoelastic energies for cubic (I) symmetry are given in Eq. \ref{eq:E_cub}, while the corresponding equations for the other supported symmetries can be found in Ref. \cite{maelas_publication2021}. By comparing the exact values and MAELAS results for $b$ and $\lambda$ we can estimate the relative error of the calculation of MAELAS as
 %%%%%%%%%%%%%%%%%%%%%%%%%%%%%%%%%%%%%%%%%%%%
\begin{eqnarray}
b^{Rel. \: Error\:}_i (\%) & = & \frac{b^{exact}_i-b^{MAELAS}_i}{b^{exact}_i}\times100,
\label{eq:error_b}
\\
\lambda^{Rel. \: Error\:}_i (\%) & = & \frac{\lambda^{exact}_i-\lambda^{MAELAS}_i}{\lambda^{exact}_i}\times100.
\label{eq:error_lmb}
\end{eqnarray}
%%%%%%%%%%%%%%%%%%%%%%%%%%%%%%%%%%%%%%%%%%%%
The estimated relative errors are shown in Table \ref{tab:accuracy} for -mode 1 (method originally implemented in version 1.0 based on the length optimization \cite{maelas_publication2021,Wu1996}, including the corrections of the trigonal (I) symmetry given in Section \ref{subsection:correct_trigonal}) and -mode 2 (new method implemented in version 2.0 that is described in Section \ref{subsection:new_method}). We observe that the new methodology (-mode 2) improves significantly the accuracy of the calculations for all magnetoelastic constants and magnetostrictive coefficients exhibiting  relative errors lower than $10^{-4}\%$. The -mode 1 gives good results for cubic symmetry, but for lower symmetries the relative error is significantly large in some cases ($>10\%$), due to the deformation gradients implemented in version 1.0 for the cell length optimization \cite{maelas_publication2021}. In  \ref{section:app_accuracy}, we analyze this problem in more detail and discuss possible ways to improve the accuracy of -mode 1 . In this appendix, we also describe an alternative potential approach called strain optimization method, which is still under evaluation and is not currently implemented in version 2.0. Based on these results, we recommend to use -mode 2 instead of -mode 1, especially for non-cubic crystals. The -mode 2 is set as the default method in version 2.0. We note that performing the calculations with these two independent methods and comparing their results can also be useful to confirm the correct execution of the program.

\begin{table}[]
\centering
\caption{Summary of the analysis of the accuracy of the methods implemented in MAELAS. The relative error for the magnetoelastic constants ($b^{Rel. \: Error\:} $) and magnetostrictive coefficients ($\lambda^{Rel. \: Error\:} $) is computed using Eqs. \ref{eq:error_b} and \ref{eq:error_lmb}, respectively.}
\label{tab:accuracy}
\resizebox{\textwidth}{!}{%
\begin{tabular}{@{}ccccccccccc@{}}
\toprule
\begin{tabular}[c]{@{}c@{}} Crystal \\ system \end{tabular} &
  $C_{ij}$ &
  \begin{tabular}[c]{@{}c@{}}$C_{ij}^{exact}$ \\ (GPa) \end{tabular} &
  \begin{tabular}[c]{@{}c@{}} $b$ \\ \end{tabular} &
  \begin{tabular}[c]{@{}c@{}} $b^{exact}$  \\ (MPa)\end{tabular} &
  \begin{tabular}[c]{@{}c@{}} $\lambda$ \\ \end{tabular} &
  \begin{tabular}[c]{@{}c@{}} $\lambda^{exact}$  \\ ($\times10^{-6}$)\end{tabular} &
  \begin{tabular}[c]{@{}c@{}}-mode 1 \\ $b^{Rel. \: Error\:} $ \\ ($\%$)\end{tabular} &
    \begin{tabular}[c]{@{}c@{}}-mode 1 \\ $\lambda^{Rel. \: Error\:} $ \\ ($\%$)\end{tabular} & \begin{tabular}[c]{@{}c@{}}-mode 2 \\ $b^{Rel. \: Error\:} $ \\ ($\%$)\end{tabular} & \begin{tabular}[c]{@{}c@{}}-mode 2 \\ $\lambda^{Rel. \: Error\:} $ \\ ($\%$)\end{tabular} \\
    \midrule 
  Cubic (I) & $C_{11}$ & 243 & $b_{1}$ & -4.1 & $\lambda_{001}$ & 26.0317 & 0.015 & 0.013 & $7\cdot10^{-6}$  & $8\cdot10^{-6}$  \\
$a=2.8293$\r{A} &   $C_{12}$ & 138 & $b_{2}$ & 10.9 & $\lambda_{111}$ & -29.7814
 & 0.005 & 0.004 & $6\cdot10^{-6}$  & $3\cdot10^{-6}$ \\ 
   & $C_{44}$ & 122 &  & &  &  &  &  &   & \\
  \midrule  
Hexagonal (I) & $C_{11}$ & 307 & $b_{21}$ & -31.9 & $\lambda^{\alpha1,2}$ & 95.0656
 & -17.6 & -17.6 & $6\cdot10^{-6}$  & $1\cdot10^{-5}$  \\  
$a=2.4561$\r{A} & $C_{12}$ & 165 & $b_{22}$ & 25.5 & $\lambda^{\alpha2,2}$ & -125.9316
 & -17.4 & -17.5 & $8\cdot10^{-6}$  & $2\cdot10^{-5}$  \\  
$c=3.9821$\r{A} & $C_{13}$ & 103 & $b_{3}$ & -8.1 & $\lambda^{\gamma,2}$ & 57.0423 & 17.0 & 17.0 & $6\cdot10^{-6}$  & $6\cdot10^{-6}$  \\  
& $C_{33}$ & 358 & $b_{4}$ & 42.9 & $\lambda^{\epsilon,2}$ & -286.0
 & 0.002 & 0.002 & $5\cdot10^{-6}$  & $7\cdot10^{-6}$  \\  
 & $C_{44}$ & 75 & &  &  &  &  &  &   &  \\  
 \midrule  
Trigonal (I) & $C_{11}$ & 428 & $b_{21}$ & 43.1 & $\lambda^{\alpha1,2}$ & -104.9605

 & 22.9 & 10.8 & $1\cdot10^{-5}$  & $1\cdot10^{-5}$  \\  
$a=3.9249$\r{A} & $C_{12}$ & 164 & $b_{22}$ & -34.2 & $\lambda^{\alpha2,2}$ & 143.1326

 & -38.6 & -16.4 & $1\cdot10^{-5}$  & $8\cdot10^{-5}$  \\  
$c=4.8311$\r{A} & $C_{13}$ & 133 & $b_{3}$ & $60.7$ & $\lambda^{\gamma,1}$ & -202.6605
 & -11.9 & -8.8 & $1\cdot10^{-5}$  & $1\cdot10^{-5}$  \\  
& $C_{14}$ & -27 & $b_{4}$ & -34.3 & $\lambda^{\gamma,2}$ & 204.2029
 & -15.8 & -42.3 & $1\cdot10^{-5}$  & $1\cdot10^{-5}$  \\  
 & $C_{33}$ & 434 & $b_{14}$ & -42.4 & $\lambda_{12}$ & -377.9282

 & -28.1 & 46.8 &   $1\cdot10^{-5}$  & $1\cdot10^{-5}$  \\  
 & $C_{44}$ & 118 & $b_{34}$ & 55.4 & $\lambda_{21}$ & 266.5791
 & 37.9 &  -34.8 & $1\cdot10^{-5}$  & $1\cdot10^{-5}$  \\  
 \midrule  
 Tetragonal (I) & $C_{11}$ & 324 & $b_{21}$ & -2.4 & $\lambda^{\alpha1,2}$ & -20.4581
 & 42.4 & 7.3 & $8\cdot10^{-6}$  & $-2\cdot10^{-5}$  \\  
$a=2.6973$\r{A} & $C_{12}$ & 67 & $b_{22}$ & -15.2 & $\lambda^{\alpha2,2}$ & 78.1888
 & 18.3 & 15.4 & $6\cdot10^{-6}$  & $-4\cdot10^{-5}$  \\  
$c=3.7593$\r{A} & $C_{13}$ & 133 & $b_{3}$ & -7.9 & $\lambda^{\gamma,2}$ & 30.7393 & -14.2 & -14.2 & $6\cdot10^{-6}$  & $-2\cdot10^{-5}$  \\  
& $C_{33}$ & 264 & $b_{4}$ & -5.6 & $\lambda^{\epsilon,2}$ & 27.7228
 & 0.002 & 0.002 & $5\cdot10^{-6}$  & $7\cdot10^{-6}$  \\  
 & $C_{44}$ & 101 & $b'_{3}$ & -7.9 & $\lambda^{\delta,2}$ & 106.7568 & 0.007 & 0.007 & $5\cdot10^{-6}$  & $6\cdot10^{-6}$ \\ 
 & $C_{66}$ & 37 &  &  &  &  &  &  &   &  \\ 
 \midrule 
 Orthorhombic & $C_{11}$ & 76 & $b_{1}$ & $43.1$ & $\lambda_{1}$ & -632.9614
 & 37.4 & 24.7 & $7\cdot10^{-6}$  & $5\cdot10^{-6}$  \\  
$a=4.0686$\r{A} & $C_{12}$ & 45 & $b_{2}$ & -34.2 & $\lambda_{2}$ & 681.5787
 & 6.9 & 7.9 & $6\cdot10^{-6}$  & $6\cdot10^{-6}$  \\  
$b=10.3157$\r{A} & $C_{13}$ & 48 & $b_{3}$ & $60.7$ & $\lambda_{3}$ & -752.4503 & 19.6 & 0.03 & $7\cdot10^{-6}$  & $7\cdot10^{-6}$  \\  
$c=3.8956$\r{A} & $C_{23}$ & 55 & $b_{4}$ & -34.3 & $\lambda_{4}$  & 471.7839
 & -11.6 & -18.5 & $6\cdot10^{-6}$  & $6\cdot10^{-6}$  \\  
 & $C_{22}$ & 102 & $b_{5}$ & -42.4 & $\lambda_{5}$ & 809.6944 & -46.8 & -10.8 & $7\cdot10^{-6}$  & $5\cdot10^{-6}$ \\ 
 & $C_{33}$ & 141 & $b_{6}$ & 55.4 & $\lambda_{6}$ & -808.9638 & -7.6 & -5.6 & $6\cdot10^{-6}$  & $6\cdot10^{-6}$ \\ 
 & $C_{44}$ & 40 & $b_{7}$ & 35.4 & $\lambda_{7}$ & -284.9353 & 47.3 & 54.3 & $6\cdot10^{-6}$  & $7\cdot10^{-6}$ \\ 
 & $C_{55}$ & 27 & $b_{8}$ & -22.6 & $\lambda_{8}$ & 253.4425 & 43.2 & 11.6 & $9\cdot10^{-6}$  & $8\cdot10^{-6}$ \\ 
 & $C_{66}$ & 39 & $b_{9}$ & 38.7 & $\lambda_{9}$   & -326.1699 & -119.8 & -85.6 &   $5\cdot10^{-6}$  & $2\cdot10^{-5}$  \\  
  \bottomrule
\end{tabular}%
}
\end{table}

\section{Tests for the new method}
\label{section:test}

In this section, we perform some calculations with the new method described in Section \ref{subsection:new_method} (-mode 2). Here, MAELAS is interfaced with VASP \cite{vasp_1,vasp_2,vasp_3} to compute the energies of each deformed state. For a fair comparison with the method implemented in version 1.0 (-mode 1), we consider the same materials (BCC Fe, FCC Ni, HCP Co, Fe$_2$Si space group (SG) 164, L1$_0$ FePd and YCo SG 63) as in the publication of version 1.0 \cite{maelas_publication2021}, and use the same relaxed unit cells, VASP settings (energy cut-off, pseudopotentials, exchange correlation,...) and elastic constants calculated with AELAS \cite{maelas_publication2021,AELAS}. A summary of the results is presented in Table \ref{tab:tests_MAELAS}. 

%%%%%%%%%%%%%%%%%%%%%%%%%%%%%%%%%%%%%%%%%%%%
\begin{table}[h!]
\centering
\caption{Anisotropic magnetostrictive coefficients and magnetoelastic constants calculated with the two methods (-mode 1 and -mode 2) available in MAELAS v2.0. In parenthesis we show the magnetostrictive coefficients with Mason's definitions \cite{maelas_publication2021,Mason}. These data correspond to the simulations with the same VASP settings, relaxed unit cell and elastic constants as in Ref. \cite{maelas_publication2021}. For Fe$_2$Si we repeated the calculations with -mode 1 as implemented in version 2.0 [v2.0], that is, including the corrections described in Section \ref{subsection:correct_trigonal}.}
\label{tab:tests_MAELAS}
\resizebox{\textwidth}{!}{%
\begin{tabular}{@{}ccc|cccc|cccc@{}}
\toprule
Material &
  Crystal system &
  \begin{tabular}[c]{@{}c@{}} DFT \\ Exchange  \\ Correlation\end{tabular}  &
  \begin{tabular}[c]{@{}c@{}}Magnetostrictive \\ coefficient\end{tabular} &
  \begin{tabular}[c]{@{}c@{}}MAELAS\\-mode 1\\ ($\times10^{-6}$)\end{tabular} & \begin{tabular}[c]{@{}c@{}}MAELAS\\-mode 2\\ ($\times10^{-6}$)\end{tabular} &
  \begin{tabular}[c]{@{}c@{}}Expt.\\ ($\times10^{-6}$)\end{tabular} &
  \begin{tabular}[c]{@{}c@{}}Magnetoelastic \\ constant\end{tabular}  &
  \begin{tabular}[c]{@{}c@{}}MAELAS\\-mode 1\\ (MPa)\end{tabular} & \begin{tabular}[c]{@{}c@{}}MAELAS\\-mode 2\\ (MPa)\end{tabular} &
  \begin{tabular}[c]{@{}c@{}}Expt.\\ (MPa)\end{tabular} \\ \midrule \hline
FCC Ni      & Cubic (I)      &  GGA & $\lambda_{001}$        & -78.4$^b$ & -72.7 & -60$^a$    & $b_1$ & 15.5$^b$  & 14.4 & 9.9$^b$  \\
            &       SG 225         &     & $\lambda_{111}$        & -46.1$^b$ & -44.0 & -35$^a$   & $b_2$ & 19.4$^b$ & 18.5 & 13.9$^b$  \\  \midrule
BCC Fe      & Cubic (I)      &  GGA & $\lambda_{001}$        & 25.7$^b$ & 29.1 & 26$^a$   & $b_1$ &  -5.2$^b$ & -5.9 & -4.1$^b$ \\
            &       SG 229         &     & $\lambda_{111}$        & 17.2$^b$ & 15.7 & -30$^a$  & $b_2$ & -5.3$^b$ & -4.9 & 10.9$^b$  \\  \midrule
HCP Co      & Hexagonal (I)  &  LSDA+U  & $\lambda^{\alpha1,2}$ ($\lambda_A$) & 111 (-109)$^b$  & 75 (-74)  & 95 (-66)$^c$  & $b_{21}$ &  -21.3$^b$  & -16.2 & -31.9$^b$ \\
           &          SG 194      &  $J=0.8$eV    & $\lambda^{\alpha2,2}$ ($\lambda_B$) & -251 (-114)$^b$  & -156 (-77) &  -126 (-123)$^c$ &   $b_{22}$ &             48.3$^b$  & 28.4  & 25.5$^b$  \\
           &                &  $U=3$eV    & $\lambda^{\gamma,2}$  ($\lambda_C$) & 4 (251)$^b$  & 2 (156) & 57 (126)$^c$ &   $b_3$  &        -0.7$^b$      & -0.4  &  -8.1$^b$  \\
            &                &     & $\lambda^{\epsilon,2}$ ($\lambda_D$) & -51 (10)$^b$ & -57 (-8)  & -286 (-128)$^c$ &             $b_4$ &  7.1$^b$  & 7.9 & 42.9$^b$  \\
            \midrule

Fe$_2$Si     & Trigonal (I)  &  GGA & $\lambda^{\alpha1,2}$  & -9$^b$ [v1.0] & -5  &  & $b_{21}$ &  3.1$^b$ [v1.0]  & 0.7 & \\
            &          SG 164      &     & $\lambda^{\alpha2,2}$  & 15$^b$ [v1.0] & 17  &                 & $b_{22}$ &   -4.2$^b$ [v1.0]  & -6.2 & \\
            &                &     & $\lambda^{\gamma,1}$   & 8$^b$ [v1.0]  & 6 &     &  $b_{3}$ &   -0.7$^b$ [v1.0] & -1.9  &  \\
            &                &     & $\lambda^{\gamma,2}$ & 29$^b$ [v1.0] & 29 &     & $b_{4}$            &    3.3$^b$ [v1.0] & -3.5  &  \\ 
            &                &     & $\lambda_{12}$   & -3$^b$ [v1.0]  & -5 &     & $b_{14}$      &   -1.4$^b$ [v1.0]   & 1.9 & \\
            &                &     & $\lambda_{21}$ & -13$^b$ [v1.0] & -14 &   & $b_{34}$      &   -0.4$^b$ [v1.0]  & 1.4 & \\ & & & & & & & & & & \\
     &  &  GGA & $\lambda^{\alpha1,2}$  & -9 [v2.0] &   &  & $b_{21}$ & 3.1 [v2.0] &  & \\
            &              &     & $\lambda^{\alpha2,2}$  & 15 [v2.0]  &   &                 & $b_{22}$ &   -4.2 [v2.0]  & & \\
            &                &     & $\lambda^{\gamma,1}$   & 8 [v2.0] &  &     &  $b_{3}$ &  -2.5 [v2.0] &   &  \\
            &                &     & $\lambda^{\gamma,2}$ & 29 [v2.0]  &  &     & $b_{4}$            &  -3.7 [v2.0]   &  &  \\ 
            &                &     & $\lambda_{12}$   & -11 [v2.0] &  &     & $b_{14}$      &  2.0 [v2.0] &  & \\
            &                &     & $\lambda_{21}$ & -13 [v2.0] &  &   & $b_{34}$      &  2.2 [v2.0] & & \\           
            \midrule            
L1$_0$ FePd & Tetragonal (I) &  GGA & $\lambda^{\alpha1,2}$  &  -21$^b$   & -41 &   &  $b_{21}$  & -2.4$^b$ & 5.1 & \\
            &        SG 123        &     & $\lambda^{\alpha2,2}$    & 79$^b$ & 82 &  & $b_{22}$             &  -15.2$^b$ &  -10.7 &  \\
            &                &     & $\lambda^{\gamma,2}$  & 31$^b$ & 27 & &   $b_3$             & -7.9$^b$ & -6.9 &  \\
            &                &     & $\lambda^{\epsilon,2}$   &  28$^b$ & 26 &     & $b_4$            & -5.6$^b$  &  -5.3 & \\
            &                &     & $\lambda^{\delta,2}$   &  106$^b$ &  106 &   & $b'_3$              &  -7.9$^b$ &  -7.9 & \\\midrule 
YCo & Orthorhombic &  LSDA+U & $\lambda_1$  &  -11$^b$ &      -36   &   &   $b_1$ & -1.7$^b$ & 0.8 & \\
            &        SG 63        &  $J=0.8$eV    & $\lambda_2$    & 32$^b$ &  57 &  & $b_2$               & 1.2$^b$  & -2.9 & \\
            &                &  $U=1.9$eV    & $\lambda_3$  & 70$^b$ & 75 &         &  $b_3$        &  -3.8$^b$ & -2.2 &  \\
            &                &     & $\lambda_4$   &  -74$^b$ & -64 &   &  $b_4$            & 4.3$^b$ & 0.6 & \\
            &                &     & $\lambda_5$   &  -30$^b$ & -35 &  &  $b_5$              &  -0.1$^b$ & 1.7  \\
 &  &  & $\lambda_6$  & 7$^b$ &  24 &  & $b_6$ &  2.3$^b$ &  -2.2 & \\
             &                &     & $\lambda_7$   &  36$^b$ & 38 &  &  $b_7$              & -4.4$^b$ & -4.2 &  \\
            &                &     & $\lambda_8$   &  -20$^b$ &  -34 &  & $b_8$              & 1.1$^b$ &  1.8 & \\
 &  &  & $\lambda_9$  & 35$^b$ & -7  &  & $b_9$ & -8.7$^b$  & -0.5 &  \\

            \bottomrule

\end{tabular}%
}
\begin{tabular}{c}
\footnotesize $^a$Ref.\cite{Handley}, $^b$Ref.\cite{maelas_publication2021}, $^c$Ref.\cite{Hubert1969}
\end{tabular}
\end{table}
%%%%%%%%%%%%%%%%%%%%%%%%%%%%%%%%%%%%%%%%%%%%%%%%%%%%%%%%

\subsection{FCC Ni}

\begin{figure}[h!]
\centering
\includegraphics[width=\columnwidth ,angle=0]{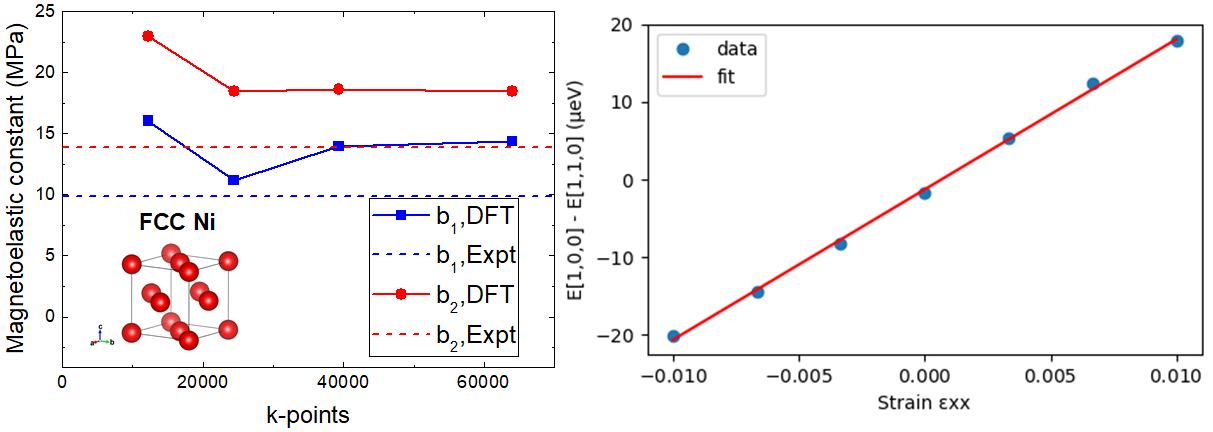}
\caption{(Left) Calculated magnetoelastic constants for FCC Ni with the new method implemented in MAELAS v2.0 (-mode 2). (Right) Calculation of $b_1$ for FCC Ni through a linear fitting of the energy difference between magnetization directions $\boldsymbol{\alpha}_1=(1,0,0)$ and $\boldsymbol{\alpha}_2=(1/\sqrt{2},1/\sqrt{2},0)$ versus strain ($\epsilon_{xx}$) data. }
\label{fig:Nifcc}
\end{figure}

The new method implemented in MAELAS v2.0 (-mode 2) gives very similar results to -mode 1 for FCC Ni (cubic (I) crystal symmetry), see Table \ref{tab:tests_MAELAS}. The analysis of the calculated magnetoelastic constants with -mode 2 as a function of the k-points in the Brillouin zone is shown in Fig. \ref{fig:Nifcc}. We see that both $b_1$ and $b_2$ converge well above $4\times10^4$ k-points and are in fairly good agreement with experiment. In Fig.\ref{fig:Nifcc} we also plot the linear fitting of the energy versus strain ($\epsilon_{xx}$) data to compute $b_1$.

%------------------------------
%------------------------------

\subsection{BCC Fe}

\begin{figure}[h!]
\centering
\includegraphics[width=\columnwidth ,angle=0]{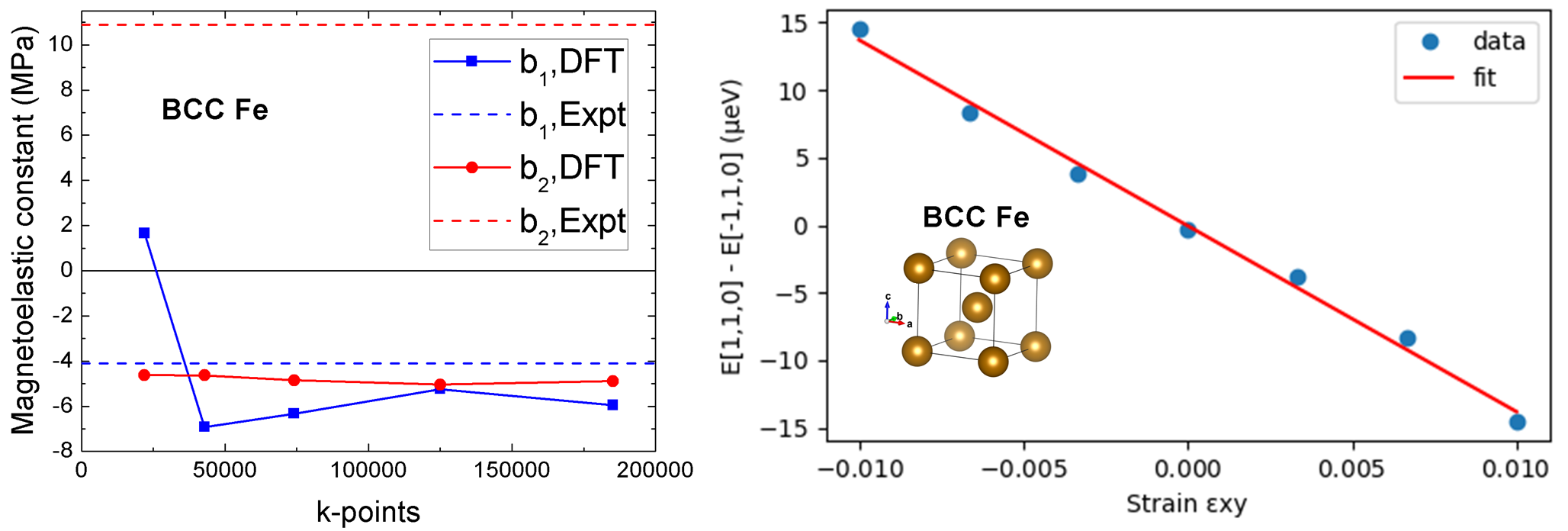}
\caption{(Left) Calculated magnetoelastic constants for BCC Fe with the new method implemented in MAELAS v2.0 (-mode 2). (Right) Calculation of $b_2$ for BCC Fe through a linear fitting of the energy difference between magnetization directions $\boldsymbol{\alpha}_1=(1/\sqrt{2},1/\sqrt{2},0)$ and $\boldsymbol{\alpha}_2=(-1/\sqrt{2},1/\sqrt{2},0)$ versus strain ($\epsilon_{xy}$) data. }
\label{fig:Febcc}
\end{figure}

The results given by -mode 2 for BCC Fe (cubic (I) crystal symmetry) are also very similar to -mode 1, see Table \ref{tab:tests_MAELAS}. The analysis of the calculated magnetoelastic constants with -mode 2 as a function of the k-points in the Brillouin zone is shown in Fig. \ref{fig:Febcc}. We see that both $b_1$ and $b_2$ converge well above $5\times10^4$ k-points. While $b_1$ is in good agreement with experiment, $b_2$ has the opposite sign to experiment due to a possible failure of DFT \cite{Jones2015,guo2002}. In Fig.\ref{fig:Febcc} we also plot the linear fitting of the energy versus strain ($\epsilon_{xy}$) data to compute $b_2$. Since the slope of the fitted linear polynomial is negative, then the calculation gives $b_2<0$.

%------------------------------
%------------------------------

\subsection{HCP Co}

\begin{figure}[h!]
\centering
\includegraphics[width=\columnwidth ,angle=0]{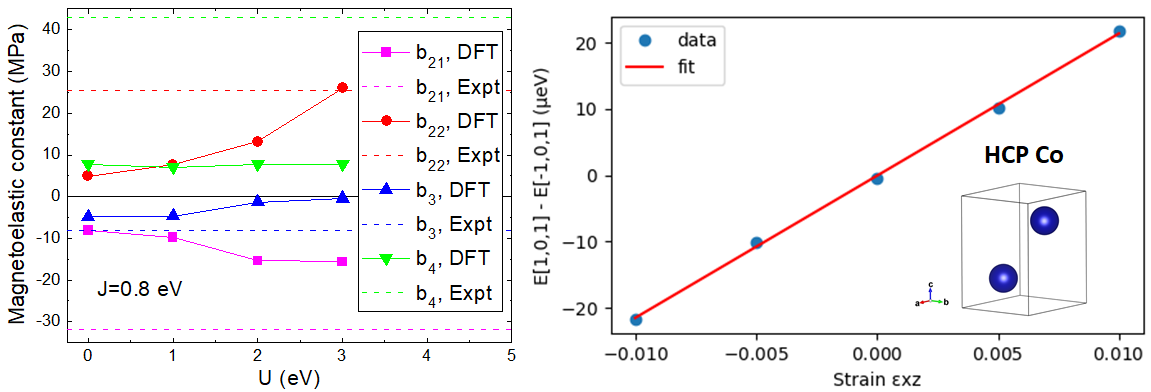}
\caption{(Left) Calculated magnetoelastic constants for HCP Co with the new method implemented in MAELAS v2.0 (-mode 2) for different values of the Hubbard U parameter. (Right) Calculation of $b_4$ for HCP Co through a linear fitting of the energy difference between magnetization directions $\boldsymbol{\alpha}_1=(1/\sqrt{2},0,1/\sqrt{2})$ and $\boldsymbol{\alpha}_2=(-1/\sqrt{2},0,1/\sqrt{2})$ versus strain ($\epsilon_{xz}$) data. }
\label{fig:Cohcp}
\end{figure}

The results of both methods for HCP Co (hexagonal (I) crystal symmetry) follow the experiment, except for $b_4$ and its corresponding magnetostrictive coefficient $\lambda^{\epsilon,2}$ which are clearly underestimated. Hence, the new method confirms the issue with $b_4$ and $\lambda^{\epsilon,2}$ that was already identified with -mode 1 in the publication of version 1.0 \cite{maelas_publication2021}. Calculated magnetoelastic constants for HCP Co with the new method implemented in MAELAS v2.0 (-mode 2) for different values of the Hubbard U parameter is shown in Fig.\ref{fig:Cohcp}. In this figure we also plot the linear fitting of the energy versus strain  data to compute $b_4$. We see that the linear fitting is quite good (R-squared=0.9989), so that the low value of $b_4$ is not related to a possible failure in the fitting procedure.

%------------------------------
%------------------------------

\subsection{Fe$_2$Si}

\begin{figure}[h!]
\centering
\includegraphics[width=\columnwidth ,angle=0]{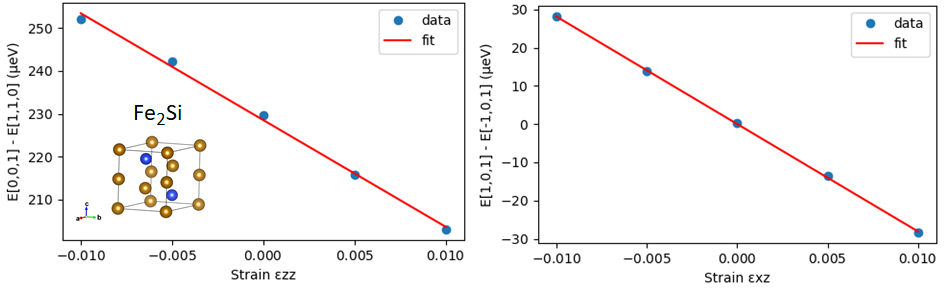}
\caption{Calculation of (left) $b_{22}$ and (right) $b_4$ for Fe$_2$Si through a linear fitting of the energy versus strain data.}
\label{fig:Fe2Si}
\end{figure}

For Fe$_2$Si (SG 164, trigonal (I) symmetry) we first recalculated the magnetostrictive coefficients and magnetoelastic constants with -mode 1 including the corrections implemented in version 2.0 for trigonal (I) symmetry that were described in Section \ref{subsection:correct_trigonal}. The corrected deformation gradient for $\lambda_{12}$ (Eq. \ref{eq:strain_trig_I_new}) in -mode 1 gives $\lambda_{12}=-11\times10^{-6}$, while the old deformation gradient (Eq. \ref{eq:strain_trig_I_old}) implemented in version 1.0 provides a negligible magnetostriction ($\lambda_{12}=-3\times10^{-6}$), as expected from the discussion about this issue in Section \ref{subsection:correct_trigonal}. Additionally, the corrections of the theoretical relations for $\lambda^{\gamma,1}$, $\lambda^{\gamma,2}$ and $\lambda_{2,1}$ given by Eq.\ref{eq:lamb_trig_correct} fix the issue with the sign of $b_4$, $b_{14}$ and $b_{34}$ found in the calculations with version 1.0. In Fig. \ref{fig:Fe2Si} we show the calculation of $b_{22}$ and  $b_4$ through a linear fitting of the energy versus strain data generated with -mode 2. 

%------------------------------
%------------------------------

\subsection{L1$_0$ FePd}
\label{section:FePd}

\begin{figure}[h!]
\centering
\includegraphics[width=\columnwidth ,angle=0]{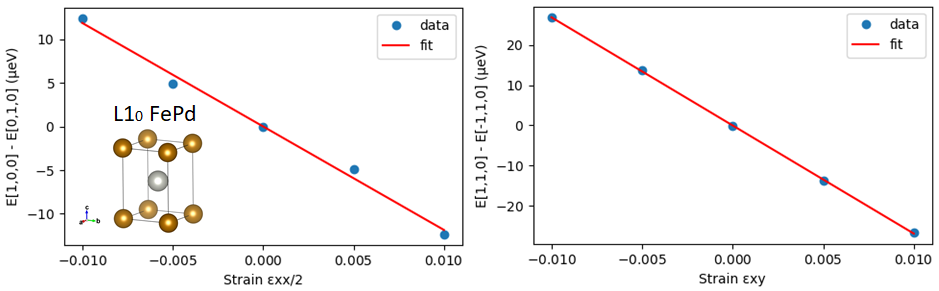}
\caption{Calculation of (left) $b_{3}$ and (right) $b'_3$ for L1$_0$ FePd through a linear fitting of the energy versus strain data.}
\label{fig:FePd}
\end{figure}

In the case of L1$_0$ FePd (tetragonal (I) symmetry), -mode 2 gives magnetoelastic constants similar to -mode 1, except for $b_{21}$ that has opposite sign due to the deviation obtained in $\lambda^{\alpha1,2}$ ($b_{21}=-(C_{11}+C_{12})\lambda^{\alpha1,2}-C_{13}\lambda^{\alpha2,2}$). Both methods also lead to similar  magnetostrictive coefficients, where the largest deviation is found in $\lambda^{\alpha1,2}$. In Fig.\ref{fig:FePd} we show the linear fitting of the energy versus strain ($\epsilon_{xy}$) data to compute $b_3$ and $b'_3$ using -mode 2.
%------------------------------
%------------------------------

\begin{figure}[h!]
\centering
\includegraphics[width=0.8\columnwidth ,angle=0]{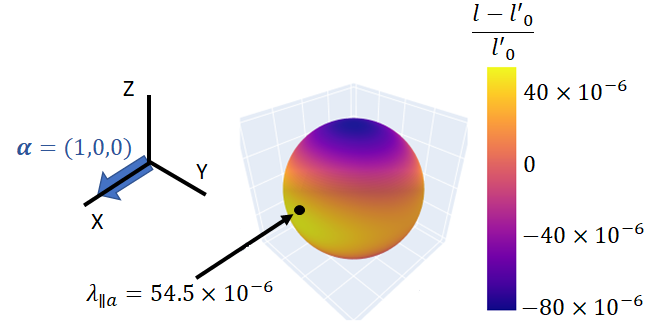}
\caption{Calculation of the fractional change in length of L1$_0$ FePd when the magnetization is saturated along the lattice vector $\boldsymbol{a}$ parallel to the x-axis $\boldsymbol{\alpha}=(1,0,0)$. As reference state, we used a demagnetized state with domains along all easy directions. Each point on the surface represent the fractional change in length along this particular crystallographic direction (measuring length direction $\boldsymbol{\beta}$). We used the  magnetostrictive coefficients calculated with -mode 2 of MAELAS which are shown in Table \ref{tab:tests_MAELAS}, as well as the calculated magnetocrystalline anisotropy constants $K_1=1.24$ MJ/m$^3$ and $K_2=0$. This figure was generated with the new update of MAELASviewer\cite{maelasviewer} that allows to select the type of reference demagnetized state.}
\label{fig:viewerFePd}
\end{figure}

In the publication of version 1.0 \cite{maelas_publication2021}, we discussed the measurements of the longitudinal magnetostriction (measuring length direction is parallel to the applied field  $\boldsymbol{H}\parallel\boldsymbol{\beta}$) for L1$_0$ FePd along the lattice vector $\boldsymbol{a}$ ($\lambda_{\parallel a}$, where $\boldsymbol{\alpha}=(1,0,0)$ and $\boldsymbol{\beta}(1,0,0)$) and lattice vector  $\boldsymbol{c}$  ($\lambda_{\parallel c}$ where ($\boldsymbol{\alpha}=(1,0,0)$ and $\boldsymbol{\beta}(1,0,0)$) performed by Shima et al.\cite{SHIMA20042173}. In such analysis, we applied Eq.\ref{eq:delta_l_tet_I} to obtain the relation between measured longitudinal magnetostriction ($\lambda_{\parallel a}$ and $\lambda_{\parallel c}$) and calculated magnetostrictive coefficient with MAELAS finding
%%%%%%%%%%%%%%%%%%%%%%%%%%%%%%%%%%
\begin{eqnarray}
\lambda_{\parallel a} & \equiv & \frac{l-l_0}{l_0}\Bigg\vert_{\boldsymbol{\beta=(1,0,0)}}^{\boldsymbol{\alpha=(1,0,0)}}=\lambda^{\alpha1,0}-\frac{\lambda^{\alpha1,2}}{3}+\frac{\lambda^{\gamma,2}}{2},\\
\lambda_{\parallel c} & \equiv & \frac{l-l_0}{l_0}\Bigg\vert_{\boldsymbol{\beta=(0,0,1)}}^{\boldsymbol{\alpha=(0,0,1)}}=\lambda^{\alpha2,0}+\frac{2\lambda^{\alpha2,2}}{3}.
\label{eq:FePd}
\end{eqnarray}
%%%%%%%%%%%%%%%%%%%%%%%%%%%%%%%%%%%
The application of Eq.\ref{eq:delta_l_tet_I} implies the assumption that the initial reference demagnetized state in the measurement of the fractional change in length corresponds to a hypothetic demagnetized state with randomly oriented atomic magnetic moments. However, an initial reference demagnetized state with magnetic domains seems a more reasonable assumption to interpret the results given by Shima et al., judging by the experimental setup \cite{SHIMA20042173}. For instance, if we assume our standard reference demagnetized state with magnetic domains along all possible easy directions, then we should applied Eq.\ref{eq:delta_l_tet_I_2} with cone angle $\Omega=0$ since L1$_0$ FePd has an easy axis. In this case, we find
%%%%%%%%%%%%%%%%%%%%%%%%%%%%%%%%%%
\begin{eqnarray}
\lambda_{\parallel a} & \equiv & \frac{l-l'_0}{l'_0}\Bigg\vert_{\boldsymbol{\beta=(1,0,0)}}^{\boldsymbol{\alpha=(1,0,0)}}=-\lambda^{\alpha1,2}+\frac{\lambda^{\gamma,2}}{2},\label{eq:FePd2a}\\
\lambda_{\parallel c} & \equiv & \frac{l-l'_0}{l'_0}\Bigg\vert_{\boldsymbol{\beta=(0,0,1)}}^{\boldsymbol{\alpha=(0,0,1)}}=0.
\label{eq:FePd2b}
\end{eqnarray}
%%%%%%%%%%%%%%%%%%%%%%%%%%%%%%%%%%%
Inserting our calculated values of $\lambda^{\alpha1,2}$ and $\lambda^{\gamma,2}$ (shown in Table \ref{tab:tests_MAELAS}) in Eq.\ref{eq:FePd2a} gives $\lambda_{\parallel a}=36.5\times10^{-6}$ and $\lambda_{\parallel a}=54.5\times10^{-6}$ with -mode 1 and -mode 2, respectively. Here, the result obtained by -mode 2 is likely to be more accurate than the one with -mode 1, based on the analysis performed in Section \ref{section:accuracy}. Shima et al. measured\cite{SHIMA20042173} $\lambda_{\parallel a}=100\times10^{-6}$ which is about two times larger than our calculation with -mode 2. This deviation could be related to the different strength of SOC in the experiment and theory. This fact could be estimated from the magnetocrystalline anisotropy energy at the unstrained state. In our calculation we obtained an uniaxial magnetocrystalline anisotropy
constant $K_{U}=[E(1,0,0)-E(0,0,1)]/(a^2c)=1.24$ MJ/m$^3$\cite{maelas_publication2021}, while in the experimental sample (L1$_0$ Fe$_{48}$Pd$_{52}$) Shima et al. measured $K_{U}=2.5$ MJ/m$^3$\cite{SHIMA20042173} at $4.2$ K, larger than other reported values \cite{Maykov,shima}, and approximately two times larger than in our calculation, as in the case of $\lambda_{\parallel a}$ with -mode 2. Concerning $\lambda_{\parallel c}$, the theoretical result given by Eq.\ref{eq:FePd2b} is in fairly good agreement to the small values measured by Shima et al. $\lambda_{\parallel c}\simeq 0$ at applied field $H=5$ kOe and $\lambda_{\parallel c}\simeq 20\times10^{-6}$ at $H=50$ kOe \cite{SHIMA20042173}.
This kind of analysis can be done easily with the visualization tool MAELASviewer\cite{maelasviewer} that now includes the possibility to select the type of demagnetized state to be used as a reference state in the calculation of the fractional change in length, see Fig.\ref{fig:viewerFePd}.
%------------------------------
%------------------------------

\subsection{YCo}

\begin{figure}[h!]
\centering
\includegraphics[width=\columnwidth ,angle=0]{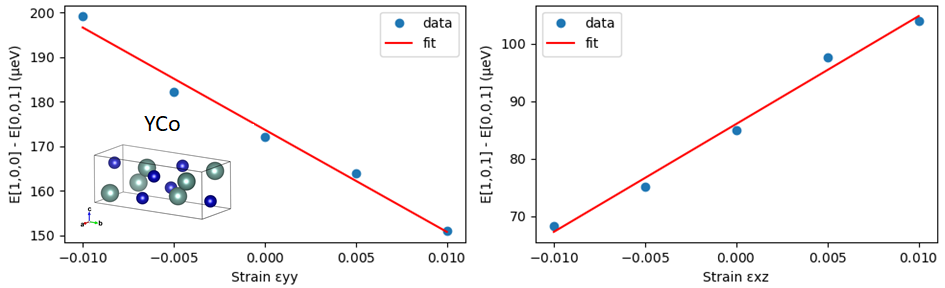}
\caption{Calculation of (left) $b_{3}$ and (right) $b_8$ for YCo through a linear fitting of the energy versus strain data.}
\label{fig:YCo}
\end{figure}

In YCo (SG 63, orthorhombic), we observe larger discrepancies between -mode 1 and -mode 2 than in the previous examples, especially in the results for the magnetoelastic constants. This fact is partially due to the lower accuracy of -mode 1 with respect -mode 2, as discussed in Section \ref{section:accuracy}.  Fig.\ref{fig:YCo} presents the linear fitting of the energy versus strain ($\epsilon_{xy}$) data to compute $b_3$ and $b_8$ using -mode 2.
%------------------------------
%------------------------------

\section{Conclusions}
\label{section:con}

In summary, version 2.0 of MAELAS includes an alternative method (-mode 2) to compute directly the anisotropic magnetoelastic constants from a linear fitting of the energy versus strain data, which is derived from the magnetoelastic energy of each crystal symmetry. The analysis of the accuracy of MAELAS reveals that this new method is more precise and convenient than the method originally implemented in version 1.0, especially for non-cubic symmetries. The fact that these two independent methods give similar results in many cases can also be a useful way of confirming the correct execution of the calculations. For instance, this analysis helped us to identify some issues in version 1.0 for the trigonal (I) symmetry, which has been fixed in version 2.0. Additionally, it also allows us to find that the accuracy of -mode 1 could be improved by using deformation gradients derived from the equilibrium magnetoelastic strain tensor, as well as potential alternative approaches like the strain optimization method.

To facilitate the comparison between theory and experiment, we implemented a new feature in the visualization tool MAELASviewer that allows to choose the type of reference demagnetized state in the calculation of the fractional change in length. This new option helped us to have better understanding of the experimental measurements of magnetostriction for L1$_0$ FePd. 

Lastly, we also derived and implemented the fractional change in length for polycrystals with trigonal, tetragonal and orthorhombic crystal symmetries, which has the same general form as the cubic and hexagonal crystals.

\section*{Acknowledgement}

This work was supported by the ERDF in the IT4Innovations national supercomputing center - path to exascale project (CZ.02.1.01/0.0/0.0/16-013/0001791) within the OPRDE.
This work was also supported by The Ministry of Education, Youth and Sports from  the Large Infrastructures for Research, Experimental Development, and Innovations project “e-INFRA CZ (ID:90140)", by the Donau project No. 8X20050, and the computational resources provided by the Open Access Grant Competition of IT4Innovations National Supercomputing Center within the projects OPEN-18-5, OPEN-18-33, and OPEN-19-14. In addition, DL, SA, and APK acknowledge the Czech Science Foundations grant No.~20-18392S and P.N., D.L., and~S.A. acknowledge support from the H2020-FETOPEN no.~863155 s-NEBULA project.

\appendix

\section{Generation of the deformed unit cells in the new method to calculate magnetoelastic constants}
\label{app_matrix}

In this appendix we present the procedure to generate the deformed unit cells in the new methodology (-mode 2 ) of MAELAS 2.0 to calculate  anisotropic magnetoelastic constants that is described in Section \ref{subsection:new_method}.  The deformed unit cells are generated by multiplying the lattice vectors of the initial unit cell $\boldsymbol{a}=(a_x,a_y,a_z)$, $\boldsymbol{b}=(b_x,b_y,b_z)$, $\boldsymbol{c}=(c_x,c_y,c_z)$ by the deformation gradient $F_{ij}$  \cite{Tadmor_2009}
%%%%%%%%%%%%%%%%%%%%%%%%%%%%%
\begin{equation}
\begin{pmatrix}
a'_{x} & b'_{x} & c'_{x}\\
a'_{y} & b'_{y} & c'_{y}\\
a'_{z} & b'_{z} & c'_{z} \\
\end{pmatrix}
= 
\begin{pmatrix}
F_{xx} & F_{xy} & F_{xz}\\
F_{yx} & F_{yy} & F_{yz}\\
F_{zx} & F_{zy} & F_{zz} \\
\end{pmatrix}\cdot
\begin{pmatrix}
a_{x} & b_{x} & c_{x}\\
a_{y} & b_{y} & c_{y}\\
a_{z} & b_{z} & c_{z} \\
\end{pmatrix}
\label{eq:deform_latt0}
\end{equation}
%%%%%%%%%%%%%%%%%%%%%%%%%%%
where $a'_i$, $b'_i$ and $c'_i$ ($i=x,y,z$) are the components of the lattice vectors of the deformed cell. In the infinitesimal strain theory, the deformation gradient is related to the displacement gradient ($\partial u_i / \partial r_j$) as $F_{ij}=\delta_{ij}+ \partial u_i/\partial r_j$, where $\delta_{ij}$ is the Kronecker delta. Since  the strain tensor ($\epsilon_{ij}$) can be expressed in terms of the displacement vector $\boldsymbol{u}$ as \cite{Landau}
%%%%%%%%%%%%%%%%%%%%%%%%%%%%%%%%%%%%%%%%%%%%
\begin{equation}
\begin{aligned}
     \epsilon_{ij}=\frac{1}{2}\left(\frac{\partial u_{i}}{\partial r_{j}}+\frac{\partial u_j}{\partial r_i}\right),\quad\quad i,j=x,y,z
    \label{eq:disp_vec}
\end{aligned}
\end{equation}
%%%%%%%%%%%%%%%%%%%%%%%%%%%%%%%%%%%%%%%%%%%%
then the strain tensor $\epsilon_{ij}$ can be written in terms of the deformation gradient as 
%%%%%%%%%%%%%%%%%%%%%%%%%%%%%
\begin{equation}
\boldsymbol{\epsilon}=
\begin{pmatrix}
\epsilon_{xx} & \epsilon_{xy} & \epsilon_{xz}\\
\epsilon_{yx} & \epsilon_{yy} & \epsilon_{yz}\\
\epsilon_{zx} & \epsilon_{zy} & \epsilon_{zz} \\
\end{pmatrix}
= 
\frac{1}{2}\begin{pmatrix}
2(F_{xx}-1) & F_{xy}+F_{yx} & F_{xz}+F_{zx} \\
F_{xy}+F_{yx}  & 2(F_{yy}-1) & F_{yz}+F_{zy} \\
F_{xz}+F_{zx}  & F_{yz}+F_{zy}  & 2(F_{zz}-1) 
\end{pmatrix}
.
\label{eq:strain_deform0}
\end{equation}
%%%%%%%%%%%%%%%%%%%%%%%%%%%
We see, that the strain tensor is symmetric ($\epsilon_{ij}=\epsilon_{ji}$, $i\neq j$), but the deformation gradient is not necessarily symmetric. For the new methodology implemented in MAELAS 2.0 we consider deformation gradients that fulfill Eq.\ref{eq:dE_tot} with the magnetization directions $\boldsymbol{\alpha}_1$ and $\boldsymbol{\alpha}_2$ given by Table \ref{tab:beta_alpha_data}. For the sake of simplicity in implementing this new approach, we use deformations that don't preserve the volume of the unit cells (non-isochoric deformation).

\subsection{Cubic (I) system}
\label{subsection:cubic_deform}

For cubic (I) systems the new method (-mode 2) generates two set of deformed unit cells using the following deformation gradients
%%%%%%%%%%%%%%%%%%%%%%%%%%%%%
\begin{eqnarray}
\boldsymbol{F}^{b_{1}}(s) & = &
\begin{pmatrix}
1+s & 0 & 0\\
0 & 1 & 0\\
0 & 0 & 1 \\
\end{pmatrix}
\label{eq:strain_cub_I_b1}
,\\
\boldsymbol{F}^{b_{2}}(s) & = & 
\begin{pmatrix}
1 & s & 0\\
s & 1 & 0\\
0 & 0 & 1 \\
\end{pmatrix}.
\label{eq:strain_cub_I_b2}
\end{eqnarray}
%%%%%%%%%%%%%%%%%%%%%%%%%%%
The parameter $s$ controls the applied deformation, and its maximum value can be specified through the command line of the program MAELAS using tag $-s$. The total number of deformed cells can be chosen with tag $-n$.

\subsection{Hexagonal (I) system}

In the case of hexagonal (I), the new method (-mode 2) generates 4 sets of deformed cells using the following deformation gradients
%%%%%%%%%%%%%%%%%%%%%%%%%%%%%
\begin{eqnarray}
\boldsymbol{F}^{b_{21}}(s) & = &
\begin{pmatrix}
1+\frac{s}{2} & 0 & 0\\
0 & 1+\frac{s}{2} & 0\\
0 & 0 & 1 \\
\end{pmatrix}
\label{eq:strain_hex_I_b21}
,\\
\boldsymbol{F}^{b_{22}}(s) & = & 
\begin{pmatrix}
1 & 0 & 0\\
0 & 1 & 0\\
0 & 0 & 1+s \\
\end{pmatrix}
\label{eq:strain_hex_I_b22}
,\\
\boldsymbol{F}^{b_{3}}(s) & = &
\begin{pmatrix}
1+\frac{s}{2} & 0 & 0\\
0 & 1-\frac{s}{2} & 0\\
0 & 0 & 1 \\
\end{pmatrix}
\label{eq:strain_hex_I_b3}
,\\
\boldsymbol{F}^{b_{4}}(s) & = &
\begin{pmatrix}
1 & 0 & s\\
0 & 1 & 0\\
s & 0 & 1 \\
\end{pmatrix}.
\label{eq:strain_hex_I_b4}
\end{eqnarray}
%%%%%%%%%%%%%%%%%%%%%%%%%%%

\subsection{Trigonal (I) system}

In the case of trigonal (I), the new method (-mode 2) generates 6 sets of deformed unit cells. The deformation gradients for $b_{12}$, $b_{22}$, $b_{3}$ and  $b_{4}$ are the same as in the hexagonal (I) case, (Eqs.\ref{eq:strain_hex_I_b21}, \ref{eq:strain_hex_I_b22}, \ref{eq:strain_hex_I_b3} and \ref{eq:strain_hex_I_b4}), while for $b_{14}$ and $b_{34} $ the deformation gradients are
%%%%%%%%%%%%%%%%%%%%%%%%%%%%%
\begin{eqnarray}
\boldsymbol{F}^{b_{14}}(s) & =
\begin{pmatrix}
1 & 0 & s\\
0 & 1 & 0\\
s & 0 & 1 \\
\end{pmatrix}
\label{eq:strain_trig_I_b14}
,\\
\boldsymbol{F}^{b_{34}}(s) & = 
\begin{pmatrix}
1 & s & 0\\
s & 1 & 0\\
0 & 0 & 1 \\
\end{pmatrix}.
\label{eq:strain_trig_I_b34}
\end{eqnarray}
%%%%%%%%%%%%%%%%%%%%%%%%%%%

\subsection{Tetragonal (I) system}

In the case of tetragonal (I), the new method (-mode 2) generates 5 sets of deformed unit cells. The deformation gradients for $b_{12}$, $b_{22}$, $b_{3}$ and  $b_{4}$ are the same as in the hexagonal (I) case, (Eqs.\ref{eq:strain_hex_I_b21}, \ref{eq:strain_hex_I_b22}, \ref{eq:strain_hex_I_b3} and \ref{eq:strain_hex_I_b4}), while for $b'_{3}$ the deformation gradient is
%%%%%%%%%%%%%%%%%%%%%%%%%%%%%
\begin{eqnarray}
\boldsymbol{F}^{b'_{3}}(s) & =
\begin{pmatrix}
1 & s & 0\\
s & 1 & 0\\
0 & 0 & 1 \\
\end{pmatrix}
\label{eq:strain_tet_I_bp3}
.
\end{eqnarray}
%%%%%%%%%%%%%%%%%%%%%%%%%%%

\subsection{Orthorhombic system}

For orthorhombic crystals the new method (-mode 2) generates 9 sets of deformed cells using the following deformation gradients
%%%%%%%%%%%%%%%%%%%%%%%%%%%%%
\begin{eqnarray}
\boldsymbol{F}^{b_{1}}(s) & = & \boldsymbol{F}^{b_{2}}(s) =
\begin{pmatrix}
1+s & 0 & 0\\
0 & 1 & 0\\
0 & 0 & 1 \\
\end{pmatrix}
\label{eq:strain_orto_b1}
,\\
\boldsymbol{F}^{b_{3}}(s) & = & \boldsymbol{F}^{b_{4}}(s) =
\begin{pmatrix}
1 & 0 & 0\\
0 & 1+s & 0\\
0 & 0 & 1 \\
\end{pmatrix}
\label{eq:strain_orto_b3}
,\\
\boldsymbol{F}^{b_{5}}(s) & = & \boldsymbol{F}^{b_{6}}(s) =
\begin{pmatrix}
1 & 0 & 0\\
0 & 1 & 0\\
0 & 0 & 1+s \\
\end{pmatrix}
\label{eq:strain_orto_b5}
,\\
\boldsymbol{F}^{b_{7}}(s) & = &
\begin{pmatrix}
1 & s & 0\\
s & 1 & 0\\
0 & 0 & 1 \\
\end{pmatrix}
\label{eq:strain_orto_b7}
,\\
\boldsymbol{F}^{b_{8}}(s) & = &
\begin{pmatrix}
1 & 0 & s\\
0 & 1 & 0\\
s & 0 & 1 \\
\end{pmatrix}
\label{eq:strain_orto_b8}
,\\
\boldsymbol{F}^{b_{9}}(s) & = &
\begin{pmatrix}
1 & 0 & 0\\
0 & 1 & s\\
0 & s & 1 \\
\end{pmatrix}.
\label{eq:strain_orto_b9}
\end{eqnarray}
%%%%%%%%%%%%%%%%%%%%%%%%%%%

\section{Analysis of the accuracy of the length optimization method and potential ways to improve it}
\label{section:app_accuracy}

In Section \ref{section:accuracy} we found that the length optimization method (-mode 1) leads to significantly large relative errors in the calculation of some magnetostrictive coefficients. In this appendix we analyze this issue in more detail. 

The implementation of the length optimization method in MAELAS contains two approximations. The first approximation is used in the formula to compute the magnetostrictive coefficients from the equilibrium lengths $l_1$ and $l_2$ along the direction $\boldsymbol{\beta}$ when magnetization points to direction $\boldsymbol{\alpha}_1$ and $\boldsymbol{\alpha}_2$, respectively \cite{maelas_publication2021}
%%%%%%%%%%%%%%%%%%%%%%%%%%%%%%%%%%%%%%%%%%%%
\begin{equation}
\begin{aligned}
\lambda\cong\frac{2(l_1-l_2)}{\rho (l_1+l_2)}.
\label{eq:l1_l2_solve0}
\end{aligned}
\end{equation}
%%%%%%%%%%%%%%%%%%%%%%%%%%%%%%%%%%%%%%%%%%%%
This is a fairly good approximation for all known magnetostrictive materials ($\lambda<10^{-2}$), and gives small relative error ($<10^{-2}$). The second approximation is related to the estimation of the equilibrium lengths $l_1$ and $l_2$ in Eq.\ref{eq:l1_l2_solve0} through a parameterized deformation gradient $\boldsymbol{F}(s)$.  Let's analyze in more detail the influence of this approximation on the calculation of $\lambda_{001}$ for Cubic (I) crystal and $\lambda^{\alpha2,2}$ for Hexagonal (I) crystal.

\subsection{Cubic (I)}

Ideally, one needs to make use of the equilibrium magnetoelastic strain tensor $\boldsymbol{\epsilon}^{eq}(\boldsymbol{\alpha})$ in order to construct the parameterized deformation gradient that contains the exact equilibrium state as a particular case of the parameter. The anisotropic part of the equilibrium strain tensor for Cubic (I) systems reads \cite{maelas_publication2021,CLARK1980531}
%%%%%%%%%%%%%%%%%%%%%%%%%%%%%%%%%%%%%%%%%%%%
\begin{equation}
\begin{aligned}
 \boldsymbol{\epsilon}^{eq} (\boldsymbol{\alpha}) = 
\begin{pmatrix}
\frac{3}{2}\lambda_{001}\left(\alpha_{x}^2-\frac{1}{3}\right) & \frac{3}{2}\lambda_{111}\alpha_{x}\alpha_{y} & \frac{3}{2}\lambda_{111}\alpha_{x}\alpha_{z} \\
\frac{3}{2}\lambda_{111}\alpha_{x}\alpha_{y} & \frac{3}{2}\lambda_{001}\left(\alpha_{y}^2-\frac{1}{3}\right) & \frac{3}{2}\lambda_{111}\alpha_{y}\alpha_{z} \\
\frac{3}{2}\lambda_{111}\alpha_{x}\alpha_{z}  & \frac{3}{2}\lambda_{111}\alpha_{y}\alpha_{z}  & \frac{3}{2}\lambda_{001}\left(\alpha_{z}^2-\frac{1}{3}\right)\\
\end{pmatrix}.
\label{eq:magelas_strain}
\end{aligned}
\end{equation}
%%%%%%%%%%%%%%%%%%%%%%%%%%%%%%%%%%%%%%%%%%%%
In -mode 1, for the calculation of $\lambda_{001}$ we selected the magnetization directions $\boldsymbol{\alpha}_1=(0,0,1)$ and $\boldsymbol{\alpha}_2=(1,0,0)$ \cite{maelas_publication2021}. Hence, the equilibrium strain tensor becomes
%%%%%%%%%%%%%%%%%%%%%%%%%%%%%%%%%%%%%%%%%%%%
\begin{equation}
\begin{aligned}
& \boldsymbol{\epsilon}^{eq}(0,0,1) = 
\begin{pmatrix}
-\frac{1}{2}\lambda_{001} & 0 & 0\\
0 & -\frac{1}{2}\lambda_{001} & 0\\
0 & 0 & \lambda_{001} \\
\end{pmatrix},\\
& \boldsymbol{\epsilon}^{eq}(1,0,0) = 
\begin{pmatrix}
\lambda_{001} & 0 & 0\\
0 & -\frac{1}{2}\lambda_{001} & 0\\
0 & 0 & -\frac{1}{2}\lambda_{001}\\
\end{pmatrix}.
\label{eq:strain_eq_cub_alpha}
\end{aligned}
\end{equation}
%%%%%%%%%%%%%%%%%%%%%%%%%%%%%%%%%%%%%%%%%%%%
The corresponding deformation gradients that give these equilibrium strain tensors are obtained using Eq. \ref{eq:strain_deform0}
%%%%%%%%%%%%%%%%%%%%%%%%%%%%%%%%%%%%%%%%%%%%
\begin{equation}
\begin{aligned}
& \boldsymbol{F}\Big\vert_{\boldsymbol{\alpha}_1=(0,0,1)}^{\lambda_{001}} = 
\begin{pmatrix}
1-\frac{1}{2}\lambda_{001} & 0 & 0\\
0 & 1-\frac{1}{2}\lambda_{001} & 0\\
0 & 0 & 1+\lambda_{001}\\
\end{pmatrix},\\
& \boldsymbol{F}\Big\vert_{\boldsymbol{\alpha}_2=(1,0,0)}^{\lambda_{001}}  = 
\begin{pmatrix}
1+\lambda_{001} & 0 & 0\\
0 & 1-\frac{1}{2}\lambda_{001} & 0\\
0 & 0 & 1-\frac{1}{2}\lambda_{001} \\
\end{pmatrix}.
\label{eq:F_eq_cub_alpha}
\end{aligned}
\end{equation}
%%%%%%%%%%%%%%%%%%%%%%%%%%%%%%%%%%%%%%%%%%%%
Therefore, if we parameterize these deformation gradients by setting the magnetostrictive coefficients as free parameters ($\lambda_i\xrightarrow{} s_i$), then they generate a set of deformed cells that contains the exact equilibrium state that corresponds to the particular case when the free parameters are equal to the magnetostrictive coefficients (minimum energy). Hence, these deformation gradients should be parameterized as 
%%%%%%%%%%%%%%%%%%%%%%%%%%%%%%%%%%%%%%%%%%%%
\begin{eqnarray}
& & \boldsymbol{F}\Big\vert_{\boldsymbol{\alpha}_1=(0,0,1)}^{\lambda_{001}} (s)= 
\begin{pmatrix}
1-\frac{1}{2}s & 0 & 0\\
0 & 1-\frac{1}{2}s & 0\\
0 & 0 & 1+s\\
\end{pmatrix},\label{eq:F_eq_cub_alpha2a}\\
& & \boldsymbol{F}\Big\vert_{\boldsymbol{\alpha}_2=(1,0,0)}^{\lambda_{001}} (s) = 
\begin{pmatrix}
1+s & 0 & 0\\
0 & 1-\frac{1}{2}s& 0\\
0 & 0 & 1-\frac{1}{2}s \\
\end{pmatrix},
\label{eq:F_eq_cub_alpha2b}
\end{eqnarray}
%%%%%%%%%%%%%%%%%%%%%%%%%%%%%%%%%%%%%%%%%%%%
where $s$ is a free parameter. Let's try to estimate the relative error in the calculation of $\lambda_{001}$ using these parameterized deformation gradients. At the minimum energy (equilibrium state), the free parameter in Eqs. \ref{eq:F_eq_cub_alpha2a} and \ref{eq:F_eq_cub_alpha2b} is equal to the magnetostrictive coefficient ($s=\lambda_{100}$). Hence, these parameterized deformation gradients give the exact equilibrium lengths $l_1$ and $l_2$ along the direction $\boldsymbol{\beta}=(0,0,1)$ (via Eq. \ref{eq:deform_latt0}) 
%%%%%%%%%%%%%%%%%%%%%%%%%%%%%%%%%%%%%%%%%%%%
\begin{equation}
\begin{aligned}
l^{exact}_1 & = \left(1+\lambda_{001}\right)c_z,\\
l^{exact}_2 & = \left(1-\frac{1}{2}\lambda_{001}\right)c_z,
\label{eq:l1_l2_cub}
\end{aligned}
\end{equation}
%%%%%%%%%%%%%%%%%%%%%%%%%%%%%%%%%%%%%%%%%%%%
where $c_z$ is the z-component of the relaxed (not deformed) lattice vector $\boldsymbol{c}=(0,0,c_z)$. Therefore, the Eq.\ref{eq:l1_l2_solve0} used in -mode 1 to compute the magnetostrictive coefficient gives \cite{maelas_publication2021}
%%%%%%%%%%%%%%%%%%%%%%%%%%%%%%%%%%%%%%%%%%%%
\begin{equation}
\begin{aligned}
\lambda_{001}^{approx}\cong\frac{2(l^{exact}_1-l^{exact}_2)}{\rho (l^{exact}_1+l^{exact}_2)}=\frac{4\lambda_{001}}{4+\lambda_{001}},
\label{eq:l1_l2_solve}
\end{aligned}
\end{equation}
%%%%%%%%%%%%%%%%%%%%%%%%%%%%%%%%%%%%%%%%%%%%
where $\rho=3/2$. The relative error using the exact value   in Table \ref{tab:accuracy} ($\lambda_{001}=26.0317\times10^{-6}$) is
%%%%%%%%%%%%%%%%%%%%%%%%%%%%%%%%%%%%%%%%%%%%
\begin{equation}
\begin{aligned}
\lambda_{001}^{Rel.Error}=\frac{\lambda_{001}-\lambda_{001}^{approx}}{\lambda_{001}}\times 100 = 0.0007\%,
\label{eq:l1_l2_solve_error_cub}
\end{aligned}
\end{equation}
%%%%%%%%%%%%%%%%%%%%%%%%%%%%%%%%%%%%%%%%%%%%
which is smaller than in the example shown in Section \ref{section:accuracy} using the  -mode 1 of MAELAS ($\lambda_{001}^{Rel.Error}=0.013\%$). The relative error is larger in MAELAS because we implemented the following deformation gradient for both magnetization directions $\boldsymbol{\alpha}_1=(0,0,1)$ and $\boldsymbol{\alpha}_2=(1,0,0)$ \cite{maelas_publication2021}
%%%%%%%%%%%%%%%%%%%%%%%%%%%%%
\begin{equation}
\begin{aligned}
\boldsymbol{F}\Big\vert_{\boldsymbol{\alpha}_1=(0,0,1)}^{\lambda_{001}} (s) = \boldsymbol{F}\Big\vert_{\boldsymbol{\alpha}_2=(1,0,0)}^{\lambda_{001}} (s) =\begin{pmatrix}
\frac{1}{\sqrt{1+s}} & 0 & 0\\
0 & \frac{1}{\sqrt{1+s}} & 0\\
0 & 0 & 1+s \\
\end{pmatrix},
\label{eq:strain_cub_I_mode1}
\end{aligned}
\end{equation}
%%%%%%%%%%%%%%%%%%%%%%%%%%%
which is not completely consistent with the parameterized deformation gradients given by Eqs.\ref{eq:F_eq_cub_alpha2a} and \ref{eq:F_eq_cub_alpha2b},  derived from the theoretical equilibrium strain tensor $\boldsymbol{\epsilon}^{eq}(\boldsymbol{\alpha})$. As a result, the deformation gradient implemented in -mode 1 of MAELAS generates a set of deformed cells that do not contain the exact equilibrium state when the magnetization is constrained along the direction $\boldsymbol{\alpha}$. Consequently, it gives approximated values of the equilibrium lengths $l^{approx}_1$ and $l^{approx}_2$ along the direction $\boldsymbol{\beta}$ when magnetization points to direction $\boldsymbol{\alpha}_1$ and $\boldsymbol{\alpha}_2$, respectively. 

\begin{figure}[h!]
\centering
\includegraphics[width=0.75\columnwidth ,angle=0]{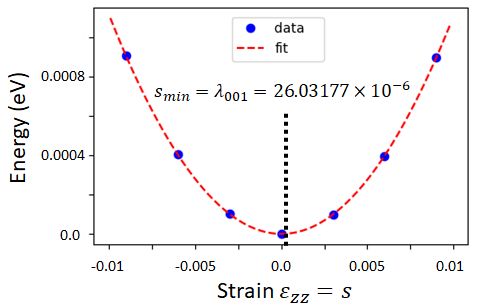}
\caption{Blue points correspond to the energy for the deformed unit cells with magnetization direction $\boldsymbol{\alpha}=(0,0,1)$ generated with the deformation gradient given by  Eq.\ref{eq:F_eq_cub_alpha2a}. The energy is evaluated using the theoretical formula for the cubic (I) symmetry (Eq.\ref{eq:E_exact}) with the elastic and magnetoelastic constants in Table \ref{tab:accuracy}. Red line is the fitting curve to a cubic polynomial. The minimum of this function corresponds to $\lambda_{001}$.}
\label{fig:strain_opt}
\end{figure}

Interestingly, this analysis reveals a potential alternative approach to the length optimization method that requires only one magnetization direction to compute $\lambda_{001}$, and without using any approximation. Namely, after a cell relaxation, a set of deformed unit cells are generated with the deformation gradient given by Eq.\ref{eq:F_eq_cub_alpha2a}. Next, the energy of these unit cells are evaluated constraining the direction of magnetization to $\boldsymbol{\alpha}=(0,0,1)$. The calculated energies versus the parameter $s$ used in the deformation gradient are fitted to a cubic polynomial (according to our preliminary tests using the exact energy Eq.\ref{eq:E_exact}, a quadratic polynomial may not be sufficient for $s\approx0.01$)
%%%%%%%%%%%%%%%%%%%%%%%%%%%%%
\begin{equation}
E(s)=A s^3 + B s^2 +C s +D,
\label{eq:fit_poli}
\end{equation}
%%%%%%%%%%%%%%%%%%%%%%%%%%%
where $A$, $B$, $C$ and $D$ are fitting parameters. The value of $s$ at the local minimum of this function ($E_{min}=E(s_{min})$) corresponds to the magnetostrictive coefficient $\lambda_{001}$
%%%%%%%%%%%%%%%%%%%%%%%%%%%%%
\begin{eqnarray}
\frac{\partial E}{\partial s} & = & 0 \: \xrightarrow[]{} \: \lambda_{001}=s_{min}=\frac{-B\pm \sqrt{B^2-3AC}}{3A}.
\label{eq:fit_poly_sol1}
\end{eqnarray}
%%%%%%%%%%%%%%%%%%%%%%%%%%%
Since the local minimum of the cubic polynomial is the critical point where the second derivative is positive, then $\lambda_{001}$ is equal to the solution in Eq.\ref{eq:fit_poly_sol1} that meets the condition
%%%%%%%%%%%%%%%%%%%%%%%%%%%%%
\begin{eqnarray}
\frac{\partial^2 E}{\partial s^2} & > & 0 \: \xrightarrow[]{} \: 6As_{min}+2B>0.
\label{eq:fit_poly_sol2}
\end{eqnarray}
%%%%%%%%%%%%%%%%%%%%%%%%%%%
In Fig.\ref{fig:strain_opt} we show the calculation of $\lambda_{001}$ with this approach. Here, the energy is  evaluated using the theoretical formula for the cubic (I) symmetry (Eq.\ref{eq:E_exact}) with the elastic and magnetoelastic constants in Table \ref{tab:accuracy}. The relative error in the calculation of $\lambda_{001}$ is $\lambda_{001}^{Rel.Error}=-0.00009\%$ which comes purely from the fitting procedure since no approximations are used in this method. Since the calculation of the magnetostrictive coefficient is done through an optimization of the strain (the magnetostrictive coefficient is equal to the strain that minimizes the energy), then we can call this approach as the strain optimization method. Currently, we are verifying the reliability of this method with first principle calculations. Thus, we have not implemented it in version 2.0 of MAELAS.

%------------------------------
%------------------------------

\subsection{Hexagonal (I)}

The calculation of $\lambda^{\alpha2,2}$ in -mode 1 is performed using the following deformation gradient for the length optimization along $\boldsymbol{\beta}=(0,0,1)$ of both magnetization directions $\boldsymbol{\alpha}_1=(0,0,1)$ and $\boldsymbol{\alpha}_2=(1,0,0)$ \cite{maelas_publication2021}
%%%%%%%%%%%%%%%%%%%%%%%%%%%%%
\begin{equation}
\begin{aligned}
\boldsymbol{F}\Big\vert_{\boldsymbol{\alpha}_1=(0,0,1)}^{\lambda^{\alpha2,2}} (s) = \boldsymbol{F}\Big\vert_{\boldsymbol{\alpha}_2=(1,0,0)}^{\lambda^{\alpha2,2}} (s) =\begin{pmatrix}
\frac{1}{\sqrt{1+s}} & 0 & 0\\
0 & \frac{1}{\sqrt{1+s}} & 0\\
0 & 0 & 1+s \\
\end{pmatrix}.
\label{eq:strain_hex_I}
\end{aligned}
\end{equation}
%%%%%%%%%%%%%%%%%%%%%%%%%%%
As in the cubic case, this deformation gradient generates a set of deformed cells that do not contain the exact equilibrium state when the magnetization is constrained along the direction $\boldsymbol{\alpha}$. Consequently, it gives approximated values of the equilibrium lengths $l_1$ and $l_2$ along the direction $\boldsymbol{\beta}$ when magnetization points to direction $\boldsymbol{\alpha}_1$ and $\boldsymbol{\alpha}_2$, respectively.

\begin{figure}[h!]
\centering
\includegraphics[width=0.7\columnwidth ,angle=0]{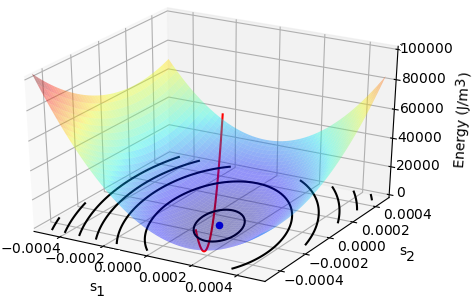}
\caption{Energy of the deformed unit cells with magnetization direction $\boldsymbol{\alpha}_1=(0,0,1)$ generated with the deformation gradients given by  Eq.\ref{eq:strain_hex_I} (red line) and Eq.\ref{eq:F_eq_hex_alpha_s} (2D surface). The energy is evaluated using the theoretical formulas for the hexagonal (I) symmetry (Eq.\ref{eq:E_exact}) with the elastic and magnetoelastic constants in Table \ref{tab:accuracy}. Black lines are the projected isoenergy contours of the 2D surface. The blue point stands for the equilibrium energy for the deformed cells generated with Eq.\ref{eq:F_eq_hex_alpha_s}, where the free parameters $s_1$ and $s_2$ are equal to the magnetostrictive coefficients $s_1=\lambda^{\alpha 1,2}=95.0656\times10^{-6}$ and $s_2=\lambda^{\alpha 2,2}=-125.9316\times10^{-6}$, see Table \ref{tab:accuracy}.}
\label{fig:F_s2}
\end{figure}

Again, one needs to make use of the equilibrium magnetoelastic strain tensor $\boldsymbol{\epsilon}^{eq}(\boldsymbol{\alpha})$ in order to find the parameterized deformation gradient that contains the exact equilibrium state when the magnetization is constrained along the direction $\boldsymbol{\alpha}$. The anisotropic part of the equilibrium strain tensor for Hexagonal (I) systems reads \cite{maelas_publication2021,CLARK1980531}
%%%%%%%%%%%%%%%%%%%%%%%%%%%%%%%%%%%%%%%%%%%%
\begin{equation}
\begin{aligned}
& \boldsymbol{\epsilon}^{eq}(\boldsymbol{\alpha}) = \\
& 
\begin{pmatrix}
\lambda^{\alpha 1,2}\left(\alpha_z^2-\frac{1}{3}\right)+ \frac{\lambda^{\gamma,2}}{2}\left(\alpha_x^2-\alpha_y^2\right)& \lambda^{\gamma,2}\alpha_x\alpha_y & \lambda^{\epsilon,2}\alpha_x\alpha_z\\
\lambda^{\gamma,2}\alpha_y\alpha_x & \lambda^{\alpha 1,2}\left(\alpha_z^2-\frac{1}{3}\right)- \frac{\lambda^{\gamma,2}}{2}\left(\alpha_x^2-\alpha_y^2\right) & \lambda^{\epsilon,2}\alpha_y\alpha_z\\
\lambda^{\epsilon,2}\alpha_z\alpha_x & \lambda^{\epsilon,2}\alpha_z\alpha_y & \lambda^{\alpha 2,2}\left(\alpha_z^2-\frac{1}{3}\right)\\
\end{pmatrix}.
\label{eq:strain_eq_hex}
\end{aligned}
\end{equation}
%%%%%%%%%%%%%%%%%%%%%%%%%%%%%%%%%%%%%%%%%%%%
In -mode 1, for the calculation of $\lambda^{\alpha2,2}$ we selected the magnetization directions $\boldsymbol{\alpha}_1=(0,0,1)$ and $\boldsymbol{\alpha}_2=(1,0,0)$ \cite{maelas_publication2021}. Hence, the equilibrium strain tensor becomes
%%%%%%%%%%%%%%%%%%%%%%%%%%%%%%%%%%%%%%%%%%%%
\begin{equation}
\begin{aligned}
& \boldsymbol{\epsilon}^{eq}(0,0,1) = 
\begin{pmatrix}
\frac{2}{3}\lambda^{\alpha 1,2} & 0 & 0\\
0 & \frac{2}{3}\lambda^{\alpha 1,2}  & 0\\
0 & 0 & \frac{2}{3}\lambda^{\alpha 2,2}\\
\end{pmatrix},\\
& \boldsymbol{\epsilon}^{eq}(1,0,0) = 
\begin{pmatrix}
-\frac{1}{3}\lambda^{\alpha 1,2}+\frac{1}{2}\lambda^{\gamma,2} & 0 & 0\\
0 & -\frac{1}{3}\lambda^{\alpha 1,2} -\frac{1}{2}\lambda^{\gamma,2} & 0\\
0 & 0 & -\frac{1}{3}\lambda^{\alpha 2,2}\\
\end{pmatrix}.
\label{eq:strain_eq_hex_alpha}
\end{aligned}
\end{equation}
%%%%%%%%%%%%%%%%%%%%%%%%%%%%%%%%%%%%%%%%%%%%
The corresponding deformation gradients that give these equilibrium strain tensors are obtained using Eq. \ref{eq:strain_deform0}
%%%%%%%%%%%%%%%%%%%%%%%%%%%%%%%%%%%%%%%%%%%%
\begin{equation}
\begin{aligned}
& \boldsymbol{F}\Big\vert_{\boldsymbol{\alpha}_1=(0,0,1)}^{\lambda^{\alpha2,2}} = 
\begin{pmatrix}
\frac{2}{3}\lambda^{\alpha 1,2} +1 & 0 & 0\\
0 & \frac{2}{3}\lambda^{\alpha 1,2} +1 & 0\\
0 & 0 & \frac{2}{3}\lambda^{\alpha 2,2}+1\\
\end{pmatrix},\\
& \boldsymbol{F}\Big\vert_{\boldsymbol{\alpha}_2=(1,0,0)}^{\lambda^{\alpha2,2}}  = 
\begin{pmatrix}
-\frac{1}{3}\lambda^{\alpha 1,2}+\frac{1}{2}\lambda^{\gamma,2}+1 & 0 & 0\\
0 & -\frac{1}{3}\lambda^{\alpha 1,2} -\frac{1}{2}\lambda^{\gamma,2} +1& 0\\
0 & 0 & -\frac{1}{3}\lambda^{\alpha 2,2}+1\\
\end{pmatrix}.
\label{eq:F_eq_hex_alpha}
\end{aligned}
\end{equation}
%%%%%%%%%%%%%%%%%%%%%%%%%%%%%%%%%%%%%%%%%%%%
Therefore, if we parameterize these deformation gradients by setting the magnetostrictive coefficients as free parameters ($\lambda_i\xrightarrow{} s_i$), then they generate a set of deformed cells that contains the exact equilibrium state that corresponds to the particular case when the free parameters are equal to the magnetostrictive coefficients (minimum energy). Hence, these deformation gradients should be parameterized as 
%%%%%%%%%%%%%%%%%%%%%%%%%%%%%%%%%%%%%%%%%%%%
\begin{eqnarray}
 \boldsymbol{F}\Big\vert_{\boldsymbol{\alpha}_1=(0,0,1)}^{\lambda^{\alpha2,2}}(s_1,s_2) & = &
\begin{pmatrix}
\frac{2}{3}s_1 +1 & 0 & 0\\
0 & \frac{2}{3}s_1 +1 & 0\\
0 & 0 & \frac{2}{3}s_2+1\\
\end{pmatrix},
\label{eq:F_eq_hex_alpha_s}
\\
\boldsymbol{F}\Big\vert_{\boldsymbol{\alpha}_2=(1,0,0)}^{\lambda^{\alpha2,2}}(s_1,s_2,s_3) & = &
\begin{pmatrix}
-\frac{1}{3}s_1+\frac{1}{2}s_2+1 & 0 & 0\\
0 & -\frac{1}{3}s_1 -\frac{1}{2}s_2+1& 0\\
0 & 0 & -\frac{1}{3}s_3+1\\
\end{pmatrix},
\label{eq:F_eq_hex_alpha_s2}
\end{eqnarray}
%%%%%%%%%%%%%%%%%%%%%%%%%%%%%%%%%%%%%%%%%%%%
where $s_1$, $s_2$ and $s_3$ are free parameters. We see that these deformation gradients are also diagonal matrices like the one implemented in -mode 1 given by Eq.\ref{eq:strain_hex_I} but with different diagonal elements. For instance, Eq.\ref{eq:strain_hex_I} is a particular case of Eq.\ref{eq:F_eq_hex_alpha_s} when the free parameters $s_1$ and $s_2$ are constrained to
%%%%%%%%%%%%%%%%%%%%%%%%%%%%%%%%%%%%%%%%%%%%
\begin{eqnarray}
s_1 & = & \frac{3}{2}\left(\frac{1}{\sqrt{1+s}}-1\right),\\
s_2 & = & \frac{3}{2}s,
\label{eq:s_1_s2}
\end{eqnarray}
%%%%%%%%%%%%%%%%%%%%%%%%%%%%%%%%%%%%%%%%%%%%
where $s$ is the single parameter used in Eq.\ref{eq:strain_hex_I}. In Fig.\ref{fig:F_s2} we show the energy of the deformed unit cells with magnetization direction $\boldsymbol{\alpha}_1=(0,0,1)$ generated with the deformation gradients given by  Eq.\ref{eq:strain_hex_I} (red line) and Eq.\ref{eq:F_eq_hex_alpha_s} (2D surface). In this figure we evaluated the energy using the theoretical formulas for the hexagonal (I) symmetry (Eq.\ref{eq:E_exact}) with the elastic and magnetoelastic constants in Table \ref{tab:accuracy}. The blue point in Fig.\ref{fig:F_s2} stands for the equilibrium energy for the deformed cells generated with Eq.\ref{eq:F_eq_hex_alpha_s}, where the free parameters $s_1$ and $s_2$ are equal to the magnetostrictive coefficients $s_1=\lambda^{\alpha 1,2}=95.0656\times10^{-6}$ and $s_2=\lambda^{\alpha 2,2}=-125.9316\times10^{-6}$, see Table \ref{tab:accuracy}. We observe that the minimum energy given by Eq.\ref{eq:strain_hex_I} (red line) is an approximation of the exact equilibrium state. Consequently, it gives an approximated value of the equilibrium length $l_1$ in the direction $\boldsymbol{\beta}=(0,0,1)$ when magnetization points to direction $\boldsymbol{\alpha}_1=(0,0,1)$.

%------------------------------
%------------------------------

Let's try to estimate the relative error in the calculation of $\lambda^{\alpha 2,2}$ induced by the deformation gradients Eqs. \ref{eq:F_eq_hex_alpha_s} and \ref{eq:F_eq_hex_alpha_s2}. At the minimum energy (equilibrium state), the free parameters in Eqs. \ref{eq:F_eq_hex_alpha_s} and \ref{eq:F_eq_hex_alpha_s2} are equal to the magnetostrictive coefficients ($s_i=\lambda_i$). Hence, these parameterized deformation gradients give the equilibrium lengths $l_1$ and $l_2$ along the direction $\boldsymbol{\beta}=(0,0,1)$ (via Eq. \ref{eq:deform_latt0}) 
%%%%%%%%%%%%%%%%%%%%%%%%%%%%%%%%%%%%%%%%%%%%
\begin{equation}
\begin{aligned}
l^{exact}_1 & = \left(\frac{2}{3}\lambda^{\alpha 2,2}+1\right)c_z,\\
l^{exact}_2 & = \left(-\frac{1}{3}\lambda^{\alpha 2,2}+1\right)c_z,
\label{eq:l1_l2}
\end{aligned}
\end{equation}
%%%%%%%%%%%%%%%%%%%%%%%%%%%%%%%%%%%%%%%%%%%%
where $c_z$ is the z-component of the relaxed (not deformed) lattice vector $\boldsymbol{c}=(0,0,c_z)$. Therefore, the formula used in -mode 1 to compute the magnetostrictive coefficient gives \cite{maelas_publication2021}
%%%%%%%%%%%%%%%%%%%%%%%%%%%%%%%%%%%%%%%%%%%%
\begin{equation}
\begin{aligned}
\lambda^{\alpha 2,2}_{approx}=\frac{2(l^{exact}_1-l^{exact}_2)}{\rho (l^{exact}_1+l^{exact}_2)}=\frac{2\lambda^{\alpha 2,2}}{\frac{1}{3}\lambda^{\alpha 2,2}+2},
\label{eq:l1_l2_solve}
\end{aligned}
\end{equation}
%%%%%%%%%%%%%%%%%%%%%%%%%%%%%%%%%%%%%%%%%%%%
where $\rho=1$. The relative error using the exact value   in Table \ref{tab:accuracy} ($\lambda^{\alpha 2,2}=-125.9316\times10^{-6}$) is
%%%%%%%%%%%%%%%%%%%%%%%%%%%%%%%%%%%%%%%%%%%%
\begin{equation}
\begin{aligned}
\lambda^{\alpha 2,2}_{Rel.Error}=\frac{\lambda^{\alpha 2,2}-\lambda^{\alpha 2,2}_{approx}}{\lambda^{\alpha 2,2}}\times 100 = -0.002\%,
\label{eq:l1_l2_solve_error}
\end{aligned}
\end{equation}
%%%%%%%%%%%%%%%%%%%%%%%%%%%%%%%%%%%%%%%%%%%%
which is clearly much smaller than in the example shown in Section \ref{section:accuracy} using the  deformation gradient with a single parameter implemented in -mode 1,  Eq.\ref{eq:strain_hex_I} ($\sim 17\%$). Hence, for hexagonal crystals deformation gradients with more than one parameter based on the equilibrium strain tensor could improve the accuracy of -mode 1 since they give more accurate values for the equilibrium lengths $l_1$ and $l_2$. Unfortunately, they will also increase the number of deformed cells needed to calculate the magnetostrictive coefficients through the sampling of the parameters in the deformation gradients, making this approach more computationally demanding. In Section \ref{section:test} we see that the implemented deformation gradients with a single parameter in -mode 1 still provide reasonable results  (similar to -mode 2) in many cases. Although, in general, we recommend to use -mode 2.

%%%%%%%%%%%%%%%%%%%%%%%%%%%%%%%%%%%%%%%



%% References


\bibliographystyle{elsarticle-num}
\bibliography{mybibfile.bib}






% \end{thebibliography}

\end{document}